\documentclass[useAMS,usenatbib,aas_macros]{mn2e}
\usepackage{aas_macros}
\usepackage{amsmath}

\usepackage{color}

\def\rf{\par\vspace{2pt plus 1pt minus 1pt}\noindent \hangindent 13pt}
\def\rff{\par\vspace{2pt plus 1pt minus 1pt}\noindent \hangindent 9pt}

\newcommand{\equ}[1]{eq.~(\ref{eq:#1})}
\newcommand{\equs}[1]{eqs.~(\ref{eq:#1})}
\newcommand{\equm}[1]{(\ref{eq:#1})}
\newcommand{\Equ}[1]{Eq.~(\ref{eq:#1})}

\newcommand{\equnp}[1]{eq.~\ref{eq:#1}}
\newcommand{\equsnp}[1]{eqs.~\ref{eq:#1}}
\newcommand{\equmnp}[1]{\ref{eq:#1}}
\newcommand{\se}[1]{\S\ref{sec:#1}}
\newcommand{\fig}[1]{Fig.~\ref{fig:#1}}

\newcommand{\Fig}[1]{Figure~\ref{fig:#1}}

\newcommand{\tab}[1]{Table~\ref{tab:#1}}
\newcommand{\be}{\begin{equation}}
\newcommand{\ee}{\end{equation}}
\newcommand{\ba}{\begin{align}}
\newcommand{\ea}{\end{align}}
\newcommand{\bad}{\begin{equation} \begin{aligned}}
\newcommand{\ead}{\end{aligned} \end{equation}}
\newcommand{\bea}{\begin{eqnarray}}
\newcommand{\eea}{\end{eqnarray}}

\def\la{\langle}
\def\ul{\underline}

\newcommand{\no}{\noindent}

\newcommand{\msun}{M_\odot}
\newcommand{\Msun}{M_\odot}

\newcommand{\ifm}[1]{\relax\ifmmode#1\else$\mathsurround=0pt #1$\fi}
\newcommand{\kms}{\ifmmode\,{\rm km}\,{\rm s}^{-1}\else km$\,$s$^{-1}$\fi}

\newcommand{\Mpc}{\,{\rm Mpc}}
\newcommand{\kpc}{\,{\rm kpc}}
\newcommand{\pc}{\,{\rm pc}}
\newcommand{\cm}{\,{\rm cm}}
\newcommand{\Gyr}{\,{\rm Gyr}}

\newcommand{\Myr}{\,{\rm Myr}}
\newcommand{\kyr}{\,{\rm kyr}}
\newcommand{\yr}{\,{\rm yr}}
\newcommand{\erg}{\,{\rm erg}}
\newcommand{\ergs}{\,{\rm erg}\,{\rm s}^{-1}}

\newcommand{\cmc}{\,{\rm cm}^{-3}}
\newcommand{\cms}{\,{\rm cm}^{-2}}

\newcommand{\ltsima}{$\; \buildrel < \over \sim \;$}
\newcommand{\lsim}{\lower.5ex\hbox{\ltsima}}
\newcommand{\gtsima}{$\; \buildrel > \over \sim \;$}
\newcommand{\gsim}{\lower.5ex\hbox{\gtsima}}
\newcommand{\prop}{\propto}

\newcommand{\rar}{\rightarrow}

\def\omm{\Omega_{\rm m}}
\def\oml{\Omega_{\Lambda}}
\def\omb{\Omega_{\rm b}}

\def\Mv{M_{\rm v}}

\def\Rv{R_{\rm v}}

\def\Mg{M_{\rm g}}
\def\Ms{M_{\rm s}}
\def\Re{R_{\rm e}}

\def\Sig1{\Sigma_1}

\def\Vsn{V_{\rm SN}}

\def\Rd{R_{\rm d}}
\def\Hd{H_{\rm d}}

\def\Mc{M_{\rm c}}
\def\Rc{R_{\rm c}}

\def\Vc{V_{\rm c}}

\def\Rf{R_{\rm f}}

\def\brho{\bar\rho}

\def\Msfdot{\dot{M}_{\rm sf}}

\def\Rs{R_{\rm s}}
\def\Vs{V_{\rm s}}
\def\Vs{V_{\rm s}}

\def\Ts{T_{\rm s}}

\def\Ssfr{\Sigma_{\rm sfr}}

\def\Smol{\Sigma_{\rm g}}
\def\drhos{\rho_{\rm sfr}}
\def\rhosfr{\rho_{\rm sfr}}
\def\rhog{\rho_{\rm g}}
\def\rhomol{\rho_{\rm g}}
\def\epsf{\epsilon_{\rm ff}}
\def\eps2{\epsilon_{-2}}
\def\mp{m_{\rm p}}
\def\tff{t_{\rm ff}}
\def\tdep{t_{\rm dep}}

\def\Rc{R_{\rm c}}
\def\tc{t_{\rm c}}
\def\Vc{V_{\rm c}}
\def\Tc{T_{\rm c}}
\def\Rf{R_{\rm f}}
\def\tf{t_{\rm f}}
\def\msn{\mu_{\rm sn}}
\def\SFR{{\rm SFR}}
\def\sfr{{\rm SFR}}

\def\fh{f}
\def\f0{f_0}
\def\nsn{n_{\rm sn}}
\def\Vsn{V_{\rm sn}}
\def\Rs{R_{\rm s}}
\def\Vs{V_{\rm s}}
\def\L{L}
\def\tb{t_{\rm b}}
\def\ts{t_{\rm s}}
\def\cs{c_{\rm s}}
\def\e51{e_{51}}
\def\l38{\ell_{38}}
\def\n0{n_0}
\def\slope{s}
\def\sloc{s_{\rm loc}}

\def\Rst{R_{\rm st}}
\def\tst{t_{\rm st}}
\def\fst{f_{\rm st}}
\def\Sst{S_{\rm st}}

\def\tlf{t_{*}}
\def\tlff{t_{*,{\rm 5}}}

\title[The global SFR law by SN feedback]
{The Global Star-Formation Law by Supernova Feedback}

\author[A. Dekel et al.]
{\parbox[t]{\textwidth}
{Avishai Dekel$^{1,2}$\thanks{E-mail: dekel@huji.ac.il},
Kartick C. Sarkar$^{1}$,
Fangzhou Jiang$^{1}$,
Frederic Bournaud$^{3}$,
Mark R. Krumholz$^{4}$,
Daniel Ceverino$^{5,6}$,
Joel R. Primack$^{7}$
}
\\ \\
$^1$Racah Institute of Physics, The Hebrew University, Jerusalem 91904 Israel\\
$^2$SCIPP, University of California, Santa Cruz, 1156 High Street,
Santa Cruz, CA 95064, USA\\ 
$^3$Laboratoire AIM Paris-Saclay, CEA/IRFU/SAp, Universite Paris Diderot, 91191, Gif-sur-Yvette Cedex, France\\
$^4$Research School of Astronomy and Astrophysics, Australian National 
University, Canberra, ACT 2612, Australia\\
$^5$Cosmic Dawn Center (DAWN)\\ 
$^6$Niels Bohr Institute, University of Copenhagen, Vibenshuset, Lyngbyvej 2, 
2100 Copenhagen, Denmark\\
$^7$Physics Department, University of California, Santa Cruz, 1156 High Street, Santa Cruz, CA 95064, USA
}

\begin{document}

\large

\pagerange{\pageref{firstpage}--\pageref{lastpage}} \pubyear{2002}

\maketitle

\label{firstpage}

\begin{abstract}
We address a simple model where the Kennicutt-Schmidt (KS) relation between the
macroscopic densities of star-formation rate (SFR, $\rhosfr$) and gas ($n$) in 
galactic discs emerges from self-regulation of the SFR via supernova feedback.  
It arises from the physics of supernova bubbles, insensitive to the microscopic
SFR recipe and not explicitly dependent on gravity. The key is that the filling
factor of SFR-suppressed supernova bubbles self-regulates to a constant, 
$\fh \sim 0.5$. Expressing the bubble fading radius and time in terms of $n$, 
the filling factor is $\fh\prop S\,n^{-\slope}$ with $\slope\simeq 1.5$, 
where $S$ is the supernova rate density.
A constant $\fh$ thus refers to $\rhosfr\prop n^{1.5}$, with a 
density-independent SFR efficiency per free-fall time $\sim 0.01$.  
The self-regulation to $\fh \sim 0.5$ and the convergence to a KS relation 
independent of the local SFR recipe are demonstrated in cosmological and 
isolated-galaxy simulations using different codes and recipes. In parallel, the
spherical analysis of bubble evolution is generalized to clustered supernovae, 
analytically and via simulations, yielding $\slope \simeq 1.5\pm 0.5$.  
An analysis of photo-ionized bubbles about pre-supernova stars yields a range 
of KS slopes but the KS relation is dominated by the supernova bubbles. 
Superbubble blowouts may lead to an alternative self-regulation by outflows and
recycling.  While the model is over-simplified, its simplicity and validity in 
the simulations may argue that it captures the origin of the KS relation.
\end{abstract}


%%%%%%%%%%%%%%%%%%%%
\section{Introduction}
\label{sec:intro}

% The global KS Relation
The {\it global}\, Kennicutt-Schmidt relation commonly refers to the observed 
correlation between the surface densities of star formation rate ($\Ssfr$) 
and molecular gas ($\Smol$), either in galactic discs as a whole or in 
macroscopic regions within the discs 
\citep{kennicutt98_araa,daddi10_ks}. 
The quantities are averaged on scales larger than the disc height, 
namely from $\sim 100\pc$ to several kiloparsecs, 
where the average gas number densities are $n \sim 1 \cmc$.
The global relation is typically 
$\Ssfr \prop \Smol^\slope$,
with the slope ranging about $\slope \simeq 1.5$ in the range $1-2$, as
summarized in \se{obs}.
%a local
This global relation may or may not be related to the {\it local}, microscopic
relation between the densities on the scales of the star-forming regions, 
typically smaller than $10\pc$, 
where the number densities are $n \sim 10^{2-4}\cmc$.

\smallskip %micro 3D
Different galaxy types and sub-galactic regions, at different redshifts and
environments and on different scales, may show somewhat different KS 
relations, which makes the overall compiled relation look rather loose. 
However, it has been demonstrated \citep{kdm12} that the local correlation 
becomes particularly tight and universal once $\Smol$ is replaced
by $\Smol/\tff$, where $\tff$ is the proper free-fall time in 
the relevant star-forming regions.
They argued that this is consistent with a universal local 3D star-formation 
law,
\be
\drhos = \epsf \frac{\rhomol}{\tff} \, ,
\label{eq:ks0}
\ee
where $\drhos$ is the star-formation rate (SFR) density
and $\rhomol$ is the local molecular gas density, averaged over
the star-forming molecular cloud. 

\smallskip % eps_micro
The microscopic SFR efficiency, $\epsf$, appears to be constant, 
independent of density, 
at a value on the order of $\epsf \sim 0.01$.
One line of evidence for this comes from direct measurements in individual 
resolved clouds
%\citep{kdm12,evans14,salim15,vuti16,heyer16,leroy17}.
\citep{kdm12,salim15,heyer16,vuti16,leroy17}.
The second line of evidence comes from the correlation of SFR with
HCN luminosity, which constrains $\epsf$ because HCN emission comes from gas at
a known density
\citep{krum_tan07,garcia12,usero15,onus18}.\footnote{One should note, however, 
that there are conflicting indications 
for variations in $\epsf$ \citep{murray11,lee16}, which are argued to be due to 
a bias in the methodology \citep{leroy17,krum18}.}

\smallskip % KS quantitative
To make \equ{ks0} more quantitative, 
the free-fall time can be expressed in terms of the density as 
$\tff = [32\, G \rho / (3\pi)]^{-1/2}$,
where we assume that the density is dominated by the gas.
With a nucleon number density 
$n=n_H=\rho/(\mu_{H} \mp)$ (we adopt $\mu_H = 1.355$ for Solar metallicity),
and denoting $n= 1\, \cmc\, \n0$
and $\epsf = 0.01 \eps2$,
\equ{ks0} becomes
\be
\drhos \simeq 0.66 \times 10^{-2} \msun \yr^{-1} \kpc^{-3}\, \eps2\, \n0^{3/2} 
\, .
\label{eq:ks}
\ee
It seems that this kind of relation can be extrapolated from the small 
scales of the 
star-forming clouds to larger macroscopic scales, and it can be considered 
as the macroscopic KS relation in 3D, the origin of which we wish to 
understand. 
In particular, we wish to figure out the origin of the slope 
$\slope \simeq 1.5$ in the macroscopic relation
\be
\drhos \prop n^{\slope} \, ,
\label{eq:kss}
\ee
which is equivalent to asking why the global $\epsf$ is not varying as a
function of $n$. 
We also wish to figure out the origin of the amplitude of the macroscopic
relation, namely what determines the global value of $\epsf \sim 0.01$, 
which seems to resemble the local value.

\smallskip % bottom-up
There are in general two types of theoretical attempts to understand the KS 
relation \citep[see][]{krum17}.
The bottom-up approach, applied to the local SFR law,
is based on the expected probability distribution function (PDF) of gas
densities in molecular clouds and the assumption that stars form only above a
constant threshold density 
\citep[e.g.][]{padoan14,burkhart18}. 
The characteristic density PDF in supersonic turbulence, largely based on 
simulations, is expected to be a lognormal distribution, while
self-gravity is expected to generate a power-law high-density tail.
Integrating above the threshold density yields the microscopic $\epsf$,
which may be subject to an uncertainty in the value and constancy of the
threshold density.
 
\smallskip % top-down
The top-down approach, applied macroscopically, attempts to model the KS 
relation
as a result of self-regulation by the combined effects of gravity, accretion, 
star formation and feedback.
These models may attempt to study the evolution of molecular clouds in a 
realistic inter-stellar medium.   
They commonly consider the balance between the 
momentum provided by feedback and the vertical self-gravity. 
They address gravitational disk instability and the driving of turbulence by 
feedback, by internal inflow in the disc driven by disk instability, 
and by accretion into the disk
\citep[e.g.][]{dsc09,ostriker11,faucher13,krum18}.
These models typically assume a form for the SFR law, e.g., \equ{ks0}
with $\epsf$ a constant independent of density, but they usually do not 
attempt to explain why it is so.  

\smallskip %Summary of our goal
Here we explore the possibility that the macroscopic 
KS relation is naturally driven by
self-regulation of the SFR by supernova (SN) feedback,
insensitive to the specific small-scale star-formation 
recipe, and not even involving gravity in an explicit way. 
The simple key hypothesis is that the mass filling factor of the gas 
in which the SFR is suppressed by feedback self-regulates to a 
constant value on the order of $\fh \sim 0.5$.
The SFR is suppressed (boosted) when $\fh$ exceeds (falls short of) this 
attractor value.
We assume that a proxy for this filling factor is the volume filling factor
of hot gas in supernova bubbles when they fade away.
If the final bubble radius and fading time can be expressed as
power laws of the ISM gas density $n$, then the hot volume filling factor can 
be expressed as
\be
\fh \prop S\, n^{-\slope} \, ,
\ee
where $S$ is the SN rate density (e.g. \se{filling}). 
In this case a constant $\fh$ would automatically imply a macroscopic 
star-formation law of the desired form in \equ{kss},
\be
\drhos \prop S \prop n^{\slope} \, .
\ee
Indeed, as we will see in \se{single}
based on the standard evolution of isolated single-SN bubbles 
\citep[e.g.,][chapter 39]{draine11}, the predicted slope is $\slope=1.48$,
suspiciously close to the macroscopic star-formation law, \equ{kss}.
Seed ideas along similar lines have
been proposed in earlier work \citep{ds86,mckee77,silk97}.

\smallskip % Hopkins
The weak dependence of the global KS relation on the assumed local SFR recipe
was indicated in hydro simulations \citep{hopkins11,hopkins13}. 
Here we explore how this self-regulation is materialized through the 
hypothesis that the hot volume filling factor of SN bubbles is self-regulated 
into a constant value.
This key hypothesis will be tested below using isolated and cosmological
simulations of galaxies, confirming the insensitivity of the global relation
to the local SFR recipe and the dominant role of SN feedback in it.
In parallel, we will analytically compute the hot filling factor as a function 
of ISM density for different sequences of co-local SNe in star-forming clusters,
which for a constant hot filling factor will provide predictions for
the KS relation. These analytic results will be tested and refined using simple 
spherical simulations.

\smallskip % assumptions
Our analytic modeling makes several simplifying assumptions, including the
following
(to be discussed below, especially in \se{ionization} and \se{disc}):
\rff $\bullet$ 
The medium outside the SN bubbles is uniform, ignoring
the complexities associated with the multi-phased ISM
and the origin of molecular hydrogen for star formation.
\rff $\bullet$
Supernova feedback is negative, such that SFR is suppressed in the gas that
has been swept by the SN bubbles (discussed in \se{positive}).
\rff $\bullet$ 
The bubbles are largely confined to the galactic disc, while the possible
effects of super-bubble blow-out are discussed in \se{blowout}.
\rff $\bullet$
The SN bubbles overwhelm the photo-ionized bubbles about the pre-SN O/B stars. 
This is argued analytically in \se{ionization}, and demonstrated in simulations
in \se{iso}.

\smallskip
The paper is organized as follows.
In \se{single} we address the idealized case of randomly distributed single
SNe, where in \se{SNR} we summarize the standard evolution of a single SN,
and in \se{filling} we introduce the concept of self-regulated hot filling
factor, compute it for single SN bubbles and derive the KS relation.
In \se{cosmo} we test the validity of self-regulation into a constant hot 
filling factor using ART hydro-gravitational simulations of discs 
in a cosmological setting. The simulations are elaborated on in \se{app_vela}. 
In \se{iso}, using RAMSES simulations of isolated galaxies, 
we reproduce the self-regulation to a constant filling factor and the
generation of a global KS relation, and show that it is insensitive to the
local SFR recipe and is determined by SN feedback. 
In \se{cluster} we return to analytic modeling, 
derive the evolution of a co-local multiple SNe in different time sequences, 
and obtain the associated filling factor and KS relation.
In \se{spheri_sim} we use spherical simulations to test and modify 
the analytic predictions for such clustered SNe. 
In \se{ionization} we address the alternative of photo-ionized bubbles.
In \se{disc} we discuss our modeling,
where 
in \se{positive} we address the assumption of negative feedback,
in \se{molecule} we comment on the dominance of molecular hydrogen,
in \se{gravity} we refer to the relevance of self gravity,
and in \se{blowout} we comment on the effects of super-bubble blowout.
In \se{conc} we summarize our results and discuss our conclusions.

%%%%%%%%%%%%%%%%%%%%%%%%%%%
\section{KS Relation - Isolated Supernovae}
\label{sec:single}

\subsection{Standard SN-Bubble Evolution}
\label{sec:SNR}

We first summarize the standard evolution of single spherical SN bubbles
\citep[e.g.,][chapter 39]{draine11}.
One assumes that a supernova of energy $E=10^{51} \erg\, \e51$
explodes in a uniform medium of Hydrogen number density $n$
and with a speed of sound (or turbulence velocity dispersion) $c=10 \kms c_1$.

%-----------
\subsubsection{The Sedov-Taylor Phase and Cooling}
\label{sec:sedov}

After a free expansion phase, dominated by the mass of the ejecta, the SN
bubble enters the Sedov-Taylor adiabatic phase, where it is approximated as
a point explosion ejecting energy into a cold medium of uniform density, 
neglecting radiative losses, the mass of the ejecta and the pressure in the 
medium.
Based on dimensional analysis, the shock radius, velocity and temperature are
\be
\Rs = 1.15 \left( \frac{E t^2}{\rho} \right) ^{1/5}
= 5.1 \pc\, \e51^{1/5} \n0^{-1/5} t_3^{2/5} \, ,
\label{eq:Rs_Sedov}
\ee
\be
\Vs = \frac{2}{5} \frac{\Rs}{t} 
= 1950 \kms\, \e51^{1/5} \n0^{-1/5} t_3^{-3/5} \, ,
\label{eq:Vs_Sedov}
\ee
\be
\Ts = \frac{3}{16} \frac{\mu \mp}{k_{\rm B}} \Vs^2
=  5.3 \times 10^7 {\rm K}\, \e51^{2/5} \n0^{-2/5} t_3^{-6/5} \, ,
\label{eq:Ts_Sedov}
\ee
where the time is $t = 10^3 \yr\, t_3$.
The internal profiles within the bubble are assumed to obey the  
Sedov-Taylor similarity solution and are obtained numerically.

\smallskip
Cooling just behind the shock front eventually makes the bubble leave the
adiabatic phase and enter the radiative phase.
In order to estimate the cooling time $\tc$,
the cooling function in the relevant temperature range is idealized by
\be
\Lambda \simeq \lambda(Z)\, T_6^{-0.7} , \quad
\lambda(Z_\odot) \simeq 1.1\times 10^{-22} \ergs \cm^3 \, .
\label{eq:cooling}
\ee
This is a fair approximation for solar-metallicity gas, our fiducial case 
below, at temperatures in the range $10^5-10^{7.3}$K.
The cooling rate is obtained by spatial integration over the bubble,
\be
\dot{E}(t) = -\int_0^{\Rs(t)} \Lambda[T(r')]\, n_e(r') n_H(r')\, 
4\pi r'^2 dr' \, ,
\label{eq:Edot}
\ee
where $n_e = \rho/(\mu_e \mp)$ and $n_H = \rho/(\mu_H \mp)$
(with $\mu_e=1.15$ and $\mu_H=1.355$ for solar metallicity).
The integral over $T^{-0.7}(r)\,n^2(r)\,r^2$ is evaluated numerically for the 
Sedov-Taylor similarity solution.
The energy loss by time $t$ is 
\be
\Delta E(t) = \int_0^t \dot{E}(t')\, dt' \, .
\label{eq:DE}
\ee
The SN bubble ends its adiabatic phase, e.g., having lost one third 
of its energy to radiation, after a time $\tc$ when the shell is at a radius 
$\Rc$ given by
\be
\tc \simeq 4.93 \times 10^4 \yr\, \e51^{0.22}\, \n0^{-0.55} \, ,
\label{eq:tc}
\ee
\be
\Rc \simeq 24.4 \pc\, \e51^{0.29}\, \n0^{-0.42} \, .
\label{eq:Rc}
\ee
The shock velocity and temperature are then
\be
\Vc = 188 \kms\, \e51^{0.066} \n0^{0.13} \, ,
\label{eq:Vc}
\ee
\be
\Tc = 4.86 \times 10^5 {\rm K}\, \e51^{0.13} \n0^{0.26} \, .
\label{eq:Tc}
\ee
%\adr{The $n^{+0.13}$ in $\Vc$ is different from Drain's $n^{-0.13}$ in eq. 
%39.22.
%It comes from Sedov $\Vs \prop n^{-0.2} t^{-0.6}$ at $\tc \prop n^{-0.55}$,
%yielding a power $-0.2 +0.6 \times 0.55 = +0.13$ for $n$.}  

%--------
\subsubsection{The Snow-plow Phase and Fading}
\label{sec:snow-plow}

At $t>\tc$, after a significant fraction of the original SN energy has been
lost to radiation,
a dense shell of cold gas is pushed by the pressure of the enclosed hot 
central volume.
The mass of the dense shell increases as it sweeps up the ambient gas, and it
slows down accordingly,
\be
\Rs \simeq \Rs(\tc) \left( \frac{t}{\tc} \right)^{2/7} \, ,
\label{eq:Rs_snow}
\ee
\be
\Vs \simeq \frac{2}{7} \frac{\Rs}{t} \, .
\label{eq:Vs_snow}
\ee
At the beginning of the snow-plow phase $\Vs \sim 150 \kms$ propagating into a
medium of $T\lsim 10^4$K, namely it is a strong shock. 

\smallskip
The snow-plow phase fades away when $\Vs$ becomes comparable to the speed of
sound of the ambient medium $c$.
This occurs after a fading time $\tf$, leaving behind a  
bubble of fade-away radius $\Rf$,
\be
\tf \simeq 1.87 \Myr\, \e51^{0.32} c_{1}^{-7/5}\, \n0^{-0.37} \, ,
\label{eq:tf}
\ee 
\be
\Rf \simeq 69.0 \pc\, \e51^{0.32} c_{1}^{-2/5}\, \n0^{-0.37} \, .
\label{eq:Rf}
\ee
The mass affected by the SN bubble, 
which is mostly in the broadened, faded away 
shell, is the mass that was initially within the volume encompassed by $\Rf$. 
As discussed in \se{intro}, we assume that the SFR in the shell is suppressed 
by mechanical effects.

%-----------------
\subsubsection{Weak Dependence on Metallicity}

Slower cooling because of a lower metallicity $Z$ would make the final bubble 
larger.
For a lower metallicity, the factor $\lambda(Z)$ in \equ{cooling} is smaller.
For example, for $Z/Z_\odot \simeq 0.1$ the value drops by a factor of
$\simeq 7$ \citep[e.g.][Fig. 34.1]{draine11}.
In the expressions above, one obtains
$\tc\prop \lambda^{-1/3}$ and $\Rc \prop \lambda^{-2/15}$.
This leads to $\Rf \prop \tf \prop \lambda^{-0.053}$.
The volume of a bubble at fading is then $\Vsn \prop \lambda^{-0.21}$.
For example, if $Z = 0.1\, Z_\odot$, 
for which $\lambda \simeq 0.14\, \lambda(Z_\odot)$, 
the volume per SN bubble becomes larger by about 50\%.
This is a relatively weak dependence on metallicity, which we will not deal 
with here. 

\smallskip
Small changes in the power-law fit to the $T$ dependence of the cooling curve,
\equ{cooling}, lead to small changes in the power indices of $n$. For example,
\citet{ds86} assumed $\Lambda \prop T^{-1}$ (instead of $T^{-0.7}$), relevant 
for a gas with lower metallicity, and obtained at the end of the radiative 
phase a somewhat larger bubble with $\tc \prop n^{-0.53}$ and
$\Rc \prop n^{-0.41}$ compared to the powers of $-0.55$ and $-0.42$ in
\equ{tc} and \equ{Rc}.
These are again relatively small effects, which we ignore here.

%%%%%%%%%%%%%%%%%%%%%%
\subsection{Bubble Filling Factor and the KS Relation}
\label{sec:filling}

We now derive the KS relation from the bubble fading radius and time and their
dependence on $n$.

%-----------------
\subsubsection{Constant filling factor as an attractor}

Consider first the idealized case where the supernovae occur at random 
positions within a uniform galactic disc (of a constant height, say),
with a SN rate per unit volume $S= 10^{-4}\kpc^{-3} \yr^{-1}\, S_{-4}$. 

\smallskip
The volume filling factor of hot bubbles of radius $\Rf$ after a fade-away time
$\tf$, tentatively neglecting overlaps between bubbles, is
\be
\f0 = \Vsn \nsn = \frac{4\pi}{3} \Rf^3\, S\, \tf \, ,
\label{eq:filling}
\ee
where $\Vsn$ is the volume of each bubble and
$\nsn$ is the number density of pre-fading bubble centers.
Substituting $\Rf$ and $\tf$ from \equ{Rf} and \equ{tf} one obtains
\be 
\f0 = 0.24\, \e51^{1.26} c_{1}^{-2.6}\, S_{-4}\, \n0^{-1.48} .
\label{eq:f0_draine}
\ee
Thus, a filling factor of a fixed value of order one half 
defines a critical line in the $S\!-\!n$ diagram,
$S \prop n^{1.48}$.
This has been demonstrated to be approximately reproduced 
in simulations \citep{li15}.
For a system above that line, most of the ISM is ``hot" within the bubbles, 
where SFR is suppressed,
while below the line it is unperturbed, cold and available for star formation.

\smallskip
% key idea -- somewhat repetitive
Our main point here is that the natural proportionality of the SN rate and 
the SFR, for a given stellar initial mass function (IMF), 
makes this line an {\it attractor}. 
When above the line, most of the ISM is ``hot", 
the SFR is suppressed, so $S$ is suppressed and decreases down toward the 
critical line.
When below the line, most of the ISM is cold, the SFR is free to grow, so $S$
increases toward the critical line.
The galaxies are thus expected to populate a line
\be
\drhos \prop S \prop  n^{1.48} \, ,
\ee
reminiscent of the KS law.

%------------------------
\subsubsection{The KS Relation for isolated SNe}

In order to obtain the normalization of the
KS relation one should translate S to SFR density, given the IMF,
\be
\drhos = \msn S ,
\ee
where $\msn=100\, \mu_2\, \msun$ is the mass in forming stars that generate 
one supernova, namely the ratio of SFR to the SN rate.
We obtain from \equ{f0_draine}
\be
\drhos = 1.26 \times 10^{-2} \msun \yr^{-1} \kpc^{-3}
 \mu_2 \e51^{-1.26} c_{1}^{2.6} \f0\, \n0^{1.48} \, .
\label{eq:ks_single}
\ee
Using the Milky Way as an example, the speed of sound is $c \simeq 6 \kms$,
the global SFR is $\sfr \simeq 2 \msun\yr^{-1}$
\citep{chomiuk11,licquia15}
and the SN rate is $1/60 \yr^{-1}$, yielding $\msn \simeq 120 \msun$.
With $\e51 = 1$ and $\f0=0.5$ we obtain 
(for isolated SNe, ignoring overlaps)
\be
\drhos \simeq 0.66 \times 10^{-2} \msun \yr^{-1} \kpc^{-3}\, \n0^{1.48} \, .
\label{eq:ks_by_sn}
\ee

\smallskip
\Equ{ks_by_sn} coincides with \equ{ks} representing the observed KS relation for
$\epsf \simeq 0.016$.
This simple analysis predicts a universal relation of the form in \equ{ks0}
with a constant $\epsf$ independent of $n$ and a value in the observed
ball park, 
though with a non-negligible dependence on the sound speed in the ISM.

%-----------------------
\subsubsection{Estimated Correction for Overlaps}

\begin{figure}
\vskip 7.25cm
\includegraphics{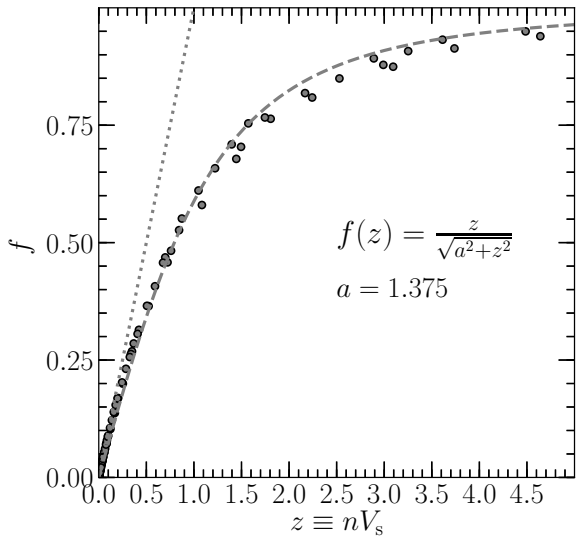}
%\special{psfile="figs/FillingFactor_vs_nVs.eps" hscale=60 vscale=60
%                     hoffset=5 voffset=-8}
\caption{
The actual filling factor $\fh$ as a function of the filling factor as derived 
by neglecting overlaps, $\f0=\nsn\Vsn$. The symbols are the results of
numerical experiments, and the fitting function is shown and quoted.
}
\label{fig:filling}
\end{figure}

The filling factor as computed above, $\f0=\nsn \Vsn$, ignored overlaps 
between bubbles.
If bubbles overlap, the actual volume filling factor $\fh$ is smaller.
At low filling factors $\fh$ approximately coincides with $\f0$, 
but at large filling factors $\f0$ may become a severe overestimate of the 
actual $\fh$.
In order to estimate $\fh(\f0)$, we tentatively consider a random distribution
of bubble centers in the volume. \Fig{filling} shows the result of a numerical
experiment, which is well fitted by the function
\be
\fh = (\phi^2+\f0^2)^{-1/2} \f0 \, , \quad \phi \simeq 1.375 \, .
\label{eq:overlap}
\ee
This gives for example $\f0 \simeq 0.60, 0.79, 1.03$
for $f=0.4, 0.5, 0.6$ respectively.
While $\fh$ is limited from above by unity, $\f0$ could in principle be larger
than unity.
With $\fh \sim 0.5$, the effect of overlap is on the order of 50\%,
keeping the filling factor in the same ball park.
\Equ{overlap} can be used to correct $\f0$ into $\fh$, especially for larger
filling factors.
Recall, however, that this is a crude approximation for single SNe at random
positions. In reality, the SNe are clustered (\se{cluster}), and the correction
for overlap between super-bubbles should be recalculated accordingly.

%1

%%%%%%%%%%%%%%%%%%%%%%%%%%%%%%%
\section{Cosmological Simulations: a Constant Hot Filling Factor}
\label{sec:cosmo}

Before generalizing the analytic estimates to cases of clustered SNe
and testing the analytic models with spherical simulations,
we turn to galactic discs in full hydro-gravitational simulations 
that incorporate star formation and supernova feedback, both in
cosmological simulations (this section) and in isolated galaxies (next section).
Our main goal here is to explore the validity of the key hypothesis
of self-regulation by feedback into a constant hot volume filling factor.
Using the isolated-galaxy simulations we will also explore the robustness to 
different local recipes for star formation, and the dominance of SN feedback.

%2
\begin{figure}
\vskip 11.9cm
\includegraphics{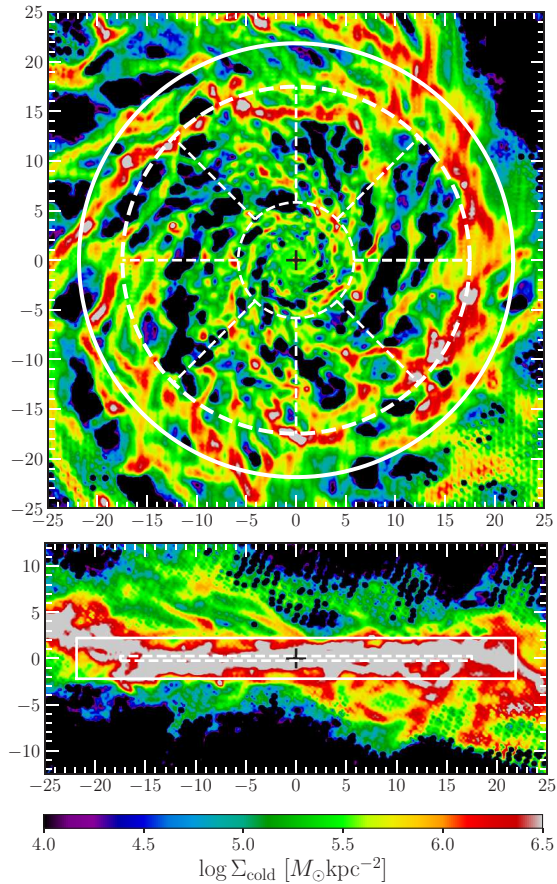} 
%\special{psfile="figs/g07_a0.5104_OneColumn.eps" 
%                   hscale=50 vscale=50 hoffset=10 voffset=-5} 
\caption{            
A cosmological-simulation VELA disc, V07 at $z=1$. 
The projected density of the cold gas is shown face-on (top) and edge-on 
(bottom). Distance in $\kpc$ is marked along the axes.
The cylindrical disc as defined in \citet{mandelker14}, with a radius $\Rd$ 
and half-height $\Hd$, is marked by the solid line,
while the volume selected for analysis here,  
a thin cylinder of radius $0.8\Rd$ and height $\pm 0.25\kpc$,
is marked by the dashed line.
Shown are the nine equal-area patches used when macroscopic sub-volumes are 
desired, consisting of eight patches covering the ring between $0.27\Rd$ and 
$0.8\Rd$ and one circular central patch of radius $0.27\Rd$.
}
\label{fig:patches}
\end{figure}

\smallskip %VELA
We first utilize the suite of 35 VELA zoom-in cosmological simulations.
Its relevant characteristics are mentioned here, while more details are
provided in \se{app_vela} and in references therein. 
The simulations are based on an Adaptive Refinement 
Tree (ART) code \citep{krav97,ceverino09}. The suite consists of 35 
galaxies that were evolved to $z\sim 1$, with a unique maximum spatial 
resolution ranging from $17.5$ to $35 \pc$ at all times. 
The dark-matter halo masses range from $10^{11}$ to $10^{12}\msun$ at $z=2$.   

\smallskip % SFR
The local SFR recipe allows stochastic star formation in grid cells where
the gas temperature is below $10^4$K and the gas density is above a threshold
of $1 \cmc$. 
If we attempt to approximate the local stochastic star formation recipe
by an expression of the sort $\drhos = \epsf \rhog/\tff$ (\equnp{ks0}), 
the efficiency would be $\epsf \sim 0.02$.
Being close to the desired global KS relation by construction through the local
SFR recipe,
these simulations by themselves do not explore 
the predicted insensitivity of the KS relation to the local SFR recipe.

\smallskip % SN feedback
Supernova feedback is implemented as a local injection of thermal energy 
\citep{ceverino09,cdb10,ceverino12}.
The energy from SN explosions (and stellar winds) is released at
a constant heating rate over the $40\Myr$ following the formation of the 
stellar particle, comparable to the age of the least massive star that 
explodes as a type-II, core collapse supernova. A velocity
kick of $\sim 10\kms$ is applied to $30\%$ of the newly formed stars to mimic
the effect of runaway from the densest region where the cooling is rapid.
The later effects of type-Ia supernovae are also included.
Naturally, the $\sim 25\pc$ grid does not resolve the main phases of the
SN-bubble evolution, making the treatment of SN feedback rather approximate.

\smallskip % rad feedback
In addition, radiation-pressure from massive stars is implemented at a 
moderate level with no infrared trapping \citep{ceverino14}.
This is incorporated through the addition of a non-thermal pressure term in
cells neighboring massive star particles younger than $5\Myr$ and whose column
densities exceed $10^{21} \cms$.

\smallskip % disk sampling
Galactic discs are selected for analysis from all the snapshots available 
in the redshift range $z=5.6-1$ in time intervals of $\sim 100\Myr$. 
The selection criterion for a disc is that the cold-gas ($T<3\times 10^4$K)
axial ratio is $\Rd/\Hd>4$, where $\Rd$ and $\Hd$ are the disc radius and 
half-height as defined in \citet{mandelker14}, see \se{app_vela}, 
yielding 25 galaxies with long periods as discs.
The radii and half-height $\Rd$ and $\Hd$ at $z=2$
span the ranges $2.5-12.6\kpc$ and $0.4-2.1\kpc$ respectively,
but these rather thick cylinders may include regions off the main bodies of the
discs, given that many discs are warped or asymmetric.
The analysis here is conservatively confined to central thin cylinders of 
radii $0.8\Rd$ and height $\pm 0.25\kpc$ as representing the main bodies of the
gas discs (\fig{patches}). 
When macroscopic sub-volumes are desired, 
each disc is divided into nine patches as in \fig{patches}, consisting of
eight equal orthogonal patches covering the ring between $0.27\Rd$ and 
$0.8\Rd$ and one circular central patch of radius $0.27\Rd$ such that it 
has the same area as the other patches.

%3
\begin{figure}
\vskip 6.8cm
\includegraphics{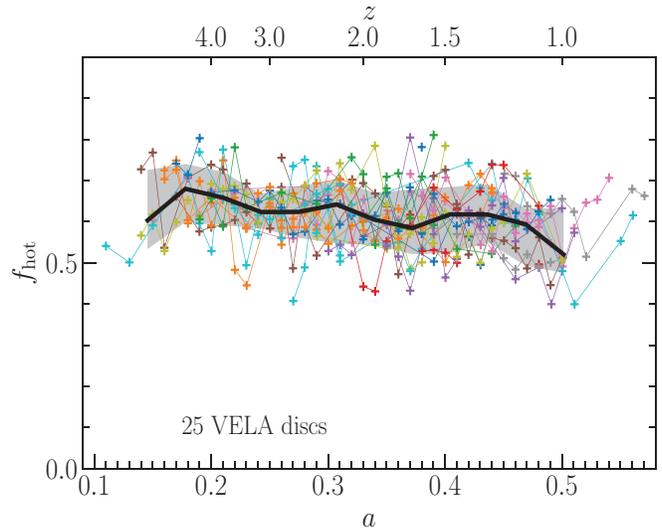}
%\special{psfile="figs/fhot_Rdhd4_line.eps" hscale=51 vscale=56
%                     hoffset=-5 voffset=0}
\caption{
The hot volume filling factor in all 25 gas discs as they evolve in time
in the VELA cosmological simulations. Each point corresponds to a whole disc in
a single snapshot and a given color corresponds to a given disc as it evolves.
The disc is confined to a thin cylinder of radius $0.8\Rd$ and height
$\pm 0.25\kpc$.
The hot phase includes all gas cells with $T>3\times 10^4$K.
The filling factor for each galaxy oscillates about a self-regulated fixed 
value,
$\fh \simeq 0.6 \pm 0.07$, same for all galaxies.
}
\label{fig:filling_cos}
\end{figure}

%4
\begin{figure}
\vskip 6.8cm
\includegraphics{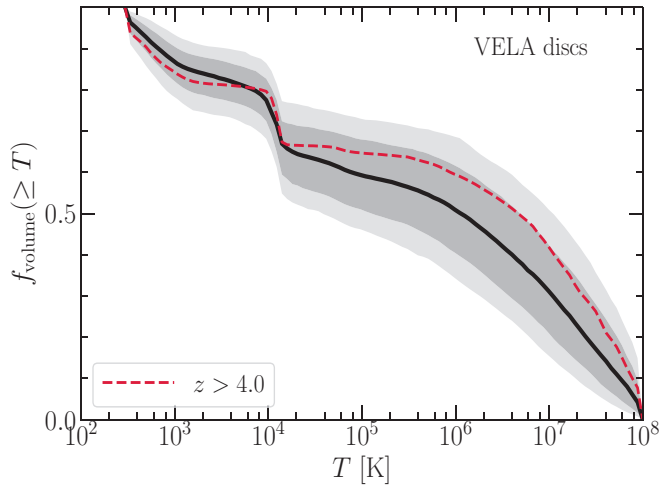}
%\special{psfile="figs/TemperatureCDF_Rdhd4.eps" hscale=50 vscale=54
%                     hoffset=-5 voffset=-3}
\caption{
The cumulative volume-weighted distribution of temperature in the VELA discs,
namely the volume filling factor of gas $>T$.
All snapshots are stacked, with the median shown as a solid line
and the scatter ($68\%$ and $95\%$) shown as shaded areas.
The plateau in the range $T=10^4-10^6$K implies that the hot filling factor is
robust to the choice of the threshold temperature in this range.
}
\label{fig:CDF_T}
\end{figure}

\smallskip % fh
Our main result from the cosmological simulations
is presented in \fig{filling_cos}, which shows the evolution 
of the hot volume filling factor in 
the whole disc in all snapshots of all galaxies in their discy phase,
where ``hot" refers to $T>3\times 10^4$K.
The snapshots of each galaxies are connected by a line of a random color.
We learn that the filling factor for each galaxy oscillates about a
self-regulated fixed value, roughly $\fh \simeq 0.6 \pm 0.07$, and that this is
similar for all galaxies.
We find that
the filling factor is largely independent of redshift and of galaxy mass.
There may be an apparent weak decline of $\fh$ with time, 
and an apparent slight increase of $\fh$ with the mass ranking at a 
fixed redshift (not explicitly shown in this figure), 
but the significance of these trends are
questionable.
This supports our basic ansatz that the feedback self-regulates the bubble
volume filling factor in the discs into a roughly constant value of order one
half.

%5
\begin{figure*}
\vskip 7.5cm
\includegraphics{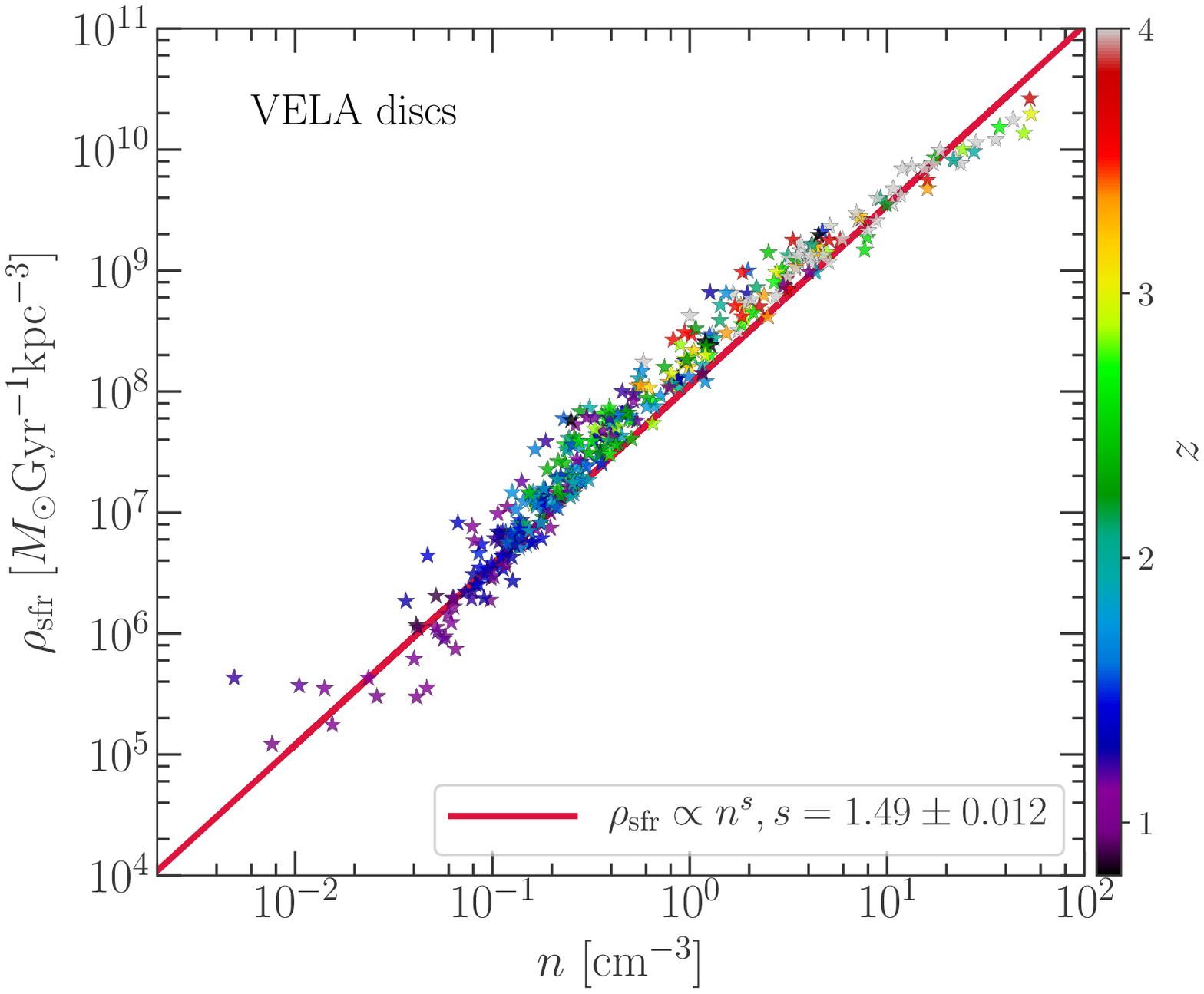}
\includegraphics{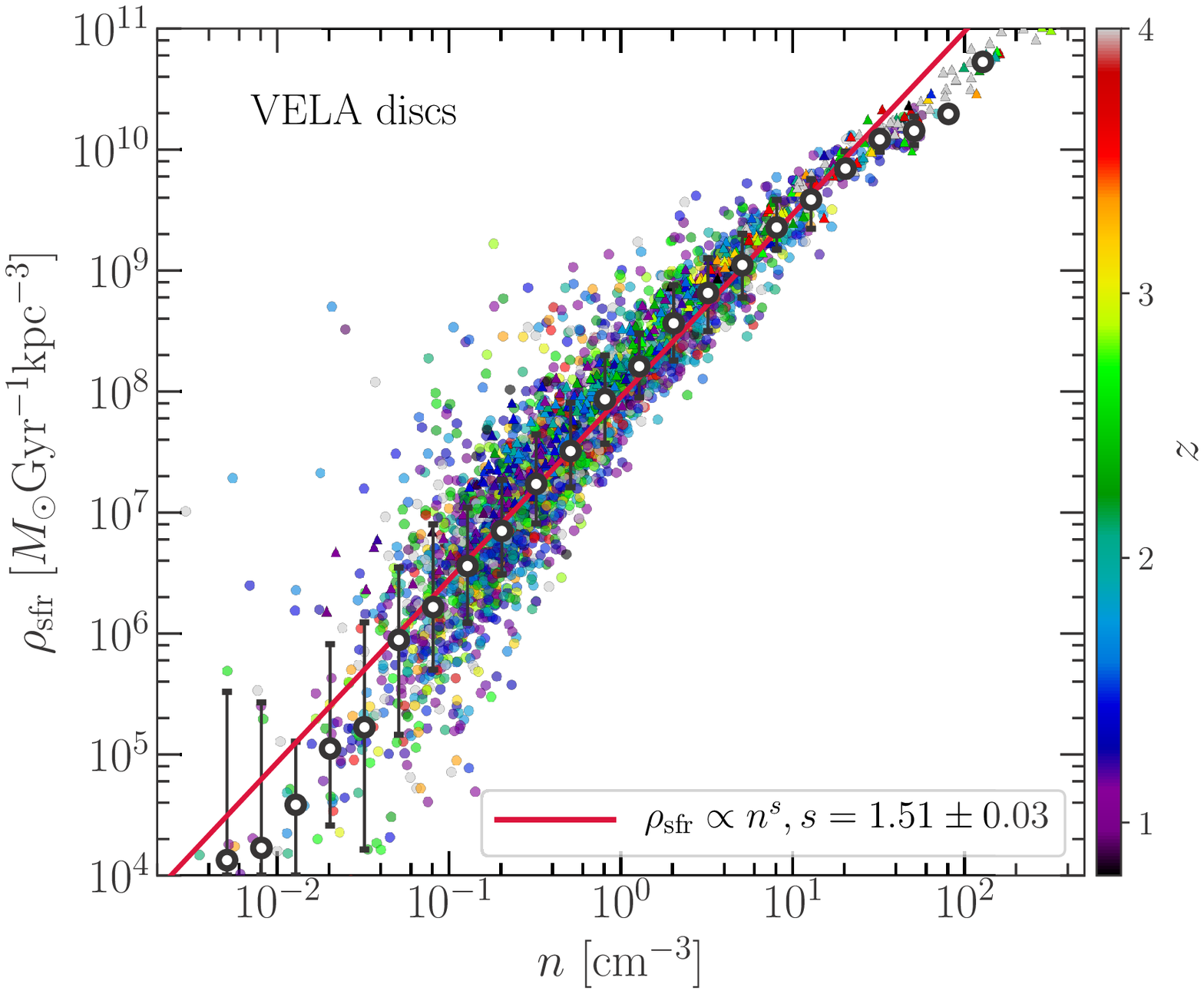}
%\special{psfile="figs/KS_Ngalaxies_Rdhd4.eps" hscale=50 vscale=50
%                     hoffset=00 voffset=-10}
%\special{psfile="figs/KS_sectors_Rdhd4_Cold.eps" hscale=50 vscale=50
%                     hoffset=260 voffset=-10}
\caption{
The 3D KS relation in the VELA simulation discs, either referring to the whole
disc (left) or to 9 patches within the disc (right) in each of the snapshots
for all the galaxies in their discy phases.
The medians in bins of density, and the $68\%$ scatter about them, are shown
in black.
There is a tight correlation with a slope $\slope \simeq 1.5$.
}
\label{fig:KS}
\end{figure*}

%6
\begin{figure}
\vskip 11.9cm
\includegraphics{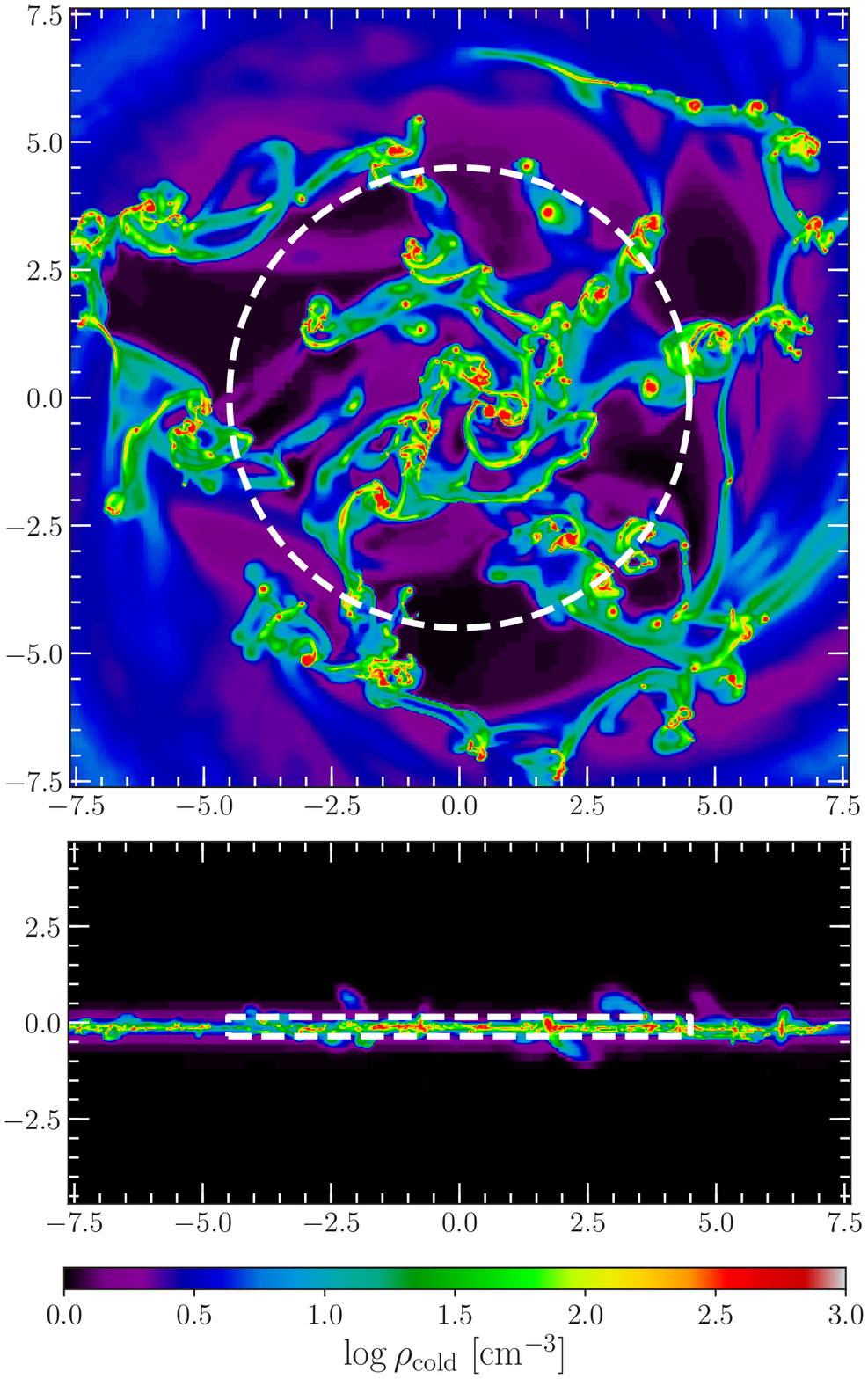}
%\special{psfile="figs/frederic_OneColumn.eps" 
%         hscale=50 vscale=50  hoffset=10 voffset=-5}
\caption{
An isolated-galaxy RAMSES simulation of 50\% gas fraction ($z\sim 2$),
run with the fiducial SFR and feedback recipes.
The mass-averaged density of the cold gas along the line of sight 
is shown face-on (top) and edge-on (bottom), at $t=300\Myr$. % Frederic?
Distance in $\kpc$ is marked along the axes.
The density is smoothed over $28\pc$ 
(while the maximum resolution is $3.6\pc$). 
The cylindrical volume used for measuring the hot gas filling factor is 
marked by the dashed line.
The KS relation is measured in sub-volumes of $1\times 1\times 0.5\kpc$.
}
\label{fig:iso_images}
\end{figure}

\smallskip % C(T)
The choice of $T>3\times 10^4$K in \fig{filling_cos} was rather arbitrary. To
test for robustness, \fig{CDF_T} shows the cumulative volume-weighted
distribution of $T$ in the disc, namely the volume filling factor for gas 
$>T$ as a function of $T$, stacked for all the discy snapshots
in the redshift range $z=5.6-1.0$.  Also shown (dashed red) is the same 
but for the high redshift discs only, $z=5.6-4.0$.
Indeed, we read $\fh \simeq 0.6$ at $T=3\times 10^4$K as in \fig{filling_cos}.
There is a step near $10^4$K, resulting from the drop in the cooling curve
near that temperature and from the fact that star formation is allowed below 
this temperature.
The resultant plateau between $10^4$ and $10^6$K (the typical virial 
temperature), which is totally flat for the high-redshift galaxies, 
implies that when the threshold temperature is varied in this  
range the hot filling factor varies by only about $\pm 0.05$.   
The self-regulated value of $\fh$ is thus robust to the choice of the
temperature threshold in the given range, with slightly smaller values of $\fh$
for higher $T$ thresholds.

\smallskip % KS
\Fig{KS} shows the global 3D KS relation for all the discs of the VELA 
simulations, either using the whole disc, or using nine patches within 
each disc.
The macroscopic gas density refers to cold gas, $T<3\times 10^4$K.
The SFR is determined using the stars younger than $40-80\Myr$, in the
minimum-bias way outlined in \se{app_vela} and in \citet{tacchella16_ms}. 
In both panels we see a tight KS relation, with a slope $\slope \simeq 1.5$. 
The slope turns out to be rather insensitive to the way the SFR is computed.
The colors, which refer to redshift, indicate that there is no significant
variation of the KS relation with time, despite the systematic evolution  
from highly perturbed discs at high redshift to more relaxed discs toward
$z \sim 1$.
No explicit lower limit is applied to the gas density in the cells (beyond the
upper limit to the temperature), motivated by the notion that the formation 
of molecular gas is not properly resolved on the grid-cell level. 
When a threshold of $n>1\cmc$ is applied, the slope of the KS relation
becomes slightly flatter, $\slope \simeq 1.2-1.3$.

\smallskip
The fact that the KS slope turns out to be almost exactly $1.5$, as predicted
in \se{single} based on the standard evolution of single SN bubbles, 
is not to be over-interpreted, as the exact evolution of SN bubbles in their
early phases is not 
resolved in these simulations, and as additional radiative feedback is
incorporated. Furthermore, one cannot (yet) reject the possibility that
the macroscopic slope reflects the local SFR recipe as incorporated in these 
simulations, although the latter imposed a density threshold, which should 
have modified the power-law relation. The robustness of the macroscopic
KS relation to the local SFR recipe should be explored via simulations where 
different local SFR recipes are incorporated (\se{iso}).
There are preliminary indications from VELA cosmological simulations
that the KS relation is robust to differences in the feedback recipes
\citep{ceverino14}, which we will revisit in \se{iso} using isolated-galaxy
simulations. 
The meaningful new finding from the cosmological simulations is the 
self-regulation to a constant hot filling factor, \fig{filling_cos},
which, for a given density dependence of the
filling factor, may be the main driver of the KS relation.

%%%%%%%%%%%%%%%%%%%%%%%%%%%%%%%%%%%%%
\section{Isolated simulations: Role of SN Feedback 
Robust to Local SFR}
\label{sec:iso}

The validity of self-regulation to a constant hot filling factor, the role of
SN feedback in it, and the insensitivity of the global KS relation to the local
SFR recipe, are tested here
via hydro-gravitational simulations of isolated galaxies. 

%==========================
\subsection{The isolated-galaxy simulations}
\label{sec:iso_method}

\smallskip % RAMSES, two fractions
Our idealized simulations of isolated galaxies are carried out with the 
RAMSES \citep{teyssier02} adaptive mesh refinement (AMR) code, including 
self-gravity, hydrodynamics, cooling and heating as well as star 
formation and stellar feedback based on subgrid recipes \citep{renaud13}. 
The two sets of simulations mimic star-forming disc galaxies at redshifts 
$z\sim0$ and $z\sim 2$, using gas fractions of 15\% and 50\%, respectively. 
They are similar to those presented in \citet{bournaud14} but with several 
combinations of star formation and feedback recipes.
Face-on and edge-on images of mass-weighted gas density along the line of 
sight in our fiducial simulation with 50\% gas fraction
are shown in \fig{iso_images} for a visual impression.

\smallskip % resolution
The simulation box size is $118\kpc$, the largest grid-cell size is $230\pc$ 
and the smallest cell after maximum refinement is $3.6\pc$. 
AMR cells are refined as soon as they contain more than 100 particles and/or 
a gas mass larger than $9 \times 10^3 \msun$ where the density is below 
$0.3 \cmc$, or when they gas mass is larger than $2.8 \times 10^3 \msun$ 
with a density above $0.3\cmc$. Gas denser than $ 0.03\cmc$ is thus refined at
$115\pc$ or better, and gas denser than $300\cmc$
is refined at the maximum, $3.6\pc$ resolution. 
The grid is also refined when the local Jeans length is smaller than 4 times 
the local cell size, and a density-dependent temperature floor keeps the Jeans 
length resolved by at least 4 cells at the highest refinement level 
\citep{teyssier10}. Cooling is
tabulated at solar metallicity and heating from a uniform UV background is 
included, as in \citet{bournaud14}.

\begin{table}
\centering
\begin{tabular}{@{}ccccc}
\multicolumn{5}{c}{{\bf Isolated-galaxy simulations}} \\
\hline
 SFR $\sloc$ & feedback & global $\slope$ & log rms & $\fh$ hot \\
                   &         & \ 15\%\ 50\%\ & \ 15\%\ 50\%\ & \ 15\%\ 50\%\ \\
\hline
\hline
 standard 1.5            &  SN+HII  & 1.41 1.44  & 0.07 0.05 & 0.38 0.35 \\
\hline
 shallow\ \ \ 1.0       &  SN+HII  & 1.28 1.23  & 0.13 0.15 & 0.42 0.41 \\
 steep\ \ \ \ \ \ 2.0  &  SN+HII  & 1.47 1.52  & 0.11 0.12 & 0.32 0.27 \\
\hline
 standard 1.5   &  SN               & 1.34 1.39  & 0.06 0.06 & 0.33 0.31 \\
 standard 1.5   &  HII              & 1.87 1.88  & 0.16 0.14 & 0.17 0.21 \\
 standard 1.5   &  none             & 2.04 1.94  & 0.23 0.19 & 0.12 0.18 \\
\hline
\end{tabular}
\caption{
Isolated-galaxy simulations.
The local SFR recipe is $\rhosfr \prop n^{\sloc}$.
The global KS slope $\slope$, the rms scatter about it,
and the hot filling factor $\fh$ are quoted for
the simulations with gas fraction 15\% and 50\% respectively.
The KS relation refers to a linear fit at $n>50\cmc$,
and $\fh$ is determined in the relaxed phase at $t>200\Myr$.
}
\label{tab:iso_sims}
\end{table}

%7
\begin{figure*}
\vskip 7.5cm
\includegraphics{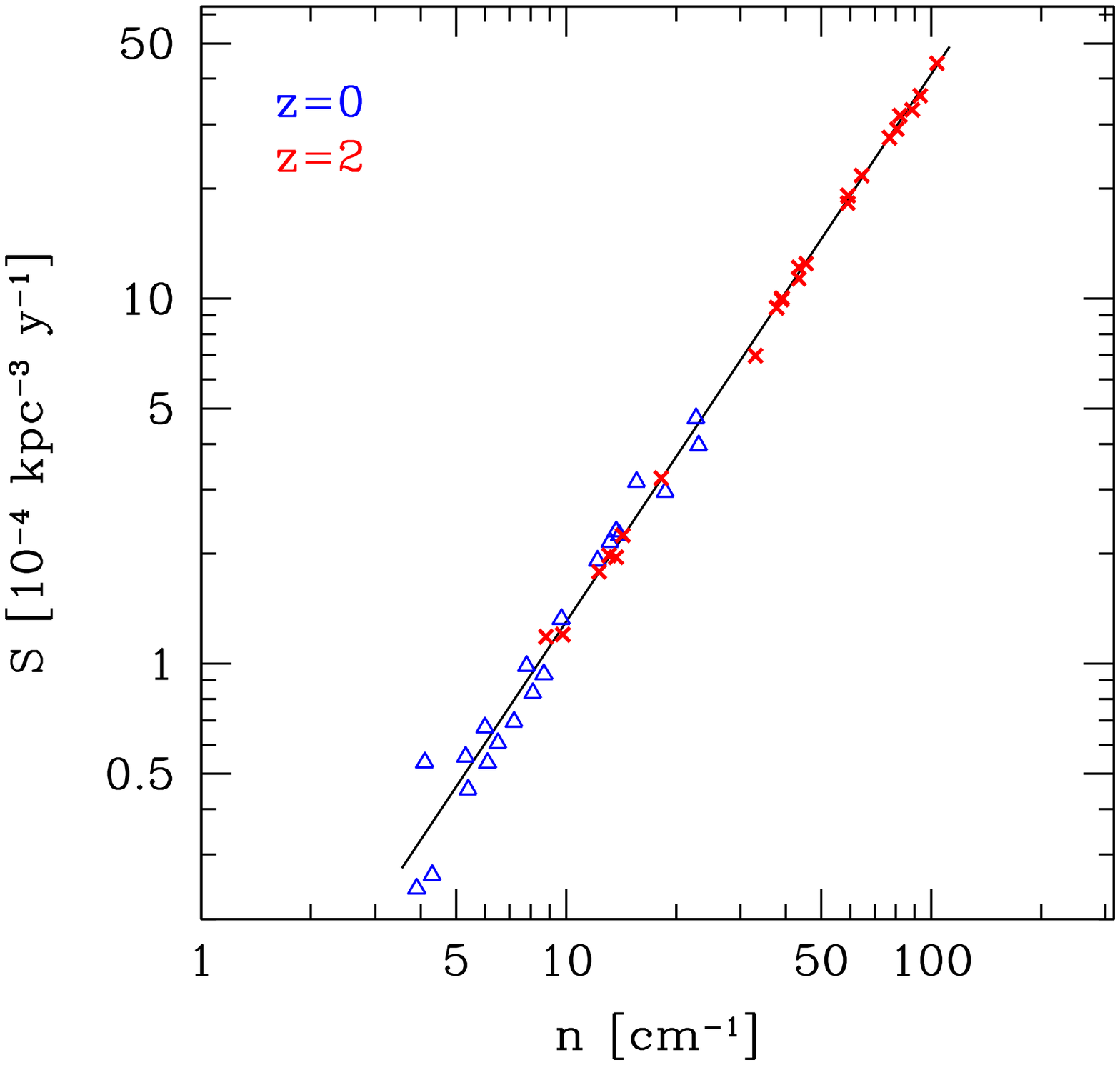}
\includegraphics{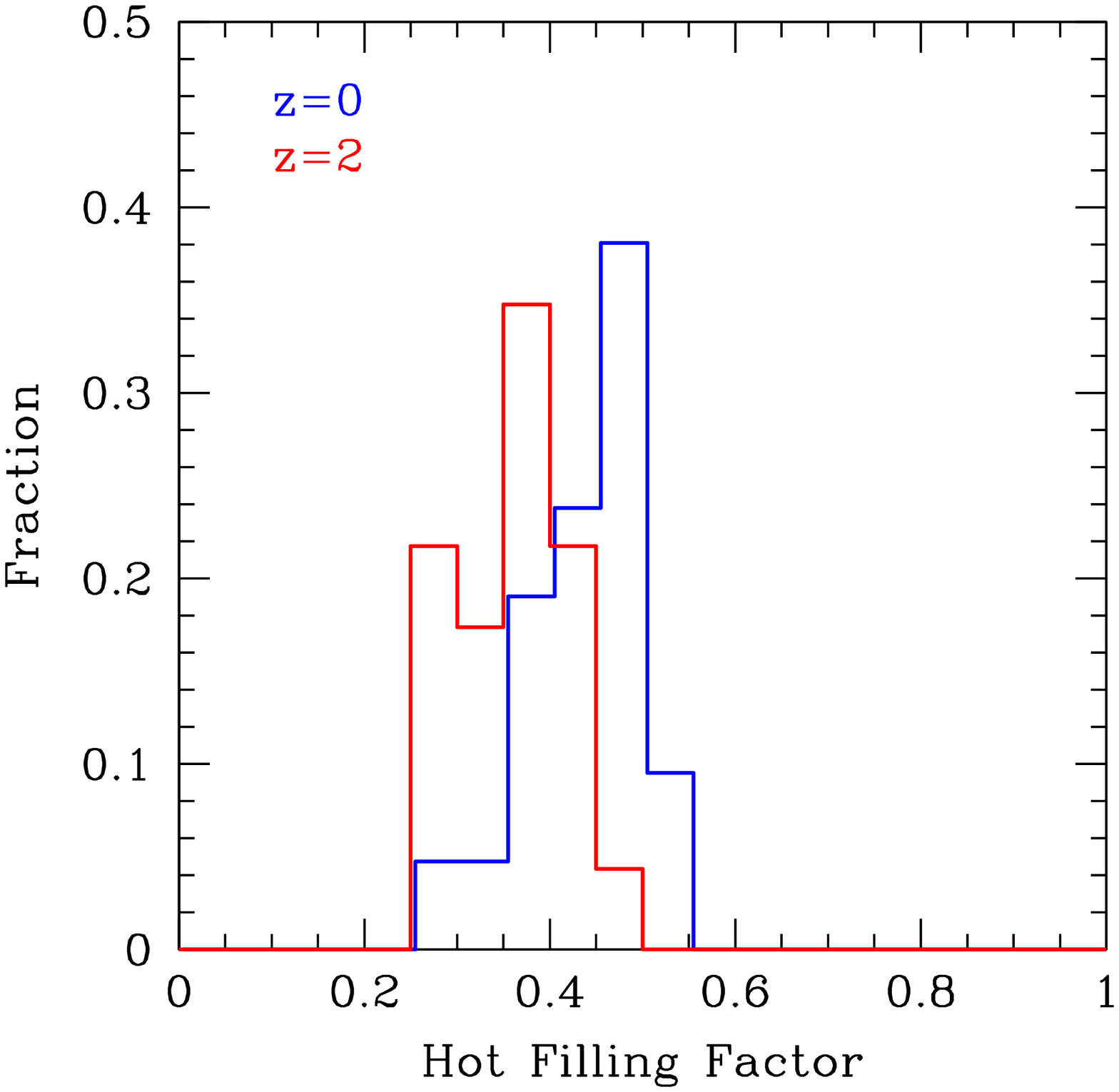}
%\special{psfile="figs/s_n.ps" hscale=43 vscale=43
%                     hoffset=0 voffset=-70}
%\special{psfile="figs/fh.ps" hscale=43 vscale=43
%                     hoffset=240 voffset=-70}
\caption{
Isolated-galaxy simulations with gas fractions of $0.15$ ($z=0$, blue)
and $0.50$ ($z=2$, red), with the fiducial SFR and feedback recipes.
The quantities are measured from four random snapshots between $t=200$ and
$400\Myr$ in boxes of $1\kpc \times 1\kpc \times h$ where
$h=0.5\kpc$ and $0.1\kpc$ for the high-$z$ and low-$z$ galaxies respectively.
{\bf Left:}
Supernova rate density versus gas density.
The solid line is $S \prop n^{1.5}$, the expected KS relation.
The larger scatter in $S$ in the low-$z$ simulations is due to small numbers of
SNe in the boxes.
{\bf Right:}
Distribution of hot volume filling factor within the galaxies,
$T>5\times 10^4$K. The filling factor is distributed over a narrow range,
$\fh \simeq 0.4 \pm 0.1$, indicating self-regulation.
The two panels demonstrate consistency with the predicted relation,
$\fh \prop S n^{-1.5} \sim$const.
}
\label{fig:iso_f}
\end{figure*}

\smallskip  % initial configuration
Each galaxy starts with a baryonic mass of $8 \times 10^{10} \msun$,
distributed in a stellar disc, a stellar bulge, and a gas disc.
The gas mass is 15\% or 50\% of the baryonic mass.
Each of the disc components has an exponential radial density profile with a
scale length of $4\kpc$, truncated at $12\kpc$.
The density profile vertical to the disc is exponential with a scale length of
$600\pc$ for the stars, and $600\pc$ or $150\pc$ for the gas in the gas-rich
and gas-poor cases, respectively.
The initial gas temperature is $5\times 10^4$K.
The bulge contains 20\% of the stellar mass.
It is spherical, obeying a \citet{hernquist90} profile with a
radial scale length of $600\pc$, truncated at $2\kpc$.
Each galaxy is embedded in a spherical dark-matter halo with a
\citet{burkert95} density profile of a characteristic radius $20\kpc$. 
The mass of the dark matter within the effective disk radius is set to be half
the baryon mass within that radius.

%8
\begin{figure*}
\vskip 9.0cm
\includegraphics{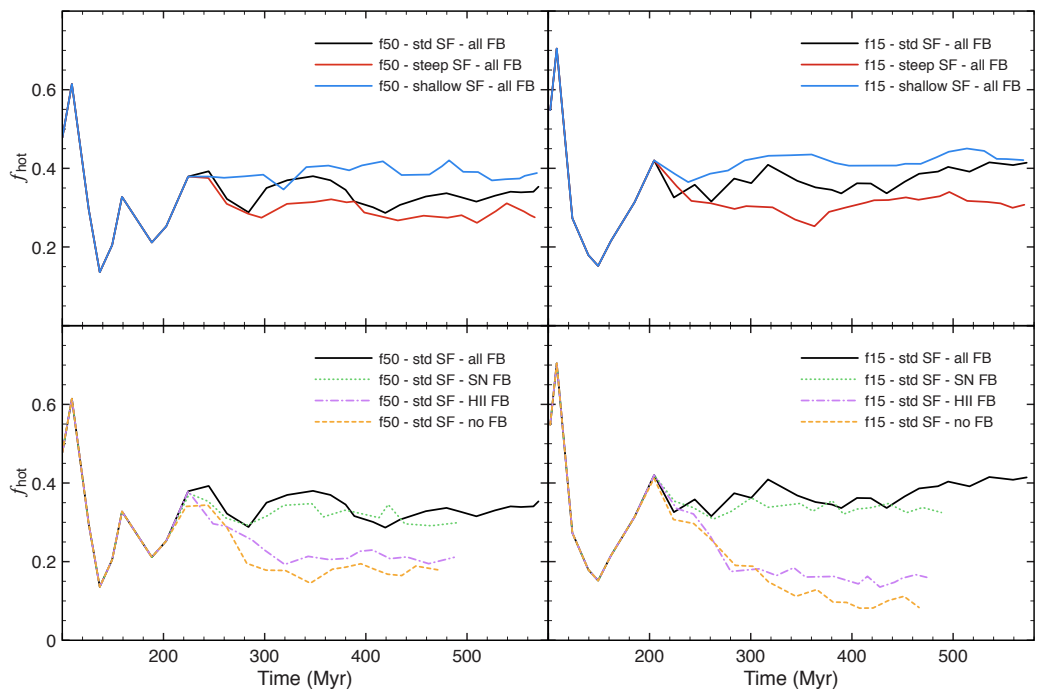}
%\special{psfile="figs/fhot_4panels.eps"
%                hscale=85 vscale=85  hoffset=40 voffset=-3}
\caption{
Time evolution of the hot filling factor in the isolated-galaxy simulations
comparing different local-SFR and feedback recipes.
Shown are the simulations with 15\% gas ($z=0$, right) and 50\% gas
($z\sim 2$, left).
{\bf Top:}
With the standard feedback implemented,
compared are local SFR recipes of the sort $\drhos \prop n^{\sloc}$ with
$\sloc=1.0,\,1.5,\,2.0$.
The deviation from the standard $\slope=1.5$ is turned on at $t=200\Myr$,
after the initial relaxation.
In each of the three cases the hot filling factor self-regulates to a
constant value in the range $0.3-0.4$.
{\bf Bottom:}
With the standard local SFR recipe, shown are four cases of feedback source:
the standard feedback of SN+HII, SN alone, HII alone, and no feedback.
The deviation from the standard feedback is applied at $t=200 \Myr$, after the
disc has relaxed from its idealized initial conditions.
The hot filling factor with no feedback is significantly smaller than with
the standard feedback, demonstrating that the hot volume is indeed
predominantly due to feedback.
The case with SN feedback alone is similar to the standard feedback including
SN and HII, demonstrating that the SN bubbles dominate the hot filling factor
over the HII bubbles (\se{ionization}).
}
\label{fig:iso_f_4panel}
\end{figure*}

%9
\begin{figure*}
\vskip 7.8cm
\includegraphics{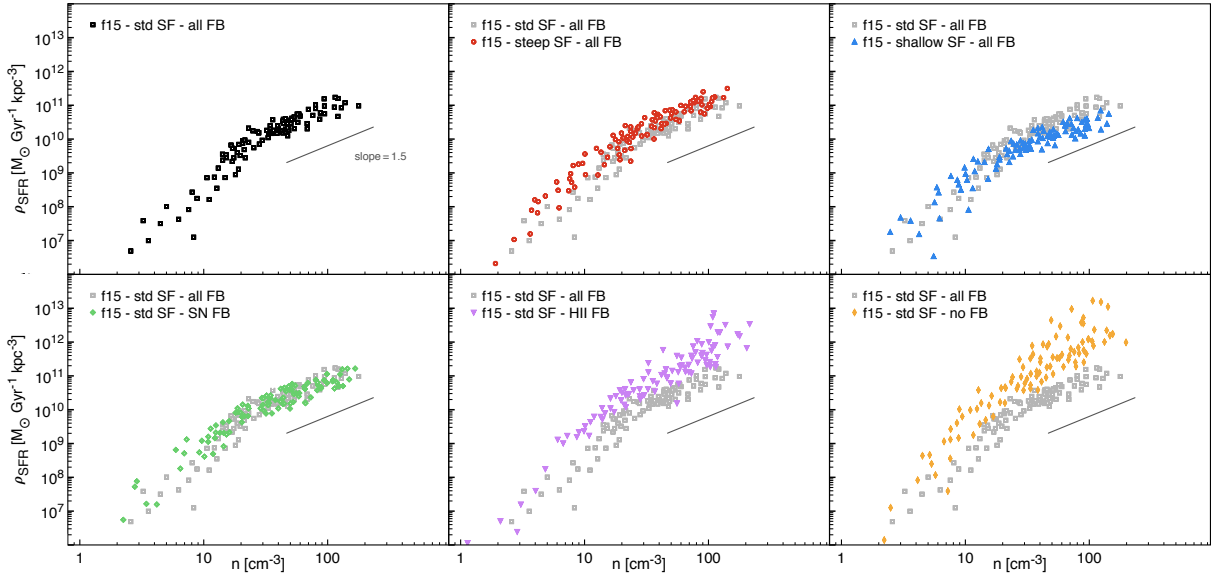}
%\special{psfile="figs/KS_f15_6pans.eps"
%                hscale=100 vscale=100  hoffset=10 voffset=-2}
\caption{
The global KS relation in the isolated-galaxy simulations with gas fraction
$0.15$. The SFR density and cold-gas density are measured in
macroscopic volumes of $1\kpc\times 1\kpc \times 0.1\kpc$ 
within the disc at different snapshots 
after $t=200\Myr$. 
{\bf Top:} Different local SFR recipes with standard feedback, 
showing from left to right the result for $\sloc=1.5$ (black or grey), 
$\sloc =2.0$ (red) and $\sloc =1.0$ (blue).  
{\bf Bottom:} Different feedback recipes with standard local SFR, showing from
left to right the results for SN-feedback only (green), HII-feedback only
(magenta), and no feedback (orange), in comparison to full feedback (grey).
The best-fit slopes are listed in \tab{iso_sims}.
The slope of the global KS relation is $\slope \simeq 1.5$, robust to 
variations in the slope of the local SFR recipe.
The KS relation is the same for the standard feedback and for SN feedback only, 
while it becomes steeper and with larger scatter with HII feedback only and
with no feedback.
}
\label{fig:iso_KS_f15}
\end{figure*}

%10
\begin{figure*}
\vskip 7.8cm
\includegraphics{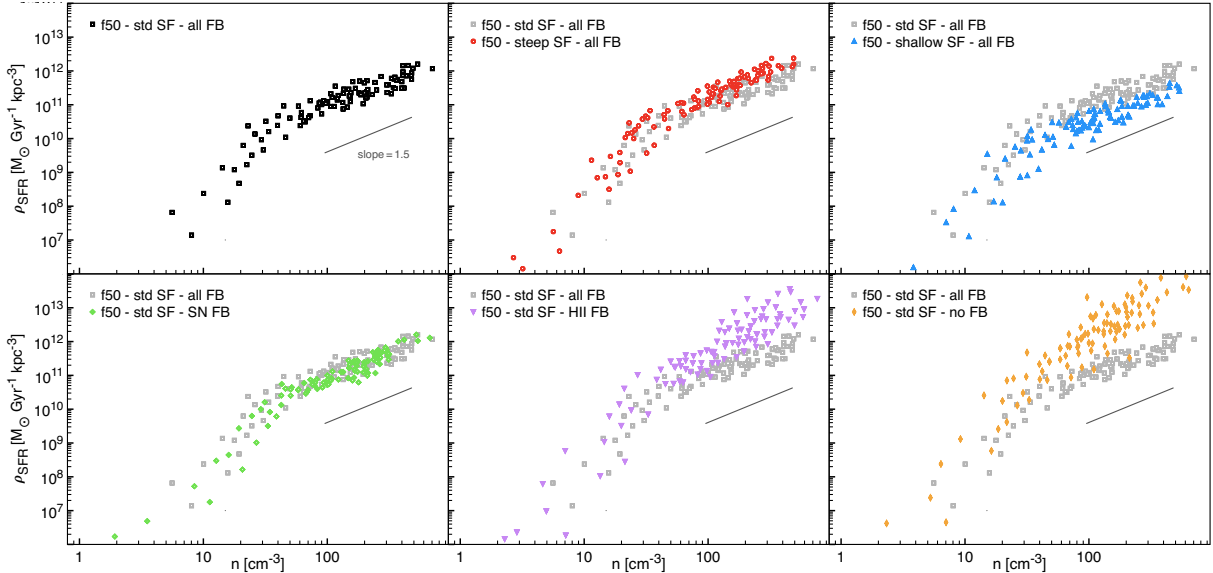}
%\special{psfile="figs/KS_f50_6pans.eps"
%                hscale=100 vscale=100  hoffset=10 voffset=-2}
\caption{
The global KS relation in the isolated-galaxy simulations with gas fraction 
$0.50$, in macroscopic volumes of $1\kpc\times 1\kpc \times 0.5\kpc$, 
otherwise the same as \fig{iso_KS_f15}, with similar conclusions.
}
\label{fig:iso_KS_f50}
\end{figure*}

\smallskip % SFR recipe 
Local star formation is allowed in cells of hydrogen number density above a 
threshold $n=10\cmc$.
Above this threshold, the SFR density is modeled as
\be
\drhos \propto \rhog^{\sloc}\, .
\label{eq:sloc}
\ee
In the ``standard" scheme, the recipe follows 
$\drhos = \epsf \rhog/\tff$ (\equnp{ks0}),
namely $\sloc=1.5$. The normalization is determined by setting the efficiency
to $\epsf = 0.025$.
To test for sensitivity of the global slope to the local SFR recipe, the same
simulations were run with a ``steep" local slope of $\sloc=2$ and with a 
``shallow" local slope of $\sloc=1$. 
The normalization in each cases was tuned to obtain the same SFR as in the 
standard case during the first $20\Myr$.

\smallskip % feedback
Feedback is modeled through a combination of supernova and radiative feedback.
The {\it supernova feedback} models the explosions of type-II supernovae, 
assumed
to explode $10\Myr$ after the star birth. Seventy five percent of the initial 
energy of each supernova, assumed to be $10^{51}\erg$, is injected in the 
$\sim 6\pc$ vicinity of the SN in the form of thermal energy (50\%) and kinetic
energy (25\%), while the rest is assumed to have radiated away before the shell
reached $6\pc$.
This is a typical energy budget for supernovae explosions once they have
expanded to $\sim 6\pc$ \citep{martizzi14}.
The {\it radiative feedback}, termed HII feedback, models the photo-ionization 
of gas by O/B stars and the radiation pressure on the gas and dust.
We assume a Str\"omgren sphere approximation around stellar particles
younger than $10\Myr$ \citep[see][for a more elaborate description]{renaud13}.
To test the role of feedback, and the contributions of SN feedback versus
HII feedback, we ran the same simulations, with the standard
local SFR recipe, but with four variants of feedback recipes;
the standard SN+HII feedback, SN feedback alone, HII feedback alone, and no
feedback.
In all cases the simulation started with the fiducial feedback recipes, to 
allow relaxation from the initial conditions to a similar realistic 
configuration with spiral arms.
Then, near $t=200\Myr$, if needed, the recipe was changed to the desired form.
%The analysis is performed after this time.

\smallskip  % method measurement
The galaxy is allowed to relax for $\sim 200\Myr$ from the initial conditions 
to a realistic configuration, involving disc instability and spiral arms. 
The analysis is performed after this time for several hundred Megayears 
(typically until $580\Myr$), after which the approximation of an isolated 
galaxy gradually becomes less valid.
The relevant quantities are measured in boxes that are spread throughout the
disc, centered on the disc central plane.
The box size is $1\kpc \times 1\kpc \times h$, where
$h=0.5\kpc$ and $0.1\kpc$ for the high-$z$ and low-$z$ galaxies respectively.
Alternatively we refer to the whole disc,
within the disc exponential scale-length and full height $h$.
In each box, we compute the SFR and supernova rate densities $\drhos$ and $S$,
the cold gas number density $n$ for $T<5\times 10^4$K,
and the volume filling factor for hot gas of $T>5\times 10^4$K.
As in the cosmological simulations, the results are found to be
insensitive to changing the temperature threshold by a factor of two.

%==========================
\subsection{Results: Role of SN Feedback and Robustness to Local SFR}

\smallskip % KS and fh at one given time
\Fig{iso_f} shows $S$ versus $n$ (left) and the distribution of $\fh$ in the
sub-volumes of each galaxy with the fiducial local SFR and feedback recipes,
measured in four random snapshots in the time range $t=200-400\Myr$.
A KS relation is demonstrated, with a tight distribution about
$S \prop n^{1.5}$. 
In principle, this might be built in by the standard local SFR recipe 
assumed in the simulations.
The non-trivial result is that 
the filling factor is distributed in a narrow range, $\fh \simeq 0.4 \pm 0.07$,
with only a small deviation between the low and high-$z$ galaxies.
This is consistent with the idea that the system self-regulates itself to
a constant filling factor of $\sim 0.5$, and points to the actual value at the
self-regulated state for the isolated-galaxy simulations.

\smallskip % fh(t)
\Fig{iso_f_4panel} 
shows the time evolution of the hot filling factor in each variant of
the two isolated galaxies (left and right columns).
The solid curves refer to the standard SFR and feedback recipes.
After an initial adjustment period of $\sim 200\Myr$, where the galaxy
relaxes from the initial conditions,
the filling factor becomes self-regulated to a rather constant value,
similar in the two galaxies independent of the gas fraction.
This self-regulation to a constant value is similar to what we saw in
the cosmological simulations, though the asymptotic value of $\fh$ is somewhat 
lower for the isolated galaxies, reflecting the different SFR and feedback 
recipes and the other differences between the two simulation types performed
with totally different codes (ART and RAMSES), initial conditions and
environments.

\smallskip  % feedback
With the feedback turned off in the bottom panels of \fig{iso_f_4panel}, 
the hot filling factor becomes much lower. The non-zero value is likely due to
shock heating within the supersonic turbulent ISM, and is possibly partly a 
remnant of the initial period with full feedback.
This demonstrates that the hot filling factor predominantly represents 
hot bubbles generated by feedback, as assumed throughout this paper.
SN-feedback alone gives rise to a hot filling factor similar to the case of
full feedback, while HII-feedback alone leads to a lower filling factor, almost
as low as with no feedback. This indicates that SN feedback dominates over HII
feedback (\se{ionization}).

\smallskip  % SFR
The top panels of \fig{iso_f_4panel} demonstrate that the convergence of the
hot filling factor to a constant value is robust to variations in the slope of
the local SFR recipe, with small variations in the value of this constant.
Together with the feedback dependence shown in the bottom panels,
this indicates that the self regulation is robust and is driven by SN feedback,
confirming the basic ansatz of our model.
The typical hot filling factor at $t>200\Myr$ is listed for each run in
\tab{iso_sims}.

\smallskip % KS panels
\Fig{iso_KS_f15} and \fig{iso_KS_f50} present the global KS relations as
produced in the isolated-galaxy simulations with gas fractions 0.15 and 0.50
respectively.
The SFR density and cold-gas density are measured in
macroscopic volumes within the disc at different snapshots after 
the initial $225\Myr$.
\tab{iso_sims} lists the slope of the global KS relation in the different
cases, as determined by a linear fit for $n>50 \cmc$, 
and the rms scatter about the linear relation.
With the standard local SFR recipe and feedback (top-left, black),
the global KS relation has a slope $s \simeq 1.43$, % 1.41-1.44, 
with a small scatter of $\pm 0.06$dex. % 0.05-0.07$dex. 
When the local slope is steeper ($\sloc=2$) or flatter ($\sloc=1$), the top
panels show that the global
slope becomes only slightly steeper or flatter, by $\simeq 0.1$, 
with a scatter $\pm 0.13$dex. % $\pm 0.11-0.15$dex.
In general, both for the simulations with low and high gas fraction,
the global slope remains roughly the same and with a small scatter, 
insensitive to the local slope.

\smallskip
The bottom panels of \fig{iso_KS_f15} and \fig{iso_KS_f50} show that, for the
standard local SFR recipe, the KS
relation is almost the same with the full feedback and with SN feedback alone.
When only HII feedback is activated, the global slope steepens to 
$\slope \simeq 1.88$, % 1.87-1.88$, 
and the scatter grows to $\pm 0.15$dex. %0.14-0.16$dex.
With no feedback, the global slope steepens further to
$\slope \simeq 2$, %1.94-2.04$, 
and the scatter grows further to $\pm 0.21$dex. %0.19-0.23$dex.
This is consistent with our basic assertion that SN feedback dominates over the 
HII feedback, and is responsible for the KS relation (see \se{ionization}).

%---------------
%\subsection{Weak dependence on the local SFR recipe}
%\label{sec:Hopkins}

\smallskip % Hopkins
The weak dependence of the large-scale KS relation on the small-scale SFR 
recipe is consistent with earlier tests using SPH hydro simulations of 
isolated galaxies \citep{hopkins11,hopkins13}.
Considering galaxies of a variety of masses at low and high redshifts, they
experimented with a variety of local star-formation recipes.
\citet{hopkins11} explored density criteria with
a SFR efficiency in the range $\epsf=0.0035-0.06$,
a power-law dependence on gas density in the range $\drhos \prop n^{1.0-2.0}$,
and a threshold gas density for star formation
in the range $n_{\rm min}=10-2500\cmc$.
\citet{hopkins13} experimented with different local physical criteria for star
formation, including self-gravity, Jeans instability, density,
temperature, molecular-gas content and cooling rate.
They found that once feedback is incorporated, the galaxy is self-regulated to
a global KS relation with only little sensitivity to the local SFR recipe.
This is in the sense that for a galaxy of given global galaxy properties,
such as gas density, the global SFR approaches roughly the same
value independent of the local SFR recipe.
\citet{hopkins14} then showed that in their FIRE cosmological simulations,
using their fiducial SFR recipe, the simulation converges to a KS relation
similar to the observed relation.
These experiments, like ours, indicate that the SFR is robustly regulated by 
feedback.
In the current paper, we address the origin of this through the
robust self-regulation of the bubble hot volume filling factor into a constant
value.
Our simulations demonstrate the convergence to a constant filling factor
both in cosmological and isolated settings, and our isolated-disc
simulations confirm the robustness to different local SFR recipes, and 
establish the dominant role of SN feedback in the self-regulation to a 
constant filling factor and the generation of the global KS relation.

%%%%%%%%%%%%%%%%%%%%%%%%%%%
\section{Clustered Supernovae - Analytic}
\label{sec:cluster}

In the previous two sections we established the validity of the concept of 
self-regulation into a constant hot filling factor in galaxy simulations that 
incorporate star-formation and SN feedback in a realistic ISM, with the 
resultant KS relation independent of the local star-formation recipe and 
apparently driven by SN feedback. 
We now return to simplified analytic modeling, making first steps
in generalizing the analytic modeling of isolated SNe (\se{single}). 
Here we address idealized cases of clustered SNe in star-forming clouds, 
where a sequence of SNe explode at the same location.
In the next section we test these analytic estimates with spherical
hydrodynamical simulations.
Then, in the following section, we consider the Str\"omgren bubbles 
photo-ionized by the pre-SN O/B stars and their interplay with the SN bubbles.
These simplified analytic attempts will help verifying the validity of the
basic model for the origin of the KS relation in somewhat more realistic 
circumstances, and physically interpreting the behavior in the full 
simulations.

%-----------
\subsection{Four characteristic times}

As star formation tends to occur in clumps (molecular clouds, star clusters), 
many SNe occur practically in the same place. This is likely to affect 
the bubble filling factor and it may change its density dependence and
therefore the resultant KS relation.
We attempt here to estimate the possible effects of clustering in a simplistic 
analytic way, to be followed by simplified simulations.
We model the sequence of SNe in a cluster with two independent parameters.
We consider first a situation where $\nu$ SNe occur in each point-like cluster,
with a constant SN rate during a burst duration $\tb$, associated with the
lifetime of the star-forming cluster.
The typical time available for a SN before the successive SN explodes is 
thus $\ts=\tb/\nu$, so the two parameters could be $\ts$ and $\tb$.
For a given cluster of SNe we consider the bubble about it.
In analogy to the case of an individual SN bubble, the cumulative bubble
will have a phase analogous to the adiabatic phase
until a cooling time $\tc$,
followed by phases analogous to the snow-plow phase in which the 
outer shell has collapsed to a thin massive shell pushed by a wind or pressure,
whose speed eventually fades away to the ISM sound speed at a fading time $\tf$.

\smallskip
The evolution of the cumulative bubble, its fading time and radius, the
resulting bubble filling factor, and the final
power index $\slope$ of the density dependence in the KS relation, 
$\drhos \prop n^\slope$, depend on the interplay between the timescales
characterizing the SN cluster, $\ts$ and $\tb$, and those of the 
cumulative bubble, $\tc$ and $\tf$.
Note that by definition $\ts < \tb$ and $\tc < \tf$.

%---------------
\subsection{Different zones in parameter space}

\begin{figure}
\vskip 6.0cm
\includegraphics{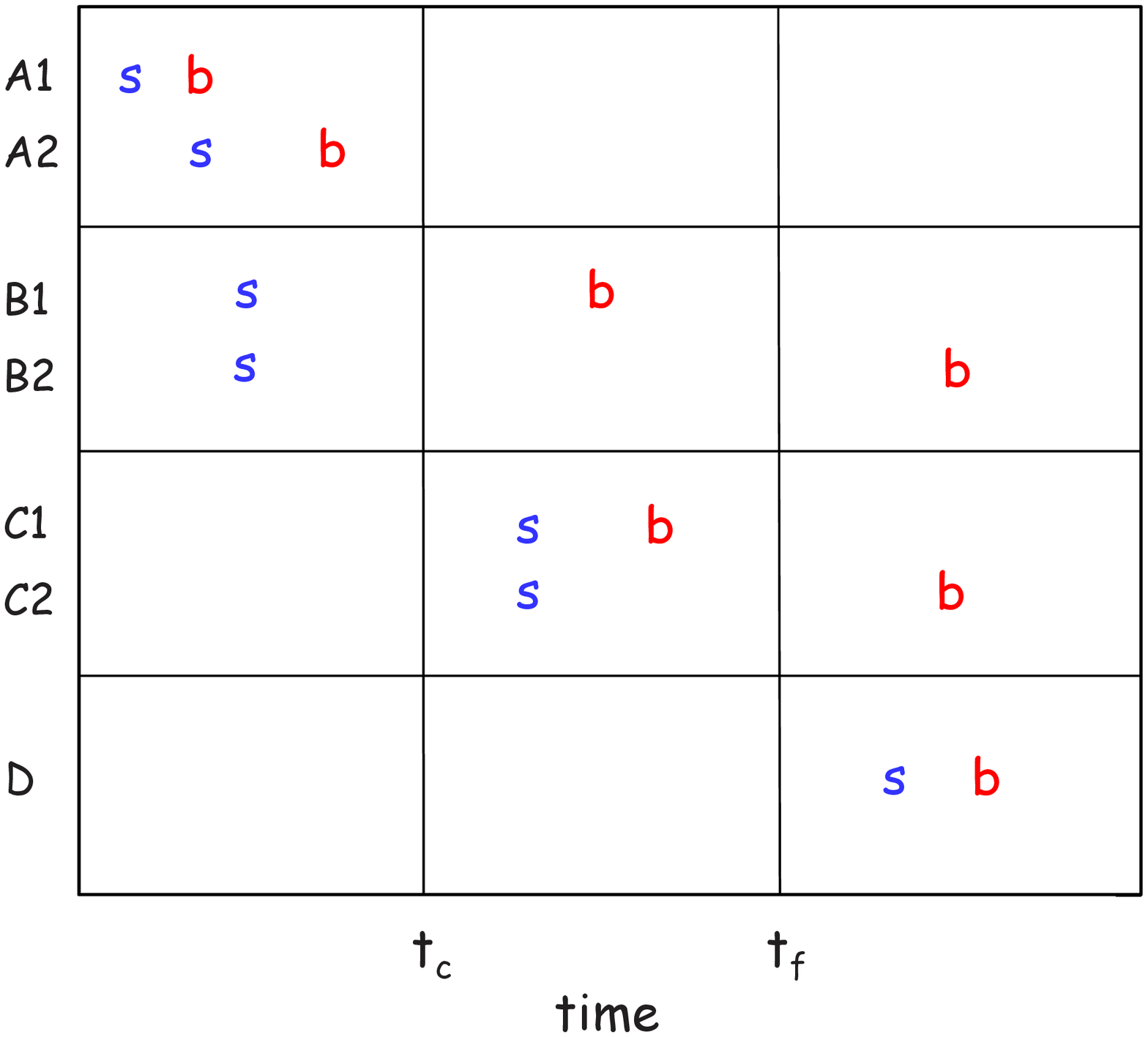}
%\special{psfile="figs/cluster_classification.eps" hscale=33 vscale=33
%                     hoffset=65 voffset=-95}
\caption{
A schematic description of the
zones in parameter space for clustered supernovae, based on the relation between
the burst times $\ts$ and $\tb$ (marked s and b) and the SN bubble times
$\tc$ and $\tf$.
}
\label{fig:classification}
\end{figure}

We divide the parameter space into the following different zones, where the
analysis of the bubble consists of different phases such that the final density
dependence may be different (see a schematic cartoon in \fig{classification}):
\rf
A. {\bf A short burst}, $\tb < \tc$ \\ 
   A1. instantaneous, $\tb \ll \tc$                         \\
   A2. non-instantaneous, $\tb \lsim \tc$
\rf
B. {\bf Long and continuous, $\ts < \tc < \tb$}\\ 
   B1. moderately long, $\tb < \tf$ \\
   B2. very long, $\tf < \tb$
\rf
C. {\bf Long and semi-continuous, $\tc < \ts < \tf$}\\
   C1. moderately long, $\tb < \tf$ \\
   C2. very long, $\tf < \tb$
\rf
D. {\bf Very long and separable, $\tf < \ts$}. 

\smallskip\no
We summarize here the expectations in each zone, and
elaborate in the following subsections.

\smallskip
% summary
% A
For a short burst, {\it zone A}, the solution can be deduced from the solution 
for an individual SN, namely a KS relation with a power index $\slope = 1.48$. 
This is obvious for a very short burst, case A1, where the SN cluster is
analogous to a single hyper-nova with the energy multiplied by $\nu$. 
We show below that the same power law with $\slope =1.48$ is expected
also in the non-instantaneous zone A2.

\smallskip
% D
A similar power index of $\slope =1.48$ is expected also
in the opposite extreme of a very long time interval between SNe, {\it zone D}.
If each individual bubble fades away well before the explosion of the 
subsequent 
SN, such that the faded bubble had time to recover the original unperturbed ISM 
environment, the cluster is expected to behave like a sequence of separable 
individual SNe. 
The solution is the single-SN solution, with a power index $\slope =1.48$ 
in the KS relation.
This may still be a sensible crude approximation when $\ts \sim \tf$, near the
border of zones D and C.

% 11

\smallskip 
% B, continuous limit
In the sub-cases of {\it zone B}, we predict analytically
slopes in the range $\slope =1.1\!-\!2.0$, and typically find values closer to 
$\slope =1.5$ in the simulations.
Here, as well as in zone A2, as long as $t<\tc$, and actually as long as
$t<\tb$,
the energy source by the frequent SNe can be treated as a {\it continuous}
energy
flux, causing a wind-driven expansion phase at a rate $\Rs \prop t^{3/5}$.
In {\it zone B2}, where the fading occurs while the burst is still on, 
$\tf < \tb$, our analytic estimate leads to $\slope =2.0$.
In {\it zone B1}, where the fading occurs well after the burst is over,
$\tb < \tf$, after the bubble went into a passive snow-plow-like phase of 
$\Rs \prop t^{2/7}$ at $t > \tb$, our analytic estimate becomes $\slope = 1.1$.
In the intermediate situation between B1 and B2, where $\tf \sim \tb$, 
we expect a value of $\slope$ between $1.1$ and $2.0$, 
growing as a function of $\tb/\tf$.

\smallskip
% C
In {\it zone C}, the continuous limit may still be valid with certain
modifications, and the trend of the slope $\slope$ with $\tb/\tf$, 
from C2 to C1, is likely to be similar to the trend in cases B. 

\smallskip
We describe below these analytic predictions, and follow in \se{spheri_sim}
with spherical simulations that confirm the convergence to $\slope \sim 1.5$ 
in most of the different cases.

%-----------------------
\subsection{A Continuous Energy Source}

Once $\ts < \tc$ (zones A and B), and possibly even as long as 
$\ts < \tf$, (zone C), while the burst is on, $t < \tb$,
we follow \citet{weaver77} in treating
the energy source from the successive SNe as continuous, 
with a constant power (``luminosity"),
\be
\L = \frac{\gamma E}{\ts} \, .
\ee
Here $\gamma$ is the efficiency by which each SN delivers energy $E$ in the
clustered environment, assumed to be of order unity.
Hereafter, we let $\e51$ actually represent $\gamma\, \e51$.
We can write
\be
\l38 = 0.32\, \e51\, {\ts}_6^{-1} \, ,
\label{eq:L38}
\ee
where $\L = 10^{38} \ergs \l38$ and $\ts = 1\Myr\, {\ts}_6$.

\smallskip
The single-SN Sedov-Taylor expansion, $\Rs \prop t^{2/5}$, is now replaced 
by a slightly faster expansion, $\Rs \prop t^{3/5}$, as the constant $E$ in 
\equ{Rs_Sedov} is replaced by $E(t)=\L\, t$ where $\L$ is a constant.
The shell properties prior to the cooling time become\footnote{The reduced 
factor of 0.88 in \equ{Rs_cont} compared to the single SN case
is because in the case of a shock driven by a constant wind only
55\% of the energy is made available to the swept-up ISM gas while the rest
is stored in the shocked wind behind it. The thermal pressure of this shocked
wind is pushing the swept-up mass, and this is somewhat less efficient than in
the Sedov-like expansion.}
\be
\Rs = 0.88 \left( \frac{\L t^3}{\rho} \right) ^{1/5} 
= 76.55\pc\, \l38^{1/5} \n0^{-1/5} t_6^{3/5} \, ,
\label{eq:Rs_cont}
\ee
\be
\Vs = \frac{3}{5} \frac{\Rs}{t} 
= 45.0 \kms \l38^{1/5} \n0^{-1/5} t_6^{-2/5} \, ,
\label{eq:Vs_cont}
\ee
\be
\Ts =\frac{3}{16} \frac{\mu m_{\rm p}}{k_{\rm B}}\Vs^2 
=2.7\times 10^4{\rm K}\, \l38^{2/5} \n0^{-2/5} t_6^{-4/5} \, ,
\label{eq:Ts_cont}
\ee
where $t = 1 \Myr\, t_6$. 

\smallskip
Proceeding in analogy to the single-SN case, \equ{Edot},
while considering the total energy injected so far, $E(t) = \L t$, 
the cooling time and the shell radius and velocity at cooling become
\be
\tc = 0.04 \Myr\, \l38^{0.29} \n0^{-0.71} \, ,
\label{eq:tc_cont}
\ee
\be
\Rc = 11 \pc\, \l38^{0.37} \n0^{-0.63} \, ,
\label{eq:Rc_cont}
\ee
\be
\Vc = 166 \kms \l38^{0.085} \n0^{0.085} \, ,
\label{eq:Vc_cont}
\ee
where the spatial integral in \equ{Edot} has been evaluated numerically for the
wind-like similarity solution.
At this time the shell exits the simple adiabatic continuous regime, 
where it expands $\prop t^{3/5}$ according to \equ{Rs_cont}, 
and it may enter another regime, depending on the zone in parameter space.

%---------------
\subsection{Zone A: Short Burst $\tb < \tc$}
\label{sec:A}

%\subsubsection{A1: Instantaneous, $\tb << \tc$}
For an instantaneous burst, $\tb \ll \tc$,
the cluster is analogous to a single SN with an energy $\nu E$.
For a given overall SN rate $S$, the rate of clustered explosions is $S/\nu$.
According to the dependence of the filling factor on energy in \equ{f0_draine},
the filling factor in the clustered case is
\be
\f0 \prop \nu^{0.26} S\, n^{-1.48}\, .
\ee
Thus, the KS relation obtained by requiring $\f0 = const.$ is
similar to that of single SNe, \equ{ks_single},
with the same density dependence $\drhos \prop n^{1.48}$,
and with the amplitude, or $\epsf$, scaled as $\nu^{-0.26} \e51^{-1.26}$.
For example, with GMCs of $10^{4}\msun$, and $\msn = 100\, \msun$
stellar mass per supernovae, one has $\nu = 100$, so 
the supernova rate at a constant $\f0$, and similarly $\epsf$, 
is smaller by a factor $\nu^{0.26} \simeq 3.3$ compared to the case of 
unclustered SNe (assuming $\gamma=1$).

%\subsubsection{A2: Non-instantaneous, $\tb < \tc$}
\smallskip
In general in zone A, $\tb < \tc$,
as long as the burst is on, $t<\tb$, the shock is driven by a continuous-energy
wind, $\Rs \prop t^{3/5}$, following \equ{Rs_cont}.

\smallskip
Once the energy input ceases but before the shell losses a significant fraction
of its energy, $\tb < t <\tc$,
the shock enters a Sedov-Taylor-like expansion,
\be
\Rs = \Rs(\tb) \left( \frac{t}{\tb} \right)^{2/5}  
= 3.84\pc\, (\nu\e51)^{1/5} \n0^{-1/5} t_3^{2/5} \, ,
\label{eq:Rs_A}
\ee
\be
\Vs = 1500 \kms (\nu\e51)^{1/5} \n0^{-1/5} t_3^{-3/5} \, ,
\ee
where \equ{L38} has been used to express $L$ and $\tb$ by $\nu$. 
Note that while the numerical values are different,
the dependence of these expressions on the variables are the same as 
the expressions for the single-SN Sedov-Taylor solution, 
\equ{Rs_Sedov} and \equ{Vs_Sedov},
once the single-SN energy $E$ is replaced by the total energy 
$\nu \gamma E$.

\smallskip
In computing the cooling time,
the energy loss to radiative cooling is the sum of the integrals of
$\dot{E}[\Rs(t)]$,
\equ{Edot}, over time in the successive intervals $0-\tb$ and $\tb - t$,
where the wind phase and the Sedov-Taylor-like phase are assumed to be
valid respectively.
In the spatial integrals of \equ{Edot} in each phase, 
the density and temperature profiles are obtained from the corresponding
self-similar solutions, and the upper bound for the integration, $\Rs(t)$,
is from \equ{Rs_A} and \equ{Rs_Sedov} respectively.
After some algebra, The cooling time, where the energy loss is one third of 
$E(t)$, turns out to be 
\be
{\tc}_{,3} = K^{0.33} \, ,
\label{eq:tc_A}
\ee
\be
K \equiv 2.22\times 10^5 (\nu\e51)^{0.67} \n0^{-1.67}
-1.51{\tb}_{,3}^{3.04} \, .
\label{eq:K}
\ee 

\smallskip
In the limit $\tb \ll \tc$, zone A1, the second term is negligible, 
and one recovers the single-SN case with the energy multiplied by $\nu\gamma$,
as expected.
In the other limit of zone A, $\tb \sim \tc$, 
an estimate is obtained by substituting $\tb = \tc$, 
\be
{\tc}_{,3} = 42.4 (\nu\e51)^{0.22} \n0^{-0.55} \, .
\ee

\smallskip
After the cooling time, at $t>\tc$, the shell is in a passive snow-plow phase, 
\be
\Rs(t) = \Rs(\tc) \left( \frac{t}{\tc} \right)^{2/7} \, ,
\ee
\be
\Vs = \frac{2}{5} \frac{\Rs(t)}{t} \, .
\ee
The shell fades away when the shock velocity is reduced to the sound speed
of the medium,
\be
\tf = 1.1\Myr\,(\nu\e51)^{0.28} \n0^{-0.28} c_1^{-1.4} K^{0.052}\, ,
\label{eq:tf_A}
\ee
\be
\Rf = 28.3\pc\,(\nu\e51)^{0.28} \n0^{-0.28} c_1^{-0.4} K^{0.052}\, . 
\label{eq:Rf_A}
\ee

\smallskip
With a rate of $S/\nu$ for the clusters of SNe,
the ``hot" volume filling factor becomes 
\be
\f0 =0.01\, \nu^{0.12} \e51^{1.12} c_1^{-2.8} S_{-4}\, \n0^{-1.12} K^{0.21} \, .
\label{eq:f0_A}
\ee
Recall that some of the $n$ dependence is in $K$, with the same sign as the
explicit $n$ dependence in \equ{f0_A}.
The resultant power of the $n$ dependence in zone A2 is thus always 
$\slope >1.12$.
When the second term in \equ{K} is small, the power is $\slope \rar 1.47$.
The cubic power of $\tb$ in $K$ makes this a good 
approximation almost all the way to $\tb \sim \tc$.
We conclude that throughout zone A the KS relation is reproduced 
with $\slope \simeq 1.5$, and with the normalization scaling as $\nu^{-0.26}$.

%--------------
\subsection{Zone B: Long, Continuous Burst $\ts < \tc < \tb$}
\label{sec:B}

As long as $t<\tc$, the shock is driven by a continuous-energy
wind, $\Rs \prop t^{3/5}$, following \equ{Rs_cont}.
The cooling time, radius and velocity are given by \equ{tc_cont},
\equ{Rc_cont} and \equ{Vc_cont}.

\smallskip
After the cooling time, 
during $\tc < t < \tb$, the shocked ISM has cooled, lost its pressure, and
collapsed to a thin shell, which is now pushed outward by the pressure
of the shocked wind region. Radiation losses in this region can be ignored as 
the typical gas velocity is $\sim 2000\kms$ and the temperature
is $5 \times 10^7$K. The total energy of the shocked wind region is 
\citep[][eq. 14]{weaver77}
\be
E_{\rm sw} = \frac{5}{11} \L t \, .
\ee
Since this energy is related to the pressure via
\be
E_{\rm sw} = \frac{4\pi}{3} \Rs^3 \cdot \frac{3}{2} p_{\rm sw} \, ,
\ee
the pressure is 
\be
p_{\rm sw} = \frac{5}{22\pi} \frac{\L t}{\Rs^3} \, .
\label{eq:psw}
\ee
The energy change in the shocked wind region includes the work done
by the expanding shell,
\be
\frac{dE_{\rm sw}}{dt} = \L - 4\pi \Rs^2 p_{\rm sw} \frac{d\Rs}{dt}.
\label{eq:dEdt}
\ee 
The shell equation of motion is then
\be
\frac{d}{dt} \left( \frac{4\pi}{3} \Rs^3 \rho \frac{d\Rs}{dt} \right)
= 4\pi \Rs^2 p_{\rm sw} \, .
\label{eq:dRdt}
\ee
\equs{psw}, \equm{dEdt} and \equm{dRdt} can be solved to obtain the shock
radius and velocity
\be
\Rs = 66.1 \pc\, \l38^{1/5}\, \n0^{-1/5}\, t_6^{3/5} \, ,
\label{eq:Rs_cont_post_cool}
\ee
\be
\Vs = 38.8 \kms \l38^{1/5}\, \n0^{-1/5}\, t_6^{-2/5} \, .
\label{eq:Vs_cont_post_cool}
\ee
This is similar to \equ{Rs_cont} and \equ{Vs_cont}, which describe the 
evolution of the shell in the
earlier phase driven by a continuous wind before cooling, at $t < \tc$.
The pre-factors are smaller here due to the collapse of the cooling shell,
during which the shock temporarily slows down.

\smallskip
In order to address the fading one should consider two different cases
where the fading occurs either before or after the end of the burst at $\tb$.

%------
\subsubsection{Zone B2: Very long burst $\tf < \tb$}
\label{sec:B2}

If the burst is very long, the shock velocity could reach the ISM speed of
sound while the burst is still on, during the active snow-plow phase. 
In this case, based on 
\equ{Rs_cont_post_cool} and \equ{Vs_cont_post_cool}, the fading would occur at
\be
\tf=16.8\Myr\, (\nu\e51)^{1/2} c_1^{-5/2} {\tb}_{,6}^{-1/2} \n0^{-1/2} \, ,
\label{eq:tf_B2}
\ee
\be
\Rf=285.9\pc\, (\nu\e51)^{1/2} c_1^{-3/2} {\tb}_{,6}^{-1/2} \n0^{-1/2} \, .
\label{eq:Rf_B2}
\ee
The ``hot" filling factor is therefore
\be
\f0=164\, \nu\, \e51^{2} c_1^{-7} {\tb}_{,6}^{-2} S_{-4}\, \n0^{-2}\, .
\label{eq:f0_B2}
\ee
Thus, in zone B2,
under the approximations made, the predicted slope for the KS relation
is $\slope =2$.
The normalization of th KS relation now scales as $\nu^{-1}$, but also
as $\tb^2$, and with a very strong dependence on the sound speed $c^7$.

%------
\subsubsection{Zone B1: Moderately long burst $\tb < \tf$}
\label{sec:B1}

If the burst is not so long, such that the fading occurs only after the burst
is over, $\tb < \tf$, the pre-faded shell eventually enters a regime, $t>\tb$,
where its evolution is no longer governed by \equ{Rs_cont_post_cool}.
In this regime, the dynamics of the shell is similar to the passive
snow-plow phase of
a single SN,\footnote{The switching off of the driving of the shell and the
transition to a snow-plow similar to a single SN can be
considered to be instantaneous with respect to $\tb$ since the information 
about the turning off of the energy source
is transferred from the center to the shell on a short timescale of
$\sim 0.1\Myr$,
traveling a distance $\Rs \sim 100\pc$ at a speed $\cs \sim 1000\kms$.}
\be
\Rs = \Rs(\tb) \left( \frac{t}{\tb} \right)^{2/7} 
    = 66.1\pc\, \l38^{1/5} {\tb}_{,6}^{11/35} \n0^{-1/5} t_6^{2/7} \, ,
\ee
\be
\Vs = \frac{2}{7} \frac{\Rs}{t} 
    = 18.5\kms\l38^{1/5} {\tb}_{,6}^{11/35} \n0^{-1/5} t_6^{-5/7} \, .
\ee
The fading, $\Vs=\cs$, thus occurs at
\be
\tf = 1.71\Myr\, (\nu\e51)^{0.28} c_1^{-1.4} {\tb}_{,6}^{0.16} \n0^{-0.28} \, ,
\label{eq:tf_B1}
\ee
\be
\Rf = 61.4\pc\, (\nu\e51)^{0.28} c_1^{-0.4} {\tb}_{,6}^{0.16} \n0^{-0.28} \, .
\label{eq:Rf_B1}
\ee
where $\l38$ was inserted using \equ{L38}.
The ``hot" volume filling factor is therefore
\be
\f0 = 0.167\, \nu^{0.12} \e51^{1.12} c_1^{-2.6} {\tb}_{,6}^{0.64} S_{-4}\, 
\n0^{-1.12}\, .
\label{eq:f0_B1}
\ee
Hence, in zone B1, under the approximations made,
the predicted slope for the KS relation is $\slope =1.12$.
The normalization scales weakly with $\nu^{-0.12}$ and more significantly
with $\tb^{-0.64}$ and $c^{2.6}$. 

%\smallskip
%\adr{How do we explain that the simulations give $\slope \sim 1.5$? 
%Kartick raised the suspicion
%that the middle regime, $\tc<t<\tb$, driven by pressure with $t^{3/5}$, 
%is actually not correct. The behavior seen is more like a snow-plow of a 
%single SN. Kartick, what should we say here?
%Mark mentions the result from Gentry+ that the Weaver solution is valid
%only in very massive clusters of $10^5$ SNe.}

\smallskip
% Zone C
We have not worked out analytic predictions for {\it zone C}, where
$\tc < \ts < \tf$. The spherical simulations described in \se{spheri_sim}
indicate that the behavior in zone C is similar to the behavior in zone B (B1
or B2 respectively). 
One limitation here is that for a given cluster of SNe the bubble may be in 
different zones of parameter space for different $n$ values, 
as the evolution depends on $\ts/\tf$, and $\tf$ varies with $n$.

%---------------------
\subsection{More Realistic Clusters of SNe}
\label{sec:clouds}

Before proceeding to simulations that may refine the analytic estimates of this
section, it is worth evaluating the possible assignment of
actual observed star-forming clusters to zones in parameter space
according to the above scheme.

%---------------
\subsubsection{A uniform time sequence}

% MW single cloud
First consider a uniform sequence of SNe, as estimated so far,
which may serve as a crude approximation for real clouds in some cases.
For example, we estimated for the Milky Way in \equ{tc} and \equ{tf}
that for a single SN $\tc \sim 0.05 \Myr$ and $\tf \sim 4 \Myr$.
The giant molecular cloud (GMC) lifetimes are typically tens of Myr,
after which they are disrupted by feedback,
so we may adopt $\tb = 50 \Myr\, {\tb}_{50}$.
This puts the GMCs in the ``long-burst" zones, B to D, where $\tc < \tb$.

\smallskip
In order do further distinguish between zones B to D, we should estimate the
time between SNe $\ts$.
Considering a cluster stellar mass $\Mc = 10^4 \msun\, M_4$,
and a stellar mass per SN $\msn = 100\, \mu_2$,
the number of SNe is $\nu = 100\, M_4 \mu_2^{-1}$,
which gives
\be
\ts \sim 0.5\Myr\, {\tb}_{50} M_4^{-1} \mu_2 \, .
\ee
If the cloud lifetime $\tb$ is much shorter than $50\Myr$, or for massive
clouds of $\sim\!10^5\msun$,
one may have $\ts \lsim \tc$, namely the cloud is in zone B.
Otherwise, for clouds that live longer or are less massive, one has
$\tc < \ts$, namely the cloud is in zone C.
In extreme cases, of long-lived, low-mass GMCs, or if $\msn$ is
somehow particularly large within GMCs, the time per SN, $\ts$, 
could be comparable to $\tf$ such that we are barely in zone D.
We thus expect the low-mass, intermediate and massive GMCs, under the
uniform-sequence approximation, to be in zone D, C and B respectively.
If $\tf < \tb$ they would likely be assigned to zone B2 or C2.

\smallskip
% high z
In high-redshift giant clumps the lifetime is expected to be longer,
possibly $\tb \sim 300 \Myr$, the characteristic migration time within the
violently unstable discs into the central bulge \citep{dsc09}.
On the other hand the clump masses are expected to be much larger,
comparable to the Toomre mass, 
e.g., $\Mc \sim 10^8 \msun$ \citep{dsc09,mandelker17}.
This gives $\ts \sim 3 \times 10^{-4} \Myr$, namely $\ts \ll \tc$, so the clump
is in zone B.
With $\tf < \tb$, the high-$z$ giant clumps are likely to be
assigned to zone B2.

%----------------------- 
\subsubsection{A Cloud of Clusters}
\label{sec:clusters}

To be more realistic,
we wish to evaluate the evolution of the cumulative super-bubble about a 
star-forming cloud of gas (g) that consists of smaller star clusters (c),
each generating a short burst.
We assume the number of clusters in the cloud to be
$\nu_{\rm g}=10\, \nu_{{\rm g},10}$.
For a cloud of mass $M_{\rm g}=10^5\msun\, M_{5}$,
assuming one SN per $100 \msun\, \mu_2$ (typically $\mu_2\sim 1.5$),
we expect a total of $\nu =10^3 M_{5} \mu_2^{-1}$ SNe in the cloud, 
namely $\nu_{\rm c}=100 M_{5} \mu_2^{-1} \nu_{{\rm g},10}^{-1}$ 
SNe per cluster.

%------------
%\subsubsection{Within a Cluster}

\smallskip
\ul{Within a cluster},
we write the duration of each burst of SNe as
$t_{\rm b,c}=1 \Myr\, t_{\rm b,c,6}$.
Dividing by $\nu_{\rm c}$, the average time between successive SNe is
\be
t_{\rm s,c}=0.01\Myr\, t_{\rm b,c,6} M_{5}^{-1} \mu_2 \nu_{{\rm g},10}.
\ee 
Recall that the individual SN cooling time, from \equ{tc}, is
\be
\tc = 0.05 \Myr\, \n0^{-0.55} \, ,
\ee
and the individual fading time, from \equ{tf}, is
\be 
\tf = 1.9 \Myr\, c_1^{-1.4} \n0^{-0.37} \, .
\ee 

\smallskip
We learn that, within each cluster,
\be
\frac{\tb}{\tc} =  20\, t_{\rm b,c,6} \n0^{0.55} \, .
\ee
This indicates that the cluster is not in zone A.

\smallskip
To test whether the cluster could be in zone D, we examine the ratio
\be
\frac{\ts}{\tf} = 0.0053\, t_{\rm b,c,6} M_{5}^{-1} \mu_2 
\nu_{{\rm g},10} c_1^{1.4} \n0^{0.37} \, .
\ee
This implies that the cluster is not in zone D unless the burst is very long,
the cloud is of much lower mass than $10^5\msun$, and the hydrogen density is 
very high.

\smallskip
To distinguish between zones B and C, we examine the ratio
\be
\frac{\ts}{\tc} = 0.2\, t_{\rm b,c,6}
M_{5}^{-1} \mu_2 \nu_{{\rm g},10} \n0^{0.55}.
\ee
This implies that for massive clouds and relatively short bursts
the individual clusters would be in zone B. 
If the SN burst is much longer than $\sim 1 \Myr$,
or the cloud is significantly less massive than $\sim 10^5\msun$, each cluster
would be in zone C.

\smallskip
If the cluster is in zone B, in order to distinguish between zones B1 and B2,
we insert $\nu_{\rm c}$ in the expression for $\tf$ in zone B1, \equ{tf_B1},
and obtain 
\be
\frac{\tf}{\tb} = 7.4\,  t_{\rm b,c,1}^{-0.84} 
M_{5}^{0.28} \mu_2^{-0.28} \nu_{{\rm g},10}^{-0.28} 
\e51^{0.28} c_1^{-1.4} \n0^{-0.28} \, .
\label{eq:tf-tb-subcl}
\ee
This is zone B1 if $t_{\rm b,c,6} < 10$ and all other factors are unity.
More so if $c_1<10 \kms$ and if $M_{5}>1$. However, it may be zone B2
if $t_{\rm b,c,6} > 10$, $M_5 \ll 1$, $\gamma \ll 1$, $\nu_{{\rm g},10} \gg 1$ 
or $\n0 \gg 1$.

% 12
\begin{figure*}
\vskip 6.8cm
\includegraphics{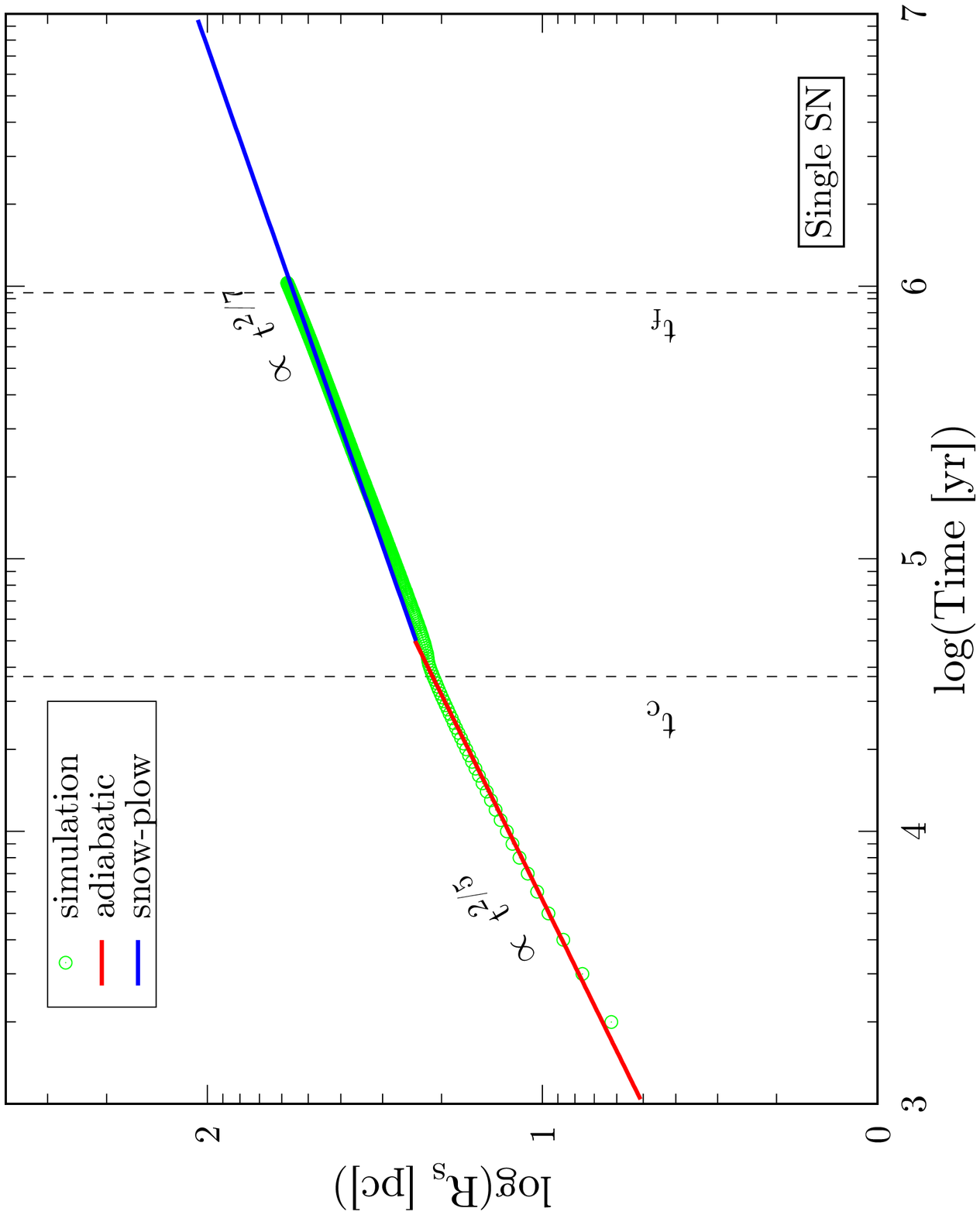}
\includegraphics{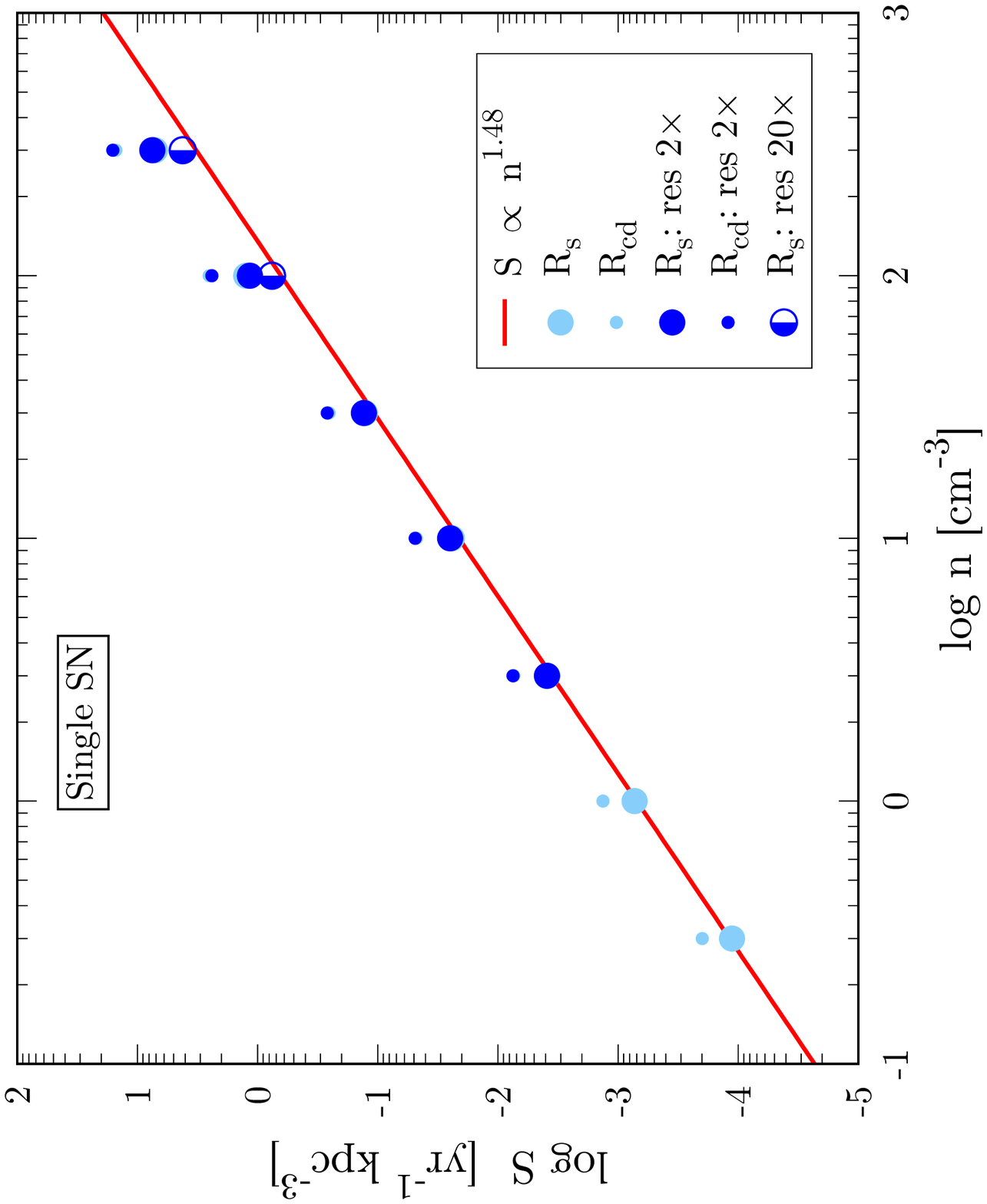}
%\special{psfile="figs/rshock_single.eps" hscale=35.5 vscale=35.5
%                  angle=-90   hoffset=-20 voffset=208}
%\special{psfile="figs/fvol_single.eps" hscale=38 vscale=38
%                  angle=-90   hoffset=230 voffset=210}
\caption{
Testing the spherical simulation with a single SN.
{\bf Left:} Shock radius as a function of time for $n=1\cmc$. The simulation
(green symbols) reproduces quite well the Sedov-Taylor phase with
$\Rs \prop t^{2/5}$ and the snow-plow phase with $\Rs \prop t^{2/7}$.
{\bf Right:} The SN rate density $S$ versus ISM hydrogen density $n$,
as derived from the fading time and radius as a function of $n$
assuming a constant bubble volume filling factor, $f_0 = 0.6$,
using \equ{filling}. 
Shown are the results for the shock radius $\Rs$ (bigger circles) and within
the inner radius of the shell $R_{\rm cd}$ (smaller circles).
Results from simulations with different resolutions are
shown in different shades of blue, as labeled.
The simulations successfully reproduce the predicted $S-n$ relation
from \equ{f0_draine}, shown as a red line.
}
\label{fig:spheri_single}
\end{figure*}

%13
\begin{figure*}
\vskip 6.8cm
\includegraphics{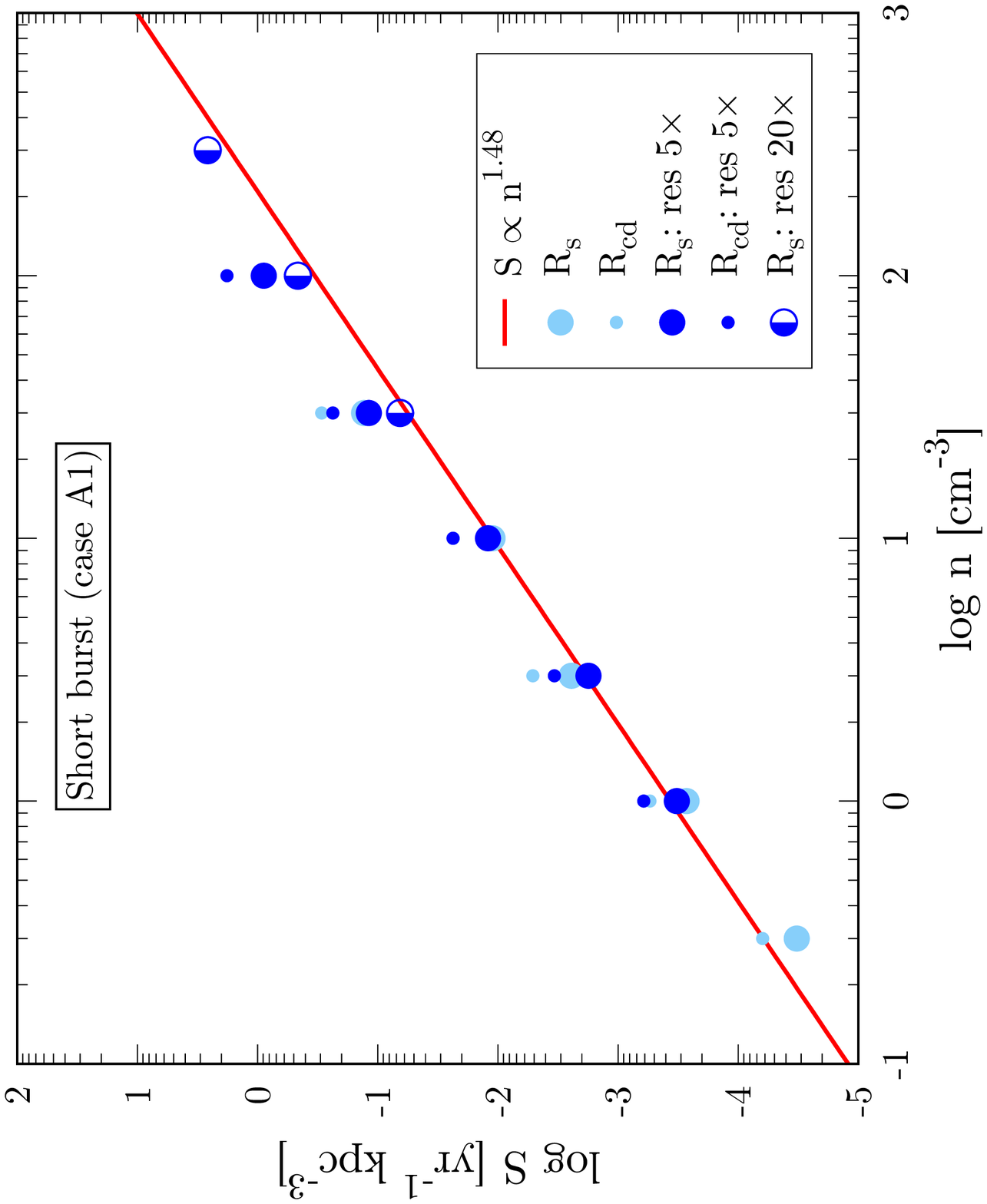}
\includegraphics{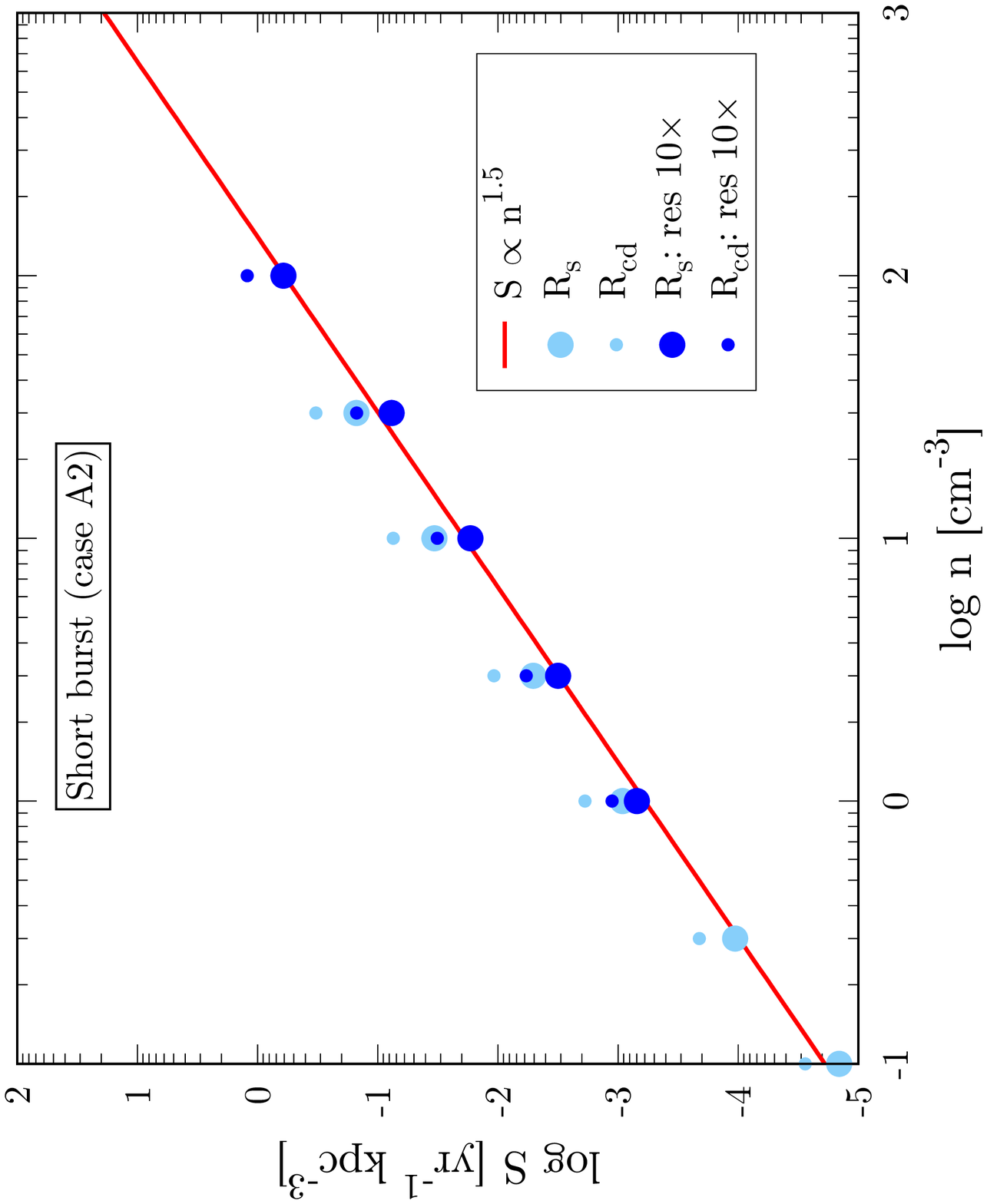}
%\special{psfile="figs/fvol_A1.eps" hscale=38 vscale=38
%                  angle=-90   hoffset=-20 voffset=210}
%\special{psfile="figs/fvol_A2.eps" hscale=38 vscale=38
%                  angle=-90   hoffset=230 voffset=210}
\caption{
Short bursts of clustered SNe in zone A.
The SN rate density $S$ required for sustaining $f_0 = 0.6$ versus ISM 
density $n$, as in the right panel of \fig{spheri_single} but 
for short bursts  
in zones A1 ($\tb \ll \tc$, left) and A2 ($\tb \lsim \tc$, right).
The simulation results (blue symbols, as labeled) are compared to the 
analytic prediction (red line).
In both cases, the simulations, at the proper resolution,
match the predicted $S \prop n^{1.5}$.
}

\label{fig:spheri_A1_A2}
\end{figure*}

%14
\begin{figure*}
\vskip 6.8cm
\includegraphics{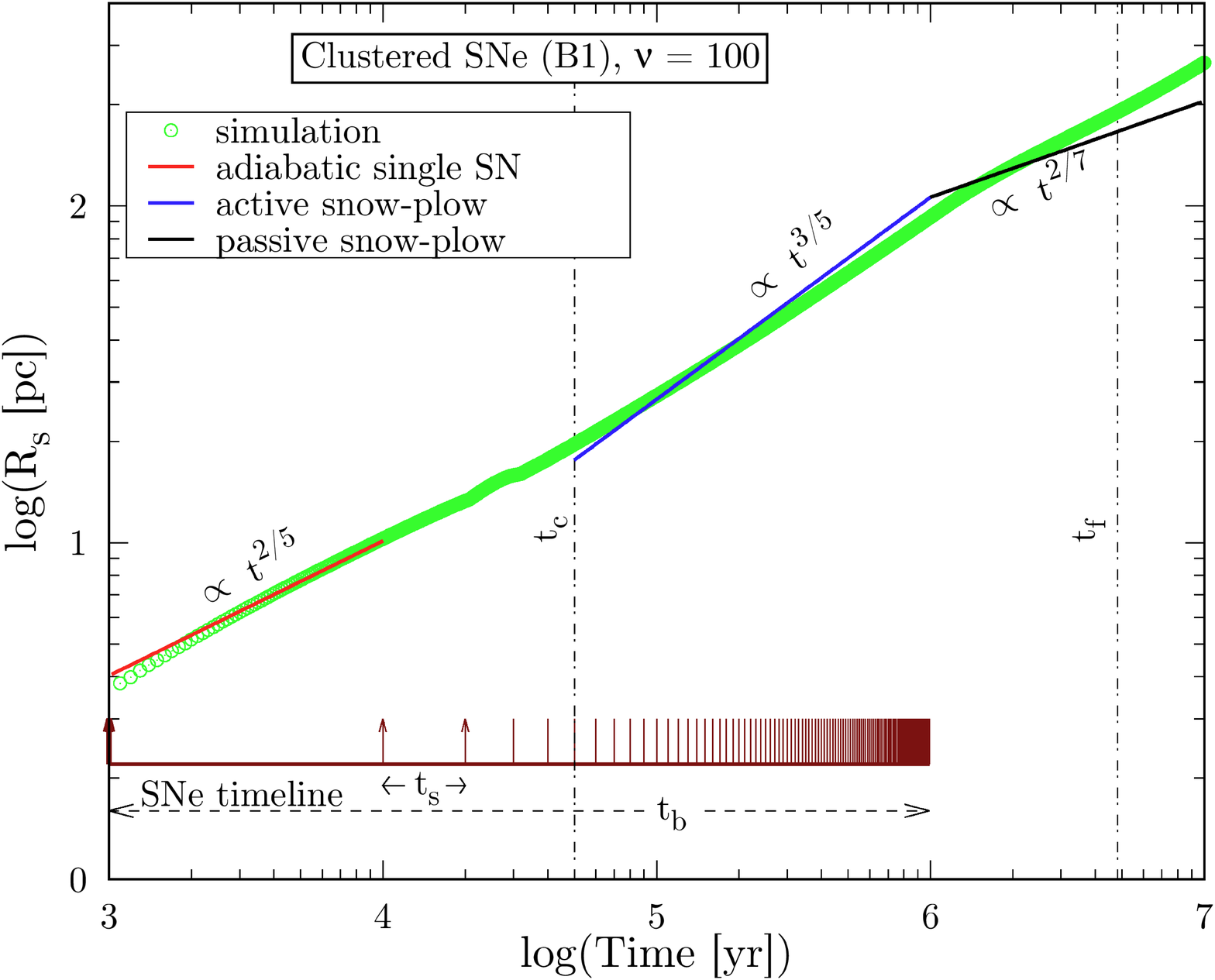}
\includegraphics{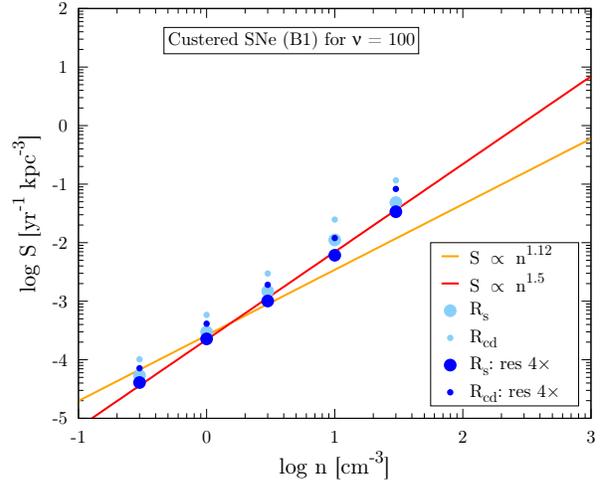}
%\special{psfile="figs/rshock_B1.eps" hscale=8.5 vscale=8.5
%                  angle=0   hoffset=-20 voffset=0.8}
%\special{psfile="figs/fvol_B1.eps" hscale=38 vscale=38
%                  angle=-90   hoffset=230 voffset=207}
\caption{         
%\adr{Revise left: Kartick comments in draft (also appendix) and his text}
Clustered SNe in zone B1 with $\nu = 100$.
{\bf Left:} Evolution of the shock radius $\Rs$ in the simulation with
$n=3\cmc$ (green symbols) compared to the analytic estimates from \se{B1}
(solid lines).
Symbols are as in \fig{spheri_single}.
The SN explosion times are are shown as brown vertical bars.
The four characteristic times are marked. 
At $t<\ts$, the simulation is closer to the solution for a single
SN ($t^{2/5}$) than for a continuous wind ($t^{3/5}$) reflecting the
discreteness of the SN explosions.
At $t\gg \ts$ and $t>\tc$, the simulation roughly follows the active snow-plow 
solution for a wind from a continuous source ($t^{3/5}$),
and when the energy source turns off at $\tb$ the numerical growth of $\Rs$
flattens toward the passive snow-plow phase but it does not fully reach the 
analytic growth rate $t^{2/7}$.
The minor wiggle near $t=3 \times 10^4\yr$ reflects the temporal slow down of
the shell when it cools and collapses to a thin shell.
{\bf Right:} The SN rate density $S$ required for maintaining $\f0=0.6$ versus
ISM density $n$, as in the right panel of \fig{spheri_single} but for the B1
case. 
The simulation results, shown in blue symbols, as labeled,
converge to $S \propto n^{1.5}$ (red line).
This is somewhat steeper from the analytic prediction, $S \prop n^{1.12}$
(orange line), and is in good agreement with the desired 3D KS relation. 
}
\label{fig:spheri_B1}
\end{figure*}

\begin{table}
\centering
\begin{tabular}{@{}lcccc}
\multicolumn{5}{c}{{\bf Spherical simulations of SN explosions}} \\
\hline
Case       & $\nu$ & $\tb$     & $\Delta r$ & $r_{\rm inj}$ \\
           &       & $\Myr$    & $\pc$      & $\pc$ \\
\hline
\hline
single SN  &  1   &  0                  & 0.024-0.0012   & 0.5 \\
A1         & 10   &  0.001              & 0.244-0.0122   & 2.0 \\
A2         & 10   & $0.5\tc$         & 0.244-0.0244   & 2.0 \\
B1         & 100  & 1                   & 0.244-0.0244   & 2.0 \\
C          & 100  & $(33\!-\!75)\tf$ & 0.122          & 2.0 \\
cloud of clusters & 10, 66 & 30, 1      & 0.244-0.122    & 2.0 \\
\hline
\end{tabular}
\caption{$\nu$ is the number of SNe in the cluster
and $\tb$ is the burst duration of the cluster.
$\Delta r$ is the grid cell size
and $r_{\rm inj}$ is the radius of the region where the SN energy is
injected.
}
\label{tab:spheri_sims}
\end{table}

%----------------
%\subsubsection{In the Cloud of Clusters}

\smallskip
\ul{In the cloud of clusters},
we assume that the duration of the burst of clusters is
$t_{\rm b,g}=50 \Myr\, t_{\rm b,g,1.7}$.
This implies a long average duration between successive clusters of
\be
t_{\rm s,g}=5 \Myr\, t_{\rm b,g,1.7} \nu_{{\rm g},10}^{-1}.
\ee
To test whether the cloud is in zone D, we compare $\ts$ with the
cumulative fading time as evaluated for each cluster in zone B1,
from \equ{tf_B1} with $\nu_{\rm c}$, and obtain
\be
\frac{\ts}{\tf} = 0.67\, t_{\rm b,g,1.7} t_{\rm b,c,6}^{-0.16}
\nu_{{\rm g},10}^{-0.72} M_5^{-0.28} c_1^{1.4} \e51^{-0.28} \n0^{0.28} \, .
\ee
This is of order unity, and can be larger (zone D) if $M_5$ is low and if $n$ 
is high.

%%%%%%%%%%%%%%%%%%%%%%%%%
\section{Clustered Supernovae - Spherical Simulations}
\label{sec:spheri_sim}

%---------
\subsection{Method of Spherical Simulations}

%Summary of simulations. 
We performed 1D spheri-symmetric hydrodynamical simulations about a fixed
center to test the
idealized analytic estimates and to extend the results to cases where analytic
modeling is more difficult.
The simulations utilize the finite volume Eulerian hydrodynamical code
\textsc{pluto}-v4.0 \citep{mignone07}, which solves the standard conservation
equations for mass, momentum and energy, with gravity turned off.
The energy is kinetic and thermal, with $\gamma=5/3$,
and it changes by input from the SN and radiative losses.
The background density is kept uniform, at $T=10^4$K, 
with a sound speed of 
$c_{\rm s}=[\gamma k_{\rm b}T/(\mu m_{\rm p})]^{1/2} 
\simeq 15.1\kms$.\footnote{This is 
assuming $\gamma=5/3$ and $\mu=0.6$, which is valid for 
$T \gsim 2\times 10^4$K, but a constant
value of $\mu$ is adopted for simplicity.}
The simulations are stopped once the shock speed is reduced to $c_{\rm s}$.
The metallicity is kept at the Solar value.
The grid resolution is uniform throughout the box and constant in time.
The default resolution is $\Delta r = 0.0244$ and $0.244\pc$ for single SNe 
and clusters respectively, and it is increased by factors up to 20 
to test for convergence (see \tab{spheri_sims}).

%Injection of SN
\smallskip
A SN is assumed to inject in the ISM gas mass of $M_{\rm sn} = 5 \msun$
and thermal energy of $E_{\rm sn} = 10^{51}\erg$, at a rate that is uniform
within a spherical volume of radius $r_{\rm inj} \sim 1 \pc$ about the center
and during a short time interval $\Delta t_{\rm inj}$, selected to be the
largest between the hydro timestep and $1\kyr$.
In the clustered SNe runs, the first SN energy is injected in a
region of $r_{\rm inj}=0.5\pc$, but the subsequent SNe are put in a larger
region of $r_{\rm inj}=2\pc$ to avoid numerical instabilities when injecting 
energy in a very low density medium.
The results are not sensitive to the exact value of $r_{\rm inj}$ and
$t_{\rm inj}$ as long as they are much smaller than the cooling radius and
time.\footnote{The small radius and time are chosen to avoid numerical effects
that could arise if the SN energy was injected in a larger volume, such as a
failure to generate a strong shock and/or having the injected gas at a 
temperature where it cools faster than it can expand.}
In order to avoid numerical instabilities due to sharp gradients of the
thermodynamic quantities at $r_{\rm inj}$, the mass and
energy injected as a function or $r$ are linearly smoothed in the shell 
$(0.9-1.1) r_{\rm inj}$.
Radiative cooling is incorporated as in \equ{Edot} and \equ{cooling}, assuming
a solar metallicity.

%----------
\subsection{Results of Spherical Simulations}

\subsubsection{Single SN}

\Fig{spheri_single} shows simulation results for a single SN, to test the
success of the simulations in reproducing the textbook analytic predictions.
The fiducial resolution is $\Delta r = 0.024\pc$ and the energy injection is
within $r_{\rm inj}=0.5\pc$.
The left panel, which shows the evolution of the shock radius $\Rs(t)$ 
for $n=1\cmc$, demonstrates that the simulation
recovers quite successfully the expected evolution in the adiabatic and 
snow-plow phases.
The right panel shows the SN rate density $S$, assuming a constant bubble
filling factor $\f0=0.6$, as a function of gas density $n$ 
for simulations with six different values of $n$.
The filling factor is computed by $f_0 = (4\pi/3) \Rf^3 \tf S$,
where $\tf$ and $\Rf$ are determined when the shock 
velocity have reduced to $15.1\kms$, the sound speed of the ISM.
Then, assuming $\f0=const.$, $S(n)$ is determined from $\Rf(n)$ and $\tf(n)$.
In addition to the shock radius that marks the outer radius of the shell,
we also determine the inner radius of the shell, $R_{\rm cd}$, encompassing 
the low-density hot bubble, defined where the shell density 
falls below half the ISM density $0.5\, n$. 
The results for the fiducial resolution are compared to the results from
simulations with twice and twenty times better resolution, indicating
convergence.
We see that the simulations roughly reproduce the predicted slope of 
$S \prop n^{1.48}$, both for $\Rs$ and $R_{\rm cd}$, as well as the predicted
normalization.
This indicates that the 1D simulations could be useful for studying the SN 
bubbles in the clustered-SN cases.

%15
\begin{figure}
\vskip 6.8cm
\includegraphics{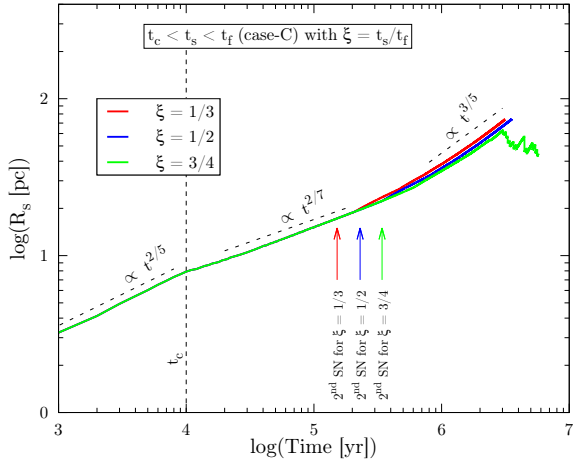}
%\special{psfile="figs/rshock_C.eps" hscale=35.5 vscale=35.5
%                  angle=-90   hoffset=-20 voffset=208}
\caption{
Clustered SNe in zone C.
The evolution of $\Rs$ for a cluster of SNe in zone C for $n=10\cmc$, with
three different values of $\xi = \ts/\tf$ (lines of different color).
The shock radius evolves as a single SN, showing the adiabatic phase
($\propto t^{2/5}$) followed by a passive snow-plow phase ($\propto t^{2/7}$)
till the second SN goes off. After a few SNe explode,
the shock evolves like a continuous wind ($\propto t^{3/5}$).
A sudden drop of $\Rs$ in case of $\xi = 3/4$ represents the fading away of the
shock.
}
\label{fig:spheri_rshock_C}
\end{figure}

%16
\begin{figure}
\vskip 6.8cm
\includegraphics{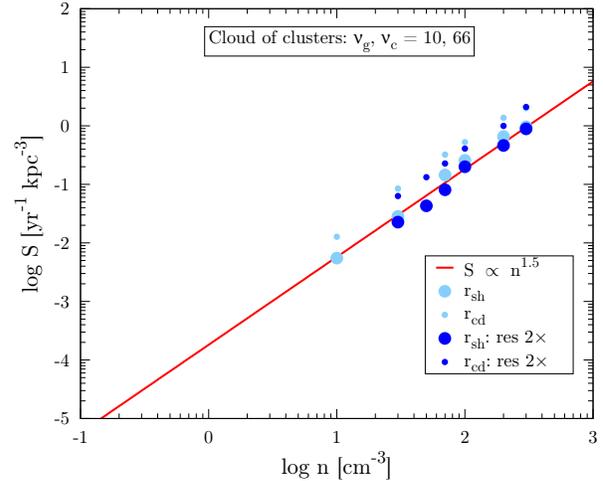}
%\special{psfile="figs/fvol_subcl.eps" hscale=38 vscale=38
%                  angle=-90   hoffset=-30 voffset=210}
\caption{
A cloud of clusters.
The SN rate density $S$ required for maintaining a constant volume bubble
filling factor of $f_0 = 0.6$ in the case of a cloud of clusters with
$\nu_{\rm g} = 10$ and $\nu_{\rm c} = 66$.
Symbols are as in \fig{spheri_single}.
The result is a good match to $S \propto n^{1.5}$.
}
\label{fig:spheri_coc_fvol}
\end{figure}

%---------------
\subsubsection{Cases A}

\Fig{spheri_A1_A2} show simulation results in the $S-n$ plane, assuming
$\f0=0.6$, for short clustered bursts in zones A1 and A2 of parameter space, 
where $\tb\ll\tc$ and $\tb\lsim\tc$ respectively.
We simulate $\nu =10$ clustered SNe in each case. 
In the A1 case the burst duration is $\tb = 10^{-3}\Myr$,
and in A2 it is chosen for each $n$ to be $0.5\tc$, given that $\tc$
depends on $n$, \equ{tc}.
We could have also changed $\nu$ as we change $n$ in order to keep a constant
luminosity, but the dependence of $\f0 \prop \nu^{0.6}$ is rather weak,
\equ{f0_A}.
The fiducial resolution is $\Delta r = 0.244\pc$ and the energy injection is
within $r_{\rm inj}=2\pc$.
The simulation results converge as the resolution is increased,
and they reproduce the analytic predictions of \se{A}, 
both in terms of slope, $S \prop n^{1.5}$, and amplitude.

%---------
\subsubsection{Case B1}

The simulation results for a clustered SNe case of type B1 are shown
in \fig{spheri_B1}, with $\nu=100$ SNe within $\tb=1\Myr$.
The fiducial resolution is again $\Delta r = 0.244\pc$ and $r_{\rm inj}=2\pc$.
Recall that the analytic model of \se{B1}
predicts an adiabatic wind phase of $\Rs \prop t^{3/5}$ until
$\tc$, followed by an active snow-plow phase of $\Rs \prop t^{3/5}$ driven by
a continuous wind as long as the energy source is on, and a transition near
$\tb$ to a passive snow-plow with $\Rs \prop t^{2/7}$ until fading.
At $t\sim\ts$, the simulated growth of $\Rs$ is closer to the single-SN
solution $t^{2/5}$ because of the discreteness of the few first SN explosions.
The simulated growth then steepens into the expected continuous wind phase,
but it shows a slightly slower 
growth rate during the active snow-plow after $\tc$\footnote{This may
partly arise from the assumption that the background pressure vanishes
and partly be a numerical effect \citep{weaver77} that is hard to overcome
both in Eulerian and Lagrangian codes.} 
and a somewhat steeper growth rate 
during the passive snow-plow, $\Rs \prop t^{0.42}$ compared to the
expected $\prop t^{2/7}$, eventually reaching fading at roughly the same 
radius as predicted.
A similar discrepancy between the simulations and the analytic model
in the passive snow-plow phase is also seen in the Lagrangian simulations 
of \citet{gentry17}, so it is likely due to an over-simplification in
the model, which is yet to be understood. 
The right panel of \fig{spheri_B1} shows that
for a constant filing factor, $\f0=0.6$, the simulations yield 
$S \prop n^{1.5}$, similar to the single-SN case and steeper than the $s=1.12$
predicted by the simplified analytic model of \se{B1}.

%-----------
\subsubsection{Case C2}

It is difficult to predict analytically the evolution of $\Rs$ in zone C where 
$\tc < \ts <\tf$ as neither a continuous wind solution can be applied 
(valid for $\ts < \tc$) nor the SNe can be considered as separable 
(valid for $\tf < \ts$). 
The evolution of $\Rs$ in a simulation in zone C is presented in 
\fig{spheri_rshock_C}, for $n=10\cmc$, with $\tc \sim 0.1\Myr$, 
and $\nu=100$ SNe.
The burst duration is chosen to be $\tb=\xi \nu \tf$ where $\xi=\ts/\tf$
varies from $1/3$ to $3/4$ 
in order to explore the behavior at $\ts \lsim \tf$. 
The simulations show, as expected, that $\Rs$ evolves like a single SN, through 
adiabatic and passive snow-plow phases, till the second SN is injected.  
We learn that after a few more SNe have been injected, the expansion rate
gradually steepens to the $\Rs \prop t^{3/5}$ expected for a continuous wind, 
as in zone B. 
Since we expect that for $\tf < \ts$ (where $\tf$ is the single-SN fading time)
the cluster would be in zone D, where the SNe can be treated as separate,
we examine the behavior in zone C as a function of proximity of 
$\ts$ to $\tf$.
We see in the figure that once $\ts > 0.7 \tf$ the shock stops expanding
and it fades away (with a density jump of only 20\%) 
after seven SNe ($t \gsim 7 \ts$) despite the fact that SNe continue to explode
at the centre. This implies that for $\ts > 0.7 \tf$ the cluster is already
in the separable zone D, where the toy model predicts $\slope = 1.5$.

\smallskip
It is difficult to study the $S-n$ relation through simulations in zone C, 
as we did in zone B,  because the behavior is expected to depend on $\ts/\tf$ 
while $\tf$ depends on $n$.
Therefore, for a fixed $\ts$ (given $\tb$ and $\nu$, namely the same $\l38$),
the bubble may be assigned to a different parameter zone for different $n$ 
values.

\smallskip
We note that a similar conclusion regarding the validity of the wind solution
has been obtained by \cite{gentry17}, using a
Lagrangian code, who found that for clusters with $\nu \sim 10$ the wind 
solution in \equ{Rs_cont_post_cool} is not valid as $\ts/\tf \gsim 1$,
while an approximate wind solution prevails when $\nu \gsim 100$.

%--------------------
\subsubsection{A cloud of clusters}

When simulating a cloud of clusters, we consider a total of 660 SNe
(corresponding to a $10^5 \msun$ super star cluster)
over a time period of $30\Myr$. The SNe are divided into $\nu_{\rm g}=10$
clusters, each containing $\nu_{\rm c} = 66$ SNe.
The mass and energy of SNe inside a cluster
are injected at time intervals of $\ts = 1/66 \Myr$, while
subsequent clusters are switched on at times separated by $3\Myr$, each lasting
for $1\Myr$.

\smallskip
\Fig{spheri_coc_fvol} shows the $S-n$ relation, for a fixed $\f0=0.6$,
from simulations of a cloud of clusters.
We find the result to be, again, a good match to $S \prop n^{1.5}$.
\Equ{tf-tb-subcl} tells us that for $n > 3\cmc$ the fading time for 
the individual clusters is $\tf < 3\Myr$ 
(considering $t_{{\rm b,c},6} = 1$ and $c = 15.1$ \kms),
which is smaller than the separation between two consecutive clusters. 
This allows the clusters to contribute to the volume filling factor 
independently following a case B1 solution, which according to the simulation
in \fig{spheri_B1} is $\slope \simeq 1.5$

%--------------
\subsection{Convergence of the simulations} 

\Fig{spheri_single}, referring to a single SN, seems to show in the left panel 
that for $n_0=1$, with the default resolution of $\Delta r = 0.244 \pc$,
the simulated evolution of the shock radius recovers the 
theoretical prediction quite accurately. However, the right panel
demonstrates that this is not the case at higher densities, where the simulated
$S$ (namely the measured filling factor) 
at the default resolution (large filled light-blue symbols, sometimes hidden
behind the blue symbols that are for a resolution twice as good) 
overestimates the theoretical prediction by $\sim 0.2$dex at $n_0=2$.
This is a resolution effect, as at high ISM densities the shell density is 
also high, requiring higher resolution for a proper treatment of 
the cooling at the shock front and at the interface between the bubble and the 
shell. 
Increasing the resolution by a factor of 20 (blue-white symbols) makes the
simulated $S$ approximate the theoretical solution to better than $0.1$dex even
at $n_0 = 300$. 
\Fig{spheri_A1_A2}, left panel, where we also have an accurate theoretical
model, shows a similar improvement of accuracy with improved resolution. 

\smallskip
We thus conclude that our simulations for a single SN converge to the
theoretical solution,
and adopt the same resolution in the simulations of clustered SNe.
It is interesting to note that the slope $\slope$ in the $S-n$ 
(namely in the KS) relation, as derived from the clustered-SN simulations,
shown in \fig{spheri_A1_A2} (right), \fig{spheri_B1} (right), and 
\fig{spheri_coc_fvol}, at $n_0\geq 1$, is rather insensitive to the resolution,
and can thus be derived from the simulations with the default resolution as
well as from the simulation with higher resolution.
We verified that 
our Eulerian simulations and the Lagrangian simulations of \citet{gentry17},
when performed on a similar zone of parameters, roughly
agree on the range of validity of the continuous wind solution .

\smallskip
We note that the convergence tested here is limited to the bubble filling 
factor, based on the
radius of the outer shock and its fading time. We do not attempt to study
convergence in other dynamical quantities such as the deposited momentum or 
the energy budget, as studied by \citet{gentry17}.

%%%%%%%%%%%%%%%%%%%%%%%%%%%%%%%%%%%%%
\section{Photo-ionized Bubbles}
\label{sec:ionization}

%=========================
\subsection{Introduction}

% intro, SN is more effective in sims
Massive O/B stars, which precede the SNe,
emit UV radiation that ionizes a Str\"omgren sphere around them, 
creating a bubble in which the SFR is suppressed, as in the SN bubble,
but without totally evacuating the bubble interior. 
On galactic scales, supernovae clearly dominate the energy and momentum budget.
Simulations show that photoionization alone cannot prevent the ionized ISM
gas from catastrophically collapsing into molecular clouds and forming stars too
rapidly, while the inclusion of SN feedback generates global turbulence in the
ISM that properly suppresses the collapse to molecular clouds and maintains a
diffuse atomic phase.
The stellar-driven bubbles may dominate the destruction of molecular gas prior
to the ignition of SNe in small star-forming clusters 
\citep{matzner02,krum09_HII,fall10},
but in massive molecular clouds photo-ionization is expected to become 
ineffective once the escape velocity exceeds $5\kms$ 
\citep[simulations by][]{dale17}.
Under any conditions where photoionization dominates,
one may apply in the photo-bubble case considerations involving the bubble 
filling factor analogous to the derivation of the KS relation by SN bubbles.

\smallskip % static
Consider an emission rate of ionizing photons ($h\nu>13.6$ eV)
$Q = 10^{49}\, Q_{49}$, and $T=10^4 {\rm K}\, T_4$ for the equilibrium 
temperature within the ionized
sphere\footnote{This temperature varies from $6\times10^3$K to $1.5\times 
10^4$K depending on the source and the medium. For a typical O star, with color
temperature of $4\times 10^4$K, in a medium of solar metallicity and density of
$100 \cmc$, the equilibrium temperature is $\sim 7\times 10^3$K.}
In a very crude approximation, assuming a 
uniform density in a static bubble \citep[e.g.][Chapter 15]{draine11},
the Str\"omgren sphere radius $\Rst$ is determined by equilibrium between 
ionization and recombination rates, of the form $Q \propto \Rst^3 n^2$, 
yielding $\Rst \propto  n_0^{-2/3}$.
If the relevant ``fading" time is the fixed lifetime of the O star,
the volume filling factor of the static Str\"omgren bubbles becomes
\be 
\fst = \frac{4\pi}{3} \Rst^3 \tst \Sst
\simeq  0.43\, Q_{0,49}\, T_4^{0.84}\, {\Sst}_{-4}\, n_0^{-2} \, ,
\label{eq:fst_stat}
\ee
where $\Sst = 10^{-4} \kpc^{-3} \yr^{-1} {\Sst}_{-4}$ is the star-formation
rate density of O stars.
This would have implied a slope of $\slope=2$ in the KS relation.

\smallskip % Dynamic photo-bubbles}
However, the above static estimate does not account for the excess pressure 
created inside the Str\"omgren sphere, which makes the bubble expand beyond the 
static estimate.  The pressure excess could range from a factor of two due to 
the doubling of the number of particles by the ionization (if the temperatures 
inside and outside the bubble are similar) to a factor of a few hundreds 
(if the outside medium is significantly cooler than the inside, as in a dense 
star-forming molecular cloud). Moreover, a spatially dependent radiation 
pressure from the star is also expected to affect the dynamics of the bubble. 
Therefore, one should appeal to a dynamical evaluation of the pressure inside 
and outside the bubble in order to properly estimate the final radius of the 
bubble, when the shell velocity has slowed down to the background speed of
sound or turbulence velocity dispersion. 

\smallskip  % gas vs radiation pressure
Following \citet{krum09_HII}, there is a characteristic radius (and time) below
which the expansion is driven by radiation pressure and above which by gas
pressure, approximated by \citep{krum09_HII,fall10}
\be
r_{\rm ch} \simeq 0.021 \pc\, Q_{49} \, ,
\label{eq:rch}
\ee
where the standard values for the other parameters have been assumed.
This implies that the radiation pressure is effective only early in the central 
region while most of the evolution is dominated by the gas pressure, so we
consider only the gas pressure below. 
Under the assumption that the shell expands slowly enough that the ionized gas 
has time to become uniform in density, which requires that sound waves be able 
to cross the ionized bubble, 
the self-similar expansion rate is
\be
\left( \frac{r}{r_{\rm ch}} \right) = 
\left[ \frac{49}{36} \left( \frac{t}{t_{\rm ch}} \right)^2 \right] ^{2/7} \, ,
\label{eq:xgas}
\ee
\be
t_{\rm ch} \simeq 2.9 \yr\, Q_{49}^{3/2}\, n_0^{1/2} \, .
\label{eq:tch}
\ee
The radius and velocity of the shell at time $t = t_6\Myr$ are thus
\be
r(t)= 30.3 \pc\, Q_{49}^{1/7}\, n_0^{-2/7}\, t_6^{4/7} \, ,
\label{eq:rt}
\ee
\be
v(t) = 16.9 \kms\, Q_{49}^{1/7}\, n_0^{-2/7}\, t_6^{-3/7} \, .
\label{eq:vt}
\ee
This similarity solution is only valid for expansion speeds below the sound 
speed in the ionized gas $\sim 10 \kms$). 
This assumption is satisfied or close to it for all real HII regions
\citep{draine11}.

%========================================
\subsection{Photo-bubbles in two regimes}

We consider a single star, or a short-duration cluster where $Q_{49} \sim \nu$,
and where the lifetime of the ionizing source is 
$\tlf = 5\Myr\,\, \tlff$.
The fiducial value of $\sim 5\Myr$ is expected for massive stars in a
cluster of $10^6\msun$ \citep{leitherer99}.

\smallskip
The stalling of the photo-bubbles can be evaluated in two different 
regimes of parameter space. 
In case A, the shell velocity slows down to the background sound speed,
$v=c$, while the continuous ionization source is still alive, $\tst < \tlf$.
In case B, $c< v(\tlf) <2c$, making the shell stall at $\tlf$ (see below).
In hypothetical case C, $v(\tlf) > 2c$, the shell could have 
continued to expand in a snow-plow phase after $\tlf$ 
until stalling with $v=c$ at $\tst > \tlf$.
However, this is not likely to happen in HII regions, where the expansion 
speed is limited to $\leq 10\kms$ while the background sound speed 
(or turbulence velocity) is typically $\gsim 5\kms$.

%-----------------------
\subsubsection{Case A: stall while the ionization source is alive}

% halt by velocity
The shell expansion halts once the speed of the ionization 
front becomes equal to the external sound speed or turbulence velocity 
dispersion, while the ionization source is still alive.
Based on \equ{vt} and \equ{rt}, the stalling time and radius are
\be
\tst = 3.22 \Myr\, Q_{49}^{1/3}\, c_1^{-7/3}\, n_0^{-2/3} \, ,
\label{eq:tst_vel}
\ee
\be
\Rst = 58.8 \pc\, Q_{49}^{1/3}\, c_1^{-4/3}\, n_0^{-2/3} \, .
\label{eq:rst_vel}
\ee
The volume filling factor of bubbles becomes\footnote{A full numerical 
solution including the radiation pressure term in \equ{xgas} indicates that  
this analytic estimate underestimates the filling factor by $\sim 30\%$  
but the dependencies on the parameters are correct.}
\be
\begin{aligned}
\fst
    &= \frac{4\pi}{3} r_{\rm stall}^3\, t_{\rm stall} \, \Sst \\
    &= 0.27\,
       Q_{49}^{4/3}\, c_1^{-19/3}\, n_0^{-8/3}\, {\Sst}_{-4} \, .
\end{aligned}
\label{eq:fst_vel}
\ee
For a constant $\fst$ this implies a KS slope $\slope \simeq 2.6$ in case A.

\smallskip
The range of validity of case A, defined by $\tst \leq \tlf$, 
translates via \equ{tst_vel} to 
\be
Q_{49} \leq 3.7\, c_1^7\, n_0^2\, \tlff^3 \, .
\ee
For a fiducial star cluster with a lifetime $\tlff=1$, embedded in a
molecular cloud of $n_0=100$ and $c_1 = 0.5$, this reads
$Q_{49} \leq 300$, which refers to a low-mass star cluster of 
$\Ms \simeq 100\, Q_{49} \leq 3\times 10^4\msun$.
Note that this upper-limit mass is very sensitive to the sound speed 
(or velocity dispersion).

%-----------------------
\subsubsection{Case B: stall when the ionization source dies}

If the shell is still expanding with a velocity well above the background speed
of sound once the ionization source shuts off, it will tend to enter a 
snow-plow expansion phase of $r \prop t^{2/7}$. 
In the transition from the expansion as $r \prop t^{4/7}$ to 
$r \prop t^{2/7}$, the velocity would drop by a factor of two.
This implies that if $c< v(t=\tlf) <2c$, the expansion will stall right at
$\tlf$ with no further expansion.

\smallskip
Substituting the condition $v(\tlf) \geq 2 c$ in \equ{vt} gives
\be
Q_{49} \leq 390\, c_1^7\, n_0^2\, \tlff^3 \, .
\label{eq:Qv}
\ee
For the fiducial star cluster above, this reads
$Q_{49} \leq 3.2 \times 10^4$, namely $\Ms \leq 3.2\times 10^6\msun$.
Again, this upper-limit mass is very sensitive to the value of $c$.

\smallskip
Substituting for the stalling time $\tst = \tlf$ in \equ{rt} gives 
\be
\Rst = 76.1\pc\, Q_{49}^{1/7}\, \tlff^{4/7}\, n_0^{-2/7} \, ,
\label{eq:rst_death}
\ee
\be
\fst = 0.92\, Q_{49}^{3/7}\, \tlff^{19/7}\, n_0^{-6/7}\, {\Sst}_{-4} \, .
\label{eq:fst_death}
\ee
This implies a KS slope $\slope=0.86$ in case B.

%-----------------------
\subsubsection{Hypothetical Case C: stall after the ionization source dies}

In the hypothetical case where $v(\tlf) > 2c$ (possible only if $c<5\kms$), 
for a massive cluster where \equ{Qv} is invalid,
when the ionizing photons shut off the shell could in principle enter a
snow-plow phase, $r \prop t^{2/7}$. The stalling time and radius and the bubble
filling would then be 
\be
\tst = 1.5\Myr\, Q_{49}^{1/5}\, c_1^{-7/5}\, \tlff^{2/5}\, n_0^{-2/5} 
\, ,
\label{eq:tst_star}
\ee
\be
\Rst = 54 \pc\, Q_{49}^{1/5}\, c_1^{-2/5}\, \tlff^{2/5}\, n_0^{-2/5} \, ,
\label{eq:rst_star}
\ee
\be
\fst  = 0.11\, Q_{49}^{4/5}\, c_1^{-13/5}\, \tlff^{8/5}\, 
n_0^{-8/5}\,  {\Sst}_{-4} \, .
\label{eq:fst_star}
\ee
This implies a KS slope $\slope = 1.6$ in this unlikely case C. 
 
\smallskip
In summary, we expect for photo-bubbles alone (ignoring the following SNe) 
three regimes, depending on $Q_{49}$, namely on the star-cluster mass.
For the fiducial values of the parameters ($\tlff=1$, $n_0=100$ and $c_1=0.5$), 
cases A, B and C are expected to be valid for cluster masses 
$\Ms < 3\times 10^4\msun$, 
$3\times 10^4 < \Ms < 3.2\times 10^6\msun$ and $\Ms > 3.2\times 10^6\msun$,
respectively, with case C being unrealistic (possibly valid only for ISM sound 
speeds lower than $5\kms$).
The KS slopes are expected to be $\slope = 2.6, 0.86, 1.6$ respectively.
However, as we will see next (and as we saw in the simulations, \se{iso}), 
the photo bubbles are expected to be overwhelmed by the SN bubbles.

%=============================
\subsection{SNe in a photo-bubble}

% photo + SN single
For a single O star followed by a single SN (or a short burst of stars),
one can first envision a Str\"omgren bubble growing about the O star. 
During this period the SFR is suppressed within the bubble
and the interior gas density is reduced due to the expansion of the shell.
Then the SN turns on, generating a bubble that grows more-or-less
following the standard evolution of a SN bubble but in a lower
medium density. 
The reduced number density inside the photo-bubble, according to eq.~3 of
\citet{krum09_HII} and \equ{xgas}, is  %Krumholz Matzner, 2009, 2010 
\be
n_{II,0} = 2.27\, Q_{49}^{2/7}\, n_0^{3/7}\, t_6^{-6/7} 
= 0.54\, Q_{49}^{2/7}\, n_0^{3/7}\, \tlff^{-6/7}
\, ,
\label{eq:nII}
\ee
assuming that the SN goes off at time $\tlf$ 
after the birth of the O/B star.
Substituting the density from \equ{nII} in \equ{f0_draine} yields for the SN 
bubble filling factor
\be
f_0 = 0.60\, e_{51}^{1.26}\, c_1^{-2.6}\, Q_{49}^{-0.42}\, \tlff^{1.27}\,
n_0^{-0.63}\, {\Sst}_{-4} \, .
\label{eq:f0_II}
\ee

\smallskip % compare filling factors
To evaluate whether the SN bubble will grow bigger than the photo bubble, we
compare the SN filling factor from \equ{f0_II}
to the photo-bubble filling factor in each of the three cases.
Assuming $e_{51} = Q_{49} = \nu$, we obtain
\be
\begin{aligned}
\frac{f_0}{\fst} 
   &= 2.2\, \nu^{-0.49}\, \tlff^{1.3}\, c_1^{3.7}\, n_0^{2.0} ,
\quad &{\rm Case\ A} \, ,\\ 
   &= 0.65\, \nu^{0.41}\, \tlff^{-1.4}\, c_1^{-2.6}\, n_0^{0.23} , %nu^{0.55}
\quad &{\rm Case\ B} \, ,\\ 
   &= 5.5\, \nu^{0.04}\, \tlff^{-0.33}\, n_0^{0.97} ,
\quad &{\rm Case\ C} \, .
\end{aligned}
\label{eq:frat}
\ee
For the fiducial choice ($\tlff=1$, $n_0=100$, $c_1=0.5$), we get
$\nu \simeq 300$ and $3\times 10^4$ for the threshold values of $Q$ separating
photo-bubble cases A from B and B from C respectively.
With $\nu < 300$ in case A we get $f_0/\fst > 104$,
with $\nu > 300$ in case B we get $f_0/\fst > 117$, and  % I had 262 
with $\nu > 3\times 10^4$ in case C  we get $f_0/\fst > 724$.  
This implies $f_0/\fst \gg 1$ in all three cases, namely the 
SN bubbles are expected to be larger than the photo-bubbles and
dominate the filling factor and therefore the KS relation. 

\smallskip %gupta + krum
Another factor that might have reduced the importance of the photo bubble 
is the accompanying stellar wind, which may take over the dynamics 
before the SN bubble takes over, as indicated in  
a 1D simulation of a $10^6\msun$ cluster by \citet{gupta16}.
They assumed an instantaneous burst of star formation and the corresponding
sequence of SNe, where the radiation and mechanical 
luminosities of winds and SNe were computed using STARBURST99 
\citep{leitherer99}. 
This cluster is in what we term zone B2.\footnote{With a SN-burst duration of 
$\sim 40\Myr$ and inter-SN interval $\ts\sim 0.004 \Myr$,
and with a cooling time for an individual SN at $n_0\sim 100$ of
$\tc \sim 0.005\Myr$ and fading time $\tf \sim 25\Myr$ for the super-bubble,
we have $\ts < \tc$ and $\tf < \tb$.}
Almost independent of background density, the radiation force turned out to be 
important only during the first Myr, after which the stellar wind takes 
over in this simulation, and the photo-ionized bubble collapses to the 
wind-driven thin shell.  Eventually, after $\sim 3 \Myr$, the dynamics of the 
super-bubble is driven by the SNe.
One should note, however, that the importance of the wind in the intermediate 
stage is controversial, as it depends criticality on how efficiently it is 
trapped in the HII region, versus how much of it leaks 
out \citep{krum19_araa}. In simulations, 
this depends sensitively on the initial conditions, and in 1D simulations 
it depends on the subgrid model used to represent wind leakage. Observations 
suggest efficient leakage, which makes winds sub-dominant compared to ionized 
gas and radiation \citep{lopez11}.

\smallskip
We conclude that for a short burst of clustered stars and SNe the 
SN filling factor is expected to dominate the bubble filling factor 
and thus determine the KS relation.
This is confirmed in the isolated-galaxy simulations presented in \se{iso},
where \fig{iso_f_4panel}, \fig{iso_KS_f15} and \fig{iso_KS_f50} demonstrate
that the SN-feedback overwhelms the HII feedback as the mechanism that
produces the constant hot filling factor and provides a tight global KS
relation with a global slope $\slope \simeq 1.5$.
It should be borne in mind, however, that
this encouraging confirmation of our theoretical considerations is naturally 
limited to the local SFR recipes and feedback processes as implemented in 
these specific simulations.

%%%%%%%%%%%%%%%%%%%%%%%%%%%%%%
\section{Discussion}
\label{sec:disc}

Here we address some of our key assumptions, discuss potential caveats, and
comment on the observed KS relation.

%----------------------------
\subsection{Negative or positive feedback?}
\label{sec:positive}

Among our simplifying assumptions,
we adopted here the notion that SN feedback (and any other stellar feedback)
is negative, suppressing the SFR in the SN bubbles.
This is motivated by the robust fact that when SN feedback is
incorporated in simulations, the timescale for star formation typically
slows down from the free-fall time ($\epsf \sim 1$) to a timescale larger
by two orders of magnitude ($\epsf \sim 0.01$).
% positive
We thus ignore here the
counter possibility that stellar or SN feedbacks may actually be positive,
triggering star formation in the bubble shells. This has been addressed in a
limited way by theory \citep{elmegreen77} and observations
\citep{samal14,egorov17}, the latter showing indications for high SFR near
feedback-driven super-bubble structures from O-type stars.
However, based on simulations, caution has been called for when interpreting
these observations in terms of ``triggering" \citep{dale15}.
% mechanical suppression
It is obvious that SFR is suppressed in the interior of a SN bubble, which is
hot and dilute. However, after a short while, most of the bubble gas has been
swept into a dense and relatively cold shell, raising the question of why the
SFR is suppressed there. The common wisdom is that the suppression is largely
by mechanical effects, such as ram pressure associated with the high
velocity of the shell with respect to the ISM, or the high turbulence within
the very perturbed and non-uniform shell and the associated strong shear.
This was partly demonstrated, e.g., in high-resolution simulations of SNe in a
single molecular cloud \citep{rogers13}.
% assume negative
While the issue of negative versus positive feedback is still an open issue
which deserves further study, as it is for AGN feedback \citep{silk13,bieri15},
our analysis was based on the assumption that the SFR is suppressed
in all the ISM gas that has been swept by the SN bubble.
The swept gas mass filling factor can
thus be approximated by the volume filling factor of the hot bubbles.

%--------------------
\subsection{Molecular gas}
\label{sec:molecule}

% dwarfs
In our model we have assumed that the portion of the ISM that is not filled by
SN bubbles is mostly star-forming molecular hydrogen. We have not included the 
possibility that a significant portion of the ISM might consist of 
non-star-forming, warm HI. As applied to weakly star-forming regions such as 
dwarf galaxies and the outer disks of modern spiral galaxies, this is in fact 
not a good assumption. In these regions warm HI dominates the neutral portion 
of the ISM by both mass and volume, and star formation is likely regulated at 
least in part by the thermal and chemical transitions between the warm, 
non-self-gravitating, non-star-forming and cold, bound, star-forming phases 
\citep{krum09_ks, ostriker10_ks, krum13_ks_lowH2, kim13, forbes16}
For such galaxies the observed KS relation is significantly more complex than 
the simple power law relation and constant $\epsilon_{\rm ff}$ that describes 
the molecular phase; the star formation rate in atomic gas shows significant 
dependence on third quantities such as metallicity and stellar density 
\citep{bolatto11, shi14, jameson16}.
The simple model we present here is not intended to apply to galaxies with a 
significant warm atomic component.

\smallskip % high z
Fortunately, this is a modest limitation when it comes to high-redshift 
galaxies. Both observations 
\citep{bigiel08, lee12, wong13}
and theory 
\citep{krum09_H2, sternberg14}
suggest that 
galaxies transition from atomic-dominated to molecule-dominated at a surface 
density $\approx 10 (Z/Z_\odot)^{-1}\, \msun\pc^{-2}$, where $Z$ is the 
metallicity. The VELA simulations that we use to test our model are all 
well above this threshold (c.f. \fig{patches}),
as are essentially all observed high-redshift galaxies 
\citep{genzel10, daddi10_gas, tacconi13}.
Thus our model applies to the majority of observable star-forming systems 
beyond the local Universe.

\smallskip % low z
For low-redshift galaxies the situation is somewhat more complex. Local dwarf 
galaxies have surface densities below the critical 
$\approx\!10 (Z/Z_\odot)^{-1} \msun \pc^{-2}$ that marks the 
HI to H$_2$ transition throughout their area, and thus are almost 
entirely HI-dominated.  
Local spirals, on the other hand, tend to cross the threshold from mostly 
HI to mostly H$_2$ at radii $\approx\!0.5\, r_{25}$ 
\citep{leroy08, schruba11}, 
a radius that is relatively close to the stellar scale length. 
Star formation follows H$_2$, so this is also the radius within which most
star formation is concentrated. Our model is therefore reasonably applicable to
the central parts of modern spirals, where a substantial fraction of their 
stars form.
% \textsc{i}

%-----------------------
\subsection{Potential effect of gravity}
\label{sec:gravity}

Our analytic modeling ignores the gravitational force acting on the expanding 
bubbles, which may, in principle, affect the expansion and even cause a 
re-collapse. 
This is discussed using sophisticated 1D models by \citet{rahner18, rahner19}.
However, their analysis of the re-collapse effect is mostly concerned with 
non-SN feedback.
For SNe, this is is unlikely to be important because of the following argument.
A single SN, even if any enhancement from clustering is ignored, reaches a 
terminal momentum of $\sim 3 \times 10^5 \msun \kms$. Averaging over an IMF 
that makes 1 SN per $100\msun$, this gives $3\times 10^3\msun\kms$ per 
$1\msun$ of stars formed. 
If $1\%$ of the cloud mass is turned into stars, the momentum budget per mass 
of gas in the GMC is reduced to only $30\kms$
This is to be compared to the typical escape speeds of GMCs, 
which are closer to $5\kms$.  
This indicates that self-gravity is not a relevant consideration for SN remnant
evolution. It might be relevant for other feedback mechanisms, which are 
working with much lower momentum budgets.

%--------------------
\subsection{Super-bubble blow-out}
\label{sec:blowout}

The analysis so far assumed that the bubbles are confined to the galactic
disc.
This is probably a valid assumption for unclustered SNe, where the fading
radius is on the order of a few tens of parsecs (\equnp{Rf}), significantly
smaller than the typical disc height. 
However, super-bubbles of multiple SNe in massive clusters (high $\nu$) 
could have a fading radius that exceeds the disc height 
(\equsnp{Rf_A}, \equmnp{Rf_B1}, \equmnp{Rf_B2}). 
In this case, the bubbles will blow out from the disc before completing their
expansion. Cold gas from the shells is expected to be ejected 
with velocities comparable to the escape velocity from 
the disc, while the over-pressured hot gas from the bubble interior is expected
to escape at higher velocities. 
This may change the considerations concerning the self-regulation
to a constant bubble filling factor, which should now consider the
balance between outflows and inflows, including recycling.  
% self-regulation by balance of inflow and outflow
Self-regulation could in principle be achieved in this case as well, 
as a high SFR would be associated with strong outflows, 
which remove (mostly cold) gas and thus decrease the SFR.
To maintain the SFR, the disc has to be fueled with cold gas, 
by cosmological inflow or by return of the outflowing cold gas.
Alternatively, it is possible that the bubble filling factor is self-regulated 
by bubble blow-out without involving the SFR and SN feedback at all, namely
once the filling factor is larger than $\sim 0.5$, it generates outflows from
the disc that keeps $f$ near $0.5$. 
We defer the analysis of these cases of major outflows and inflows
to future work.

\smallskip %non-negligible outflows
If the outflows are non-negligible but they do not change the overall picture
where the bubbles are largely confined to the disc,
one may repeat the analysis with the fading radius replaced by 
the half-height of the disc, $h$, assuming that the bubble stops growing within
the disc plane once it blows out of the disc in the vertical direction. 
In this case the relevant bubble volume is $\Vsn \sim h^3$.
The disc height is determined by the vertical balance between gravity 
and pressure gradient, $dP/dz \prop G \rho \Sigma$, where $\Sigma$ is the
surface density in the disc. With $dP/dz \prop \rho c^2/h$, this gives
$h \prop c n^{-1/2}$, or  
\be
\Vsn \propto c^3\, n^{-3/2} .
\ee
If we crudely assume that the time for the bubble to reach the disc edge 
scales like the single-SN fading time, $t(h) \prop n^{-0.37}$,
the bubble filling factor would be 
\be
\fh \propto \Vsn\, t(h)\, S \propto n^{-1.87}\, S \, .
\ee 
With a constant $\fh$ we therefore expect a KS relation $\drhos \prop n^\slope$
with $\slope$ in the range $1.5-2.0$.
Alternatively, for clustered SNe, one can 
use the shell expansion rate in the continuous source phase,
$\Rs \propto t^{3/5}\, n^{-1/5}$ (\equnp{Rs_cont} or \equnp{Rs_cont_post_cool}).
Assuming as above $\Rs=h$ and $h \prop c n^{-1/2}$, one obtains
$t(h) \propto c^{1.67}\, n^{-0.5}$, and the bubble filling factor becomes
\be
\fh \propto n^{-2}\, S \, ,
\ee
namely $\slope=2$ in the KS relation.

\smallskip % timescales 
The importance of outflows can be evaluated by comparing the timescale for
outflows with the timescale for self-regulation of the bubble filling factor.
For the latter we take as a reference the bubble fading time, which is  
$\tf \sim 1 \Myr$ for single SNe,
$\tf \sim 10 \Myr$ for SNe in low-mass clusters,
and $\tf \sim 100 \Myr$ for massive clusters.
% MW
Consider Milky-Way-like galaxies as an example,
where the outflow rate, if comparable to the SFR, is 
$\dot{M} \lsim 10 \Msun\yr^{-1}$, and the gas mass is $M\sim 10^{10}\msun$. 
The timescale for gas depletion by outflow is thus 
$t_{\rm out} = M/\dot{M} \gsim 1\Gyr$ (or even longer for the large-scale 
outflow, given that it contributes only a fraction of the total outflow). 
This is much longer than the SN fading time, indicating that outflows from 
Milky-Way-type discs are not expected to have a strong effect on the 
filling-factor considerations.
% iso sims
Our isolated-galaxy simulations indeed show similar outflow timescales.
For example, the galaxy with 15\% gas fraction (mimicking $z=0$) has 
$t_{\rm out} \simeq 1.6\Gyr$ (and $\dot{M}_{\rm out} \simeq 0.7\, \SFR$),
while the galaxy with 50\% gas ($z \sim 2$) has $t_{\rm out} \simeq 300 \Myr$.
%\adr{More details on isolated sims?}

\smallskip %VELA
In the VELA cosmological simulations at $z \sim 1$, we measure the typical 
outflow timescales to be $t_{\rm out} \sim 100 \Myr$.  This is consistent with
the typical SFR and cosmological accretion rate being higher by an order of 
magnitude at $z \sim 1$ compared to $z=0$. 
This is still longer than the fading time for single SNe and for
SNe in low-mass clusters, but it is comparable to $\tf$ for massive star
clusters, indicating that SN blow-outs should be considered in the case of 
massive star-forming clusters.

\smallskip
One may ask whether accretion onto the disc may affect the considerations.
The typical cosmological accretion timescale (in the EdS regime, $z>1$ say) 
is \citep{dekel13}
\be
t_{\rm acc} = \frac{M}{\dot{M}}\sim 30 \Gyr\, (1+z)^{-5/2} .
\ee
This is larger than $1\Gyr$ for $z<3$, namely it is much longer than
the SN fading time, indicating that the effect of cosmological accretion 
is not directly relevant for our purposes.
On the other hand, the effect of recycled gas, returning after it outflew from
the disc, may be more important,
as indicated by comparing theoretical and observational specific SFR at
$z = 1-4$ \citep[e.g.][]{dm14}. 
Assuming an initial vertical outflow velocity on the order of the circular 
velocity, the return time would be comparable to the disc dynamical time,
typically on the order of $100 \Myr$. 
This indicates that the recycling should affect the disc gas budget 
in the case of SNe in massive star clusters.

%--------------------
\subsection{On the observed KS relation}
\label{sec:obs}

While our target seemed to be the slope $\slope \simeq 1.5$ of the macroscopic
relation, we note that
the observed value of the slope may deviate from $1.5$, 
in the range $\slope = 1-2$.
For example,
in patches of $\sim 1\kpc^2$ in nearby galaxies, 
the typical obtained slopes are $s \simeq 1$
\citep{leroy13,leroy17,utomo18}.
On the other hand,
referring to whole-galaxy averages, and including high-redshift and starburst
galaxies, the estimates range from $\slope \sim 1$ \citep{genzel11} 
to $\slope \sim 2$ \citep{faucher13}, partly depending on the assumed
$\alpha_{\rm CO}$, the CO-to-H$_2$ conversion factor.
A more complete data compilation consistently yields slopes in the range 
$\slope = 1-2$ \citep{krum18}.

\smallskip
As another example,
\citet{tacconi18}, who evaluated gas masses $\Mg$ for hundreds of
galaxies in the redshift range $z=0-4$, obtained a gradient of the depletion
time across
the Main Sequence of star-forming galaxies, at given stellar mass and redshift,
\be
\tdep = \frac{\Mg}{\Msfdot} \prop \Msfdot^{-\tau} \, ,
\ee
where $\Msfdot$ is the star-formation rate.
The best-fit slope from the data is $\tau = 0.44$
but the minimum uncertainty of $\pm 0.16$ dex, largely
reflecting systematics in the different analyses, allows values of
$\tau$ from below $0.3$ to above $0.6$.
The deduced KS relation is $\Msfdot \prop \Mg^{\slope}$ with
$\slope=1/(1-\tau)$, namely,
the best-fit KS slope is $\slope=1.79$, but values from below
$1.4$ to above $2.0$ can be accommodated.
We note that this data may mix relatively relaxed discs and (merger-induced?)
starbursts.

%%%%%%%%%%%%%%%%%%%%%%%%%
\section{Conclusion}
\label{sec:conc}

% KS from SN
We suggest that the relation between the macroscopic densities
of SFR and gas mass, the KS relation, 
may naturally arise from the way supernova bubbles evolve, 
independent of the details of the microscopic star-formation law, and with no
explicit dependence on gravity.

\smallskip %f=const
The key idea is that the filling factor $\fh$ of the SN bubbles in which SFR is 
suppressed is self-regulated into a constant value of order one half; a larger
(smaller) filling factor causes a slowdown (speedup) of the SFR and hence the 
SN rate, which reduces (increases) the filling factor. 
This has been demonstrated using both zoom-in cosmological simulations and
isolated-galaxy simulations, using different codes, subgrid recipes and
initial/boundary conditions. 

\smallskip % fn=const --> KS
Given the bubble fading radius and time, $\Rf(n)$ and $\tf(n)$, 
as a function of hydrogen number density in the ISM, $n$,
the filling factor of gas where SFR
is suppressed can be expressed in terms of the SN rate density, $S$,
as $\fh \prop S\, n^{-\slope}$.
With $\fh$ fixed at a constant value, this implies for the macroscopic
SFR density $\drhos \prop S \prop n^{\slope}$.

\smallskip % bubble physics -- > s~1.5
An analytic toy model based on the standard evolution of SN bubbles,
assumed to be spread at random in space, predicts a slope of $\slope=1.48$,
suspiciously close to the observed value.
We generalized the toy-model analysis to a sequence of SNe coincidental in 
space and of different patterns in time,
exploring different regimes in parameter space, using analytic modeling
and spherical simulations. While the analytic predictions for the slope
range from $\slope=1$ to $\slope=2$ in different cases, the slopes in the
simulations tend to be closer to $\slope \simeq 1.5$ in most cases.

\smallskip % epsf
When expressed in terms of the free-fall time in macroscopic volumes,
$\drhos=\epsf \rhog/\tff$, 
the model predicts that the efficiency factor $\epsf$ is independent of $n$,   
and is of order $0.01-0.02$, in the ball park of the observed values, both on
microscopic and macroscopic scales.

\smallskip % photo-bubbles
An analogous analytic toy model for suppressed SFR in photo-ionized bubbles 
around the massive stars that precede the SNe indicates slopes in a broader
range depending on the mass of the star-forming cluster. 
However, the ionized spheres collapse to the wind-driven 
super-bubbles after the first Gigayear, and the super-baubles are eventually
dominated by the SN energy for most of their lifetimes, indicating that the 
SN super-bubbles dominate the filling factor and determine the KS relation.

\smallskip % robustness to SF recipe
Our zoom-in cosmological simulations as well as our isolated-disc simulations
convincingly demonstrate the self-regulation to a constant hot filling factor 
in realistic galactic discs, and the associated generation of a KS relation.
The isolated-disc simulations, run with different local SFR recipes, 
demonstrate that the self-regulation to a constant hot filling
factor is indeed insensitive to the local SFR law, and so is the resultant
KS relation. 
The latter is consistent with the simulated results of 
\citet{hopkins11} and \citet{hopkins13} that the global SFR is robust to 
variations in the local SFR recipes over a broad range of local recipes.
Our simulations also indicate that the self-regulation is due to feedback, and
that SN feedback dominates over photo-ionization feedback.

\smallskip % too simple?
Our analytic modeling of the bubbles and their filling factor
is clearly a crude over-simplification in many ways. 
For example, we ignore the complex spatial structure of star-forming clouds,
the inhomogeneous nature of the ISM 
and the effects of neighboring bubbles running into each other. 
We address only in a crude way the interplay between the effects of 
pre-SN massive stars, through winds, photo-ionization and radiation pressure, 
and the evolution of SN bubbles. 
Among other shortcomings, we do not deal with the formation of molecular gas, 
ignore the ejection of gas from the galactic disc, and neglect the possible 
positive aspects of the feedback.
For the global evolution of the discs, we do not consider the accretion onto
the disc, the self-regulation of disc instability, the associated
instability-driven inflow within the disc, the generation of turbulence by the
above and by feedback, and the vertical balance between turbulence and gravity.
Despite all these shortcomings, the simple concept of self-regulation 
to a constant filling factor for the gas where star formation is suppressed,
combined with the evolution of SN bubbles based on common physics, 
gives rise in a trivial
way to a universal relation between the macroscopic SFR and gas content 
in galaxies or in patches within galaxies, similar to the observed KS relation.
Given the severe shortcomings of our model one may argue that it is too simple 
to be true.  On the other hand, its simplicity and robustness may indicate 
that it captures the essence of the origin of the KS relation.

\smallskip % confirmed by simulations
The confirmed validity of the self-regulation into a constant filling factor 
in the simulations, cosmological and isolated, and the demonstrated robustness 
of the macroscopic relation to the local star-formation recipe, provide
encouraging evidence for the validity of this basic scenario. It argues that
supernova feedback is the key factor in the origin of the KS relation.

\smallskip % future work
Future work may proceed in several routes.
First, one should improve the idealized analytic modeling for better agreement 
between the analytic predictions and the spherical simulations. 
Then, one may attempt to generalize the modeling to address some of the 
effects that were ignored in the current simplified analysis.
In particular, one should consider the alternative possibility of 
self-regulation by outflows and inflows.
Finally, one should analyze simulated galaxies in greater detail, e.g.,
to follow the evolution of super-bubbles and the way the self-regulation to a
constant filling factor is achieved.

%%%%%%%%%%%%%
\section*{Acknowledgments}

This work was inspired by an interaction with Jerry Ostriker. 
We are grateful for stimulating discussions 
with Siddhartha Gupta, Miao Li, Chris McKee, Brant Robertson and 
Amiel Sternberg.
This work was partly supported by the grants France-Israel PICS,
Germany-Israel GIF I-1341-303.7/2016, Germany-Israel DIP STE1869/2-1 
GE625/17-1,
I-CORE Program of the PBC/ISF 1829/12, ISF 857/14,
US-Israel BSF 2014-273, and NSF AST-1405962.
The cosmological VELA simulations were performed at the National Energy 
Research Scientific Computing Center (NERSC) at Lawrence Berkeley National 
Laboratory, and at NASA Advanced Supercomputing (NAS) at NASA Ames Research 
Center. Development and analysis have been performed in the astro cluster at 
HU.
The isolated-galaxy simulations were performed using the HPC resources of 
CINES and TGCC under the allocations A0030402192 and A0050402192 made by GENCI. 

%%%%%%%%%%%%%%%%%%
\bibliographystyle{mn2e}
\bibliography{ks}
%\bibliography{bn}

%%%%%%%%%%%%%%%%%%%%%%%%%%%%%%%%%%
\appendix

%%%%%%%%%
\section{The VELA Cosmological Simulations}
\label{sec:app_vela}

The VELA suite consists of hydro-cosmological simulations of 35 moderately 
massive galaxies. Full details are presented in \citet{ceverino14,zolotov15}.
This suite has been used to study central issues in the evolution of galaxies
at high redshifts, including compaction to blue nuggets and the trigger of
quenching
\citep{zolotov15,tacchella16_ms,tacchella16_prof}, 
evolution of global shape
\citep{ceverino15_e,tomassetti16},
violent disc instability \citep{mandelker14,mandelker17},
OVI in the CGM \citep{roca19},
and galaxy size and angular momentum \citep{jiang19_spin}.
Additional analysis of the same suite of simulations are discussed in 
\citet{moody14,snyder15}. 
In this appendix we give an overview of the key aspects of the simulations 
and their limitations.

%--------------
\subsection{The Cosmological Simulations}

The VELA simulations make use of the Adaptive Refinement Tree (ART) code 
\citep{krav97,krav03,ceverino09}, which accurately follows the 
evolution of a gravitating N-body system and the Eulerian gas dynamics using 
an adaptive mesh refinement approach. The adaptive mesh refinement maximum
resolution is $17-35\pc$ at all times, which is achieved at densities of 
$\sim10^{-4}-10^3\cmc$. 
Beside gravity and hydrodynamics, the code incorporates physical process 
relevant for galaxy formation such as gas cooling by atomic hydrogen and 
helium, metal and molecular hydrogen cooling, photoionization heating by the 
UV background with partial self-shielding, star formation, stellar mass loss, 
metal enrichment of the ISM and stellar feedback. Supernovae and stellar winds 
are implemented by local injection of thermal energy as described in 
\citet{ceverino09,cdb10} and \citet{ceverino12}. Radiation-pressure
stellar feedback is implemented at a moderate level, following 
\citet{dekel13}, as described in \citet{ceverino14}.

\smallskip
Cooling and heating rates are tabulated for a given gas density, temperature, 
metallicity and UV background based on the CLOUDY code \citep{ferland98}, 
assuming a slab of thickness $1\kpc$. A uniform UV background based on the 
redshift-dependent \citet{haardt96} model is assumed, except at gas densities 
higher than $0.1\cmc$, where a substantially suppressed UV 
background is used 
($5.9\times10^6 \erg\, {\rm s}^{-1} \cms\, {\rm Hz}^{-1}$) 
in order to mimic the partial self-shielding of dense gas, allowing dense gas 
to cool down to temperatures of $\sim300$K. The assumed equation of 
state is that of an ideal mono-atomic gas. Artificial fragmentation on the cell
size is prevented by introducing a pressure floor, which ensures that the Jeans 
scale is resolved by at least 7 cells \citep[see][]{cdb10}.

\smallskip
Star particles form in timesteps of $5 \Myr$ in cells where the gas density 
exceeds a threshold of $1~\cmc$ and the temperatures is below 
$10^4$K. Most stars ($>90\%$) end up forming at temperatures well 
below $10^3$K, and more than half of the stars form near  
$300$K in cells where the gas density is higher than $10~\cmc$. 
The code implements a stochastic star-formation where a star particle with a
mass of $42\%$ of the gas mass forms with a probability
$P=(\rhog/10^3\cmc)^{1/2}$ but not higher than $0.2$. 
This corresponds to a local SFR that crudely mimics 
$\drhos \epsf \rhog/\tff$ with $\epsf \sim 0.02$. 
A stellar initial mass function of \citet{chabrier03} is assumed. 

\smallskip
Thermal feedback that mimics the energy release from stellar winds and 
supernova explosions s incorporated as a constant heating rate over 
the $40~\Myr$ following star formation. 
A velocity kick of $\sim10\kms$ is applied 
to $30~\%$ of the newly formed stellar particles -- this enables SN explosions
in lower density regions where the cooling may not overcome the heating without
implementing an artificial shutdown of cooling \citep{ceverino09}.
The code also incorporates the later effects of Type Ia supernova and 
stellar mass loss, and it follows the metal enrichment of the ISM. 

\smallskip
Radiation pressure is incorporated through the addition of a non-thermal 
pressure term to the total gas pressure in regions where ionizing photons 
from massive stars are produced and may be trapped. This ionizing radiation 
injects momentum in the cells neighbouring massive star particles younger than 
$5\Myr$, and whose column density exceeds 
$10^{21}\cms$, isotropically pressurizing the star-forming 
regions \citep[see more details in][]{agertz13,ceverino14}.

\smallskip
The initial conditions for the simulations are based on dark-matter haloes that
were drawn from dissipationless N-body simulations at lower resolution in 
cosmological boxes of $15-60\Mpc$. The $\Lambda$CDM cosmological model was
assumed with the WMAP5 values of the cosmological parameters, 
$\omm=0.27$, $\oml=0.73$, $\omb=0.045$, $h=0.7$ and 
$\sigma_8=0.82$ \citep{komatsu09}. Each halo was selected to have a
given virial mass at $z = 1$ and no ongoing major merger at $z=1$.
This latter criterion eliminated less than $10~\%$ of the haloes, those
that tend to be in a dense, proto-cluster environment at $z\sim1$. 
The virial masses at $z=1$ were chosen to be in the range 
$\Mv=2\times10^{11}-2\times10^{12}~M_{\odot}$, about a
median of $4.6\times10^{11}~M_{\odot}$. If left in isolation, the median mass 
at $z=0$ was intended to be $\sim10^{12}~M_{\odot}$. 

\smallskip % limitations
The VELA cosmological simulations are state-of-the-art in terms 
of high-resolution adaptive mesh refinement hydrodynamics and the treatment of 
key physical processes at the subgrid level. 
In particular, they trace the cosmological streams that feed galaxies at high 
redshift, including mergers and smooth flows, and they resolve the violent disc
instability that governs high-$z$ disc evolution and bulge formation 
\citep{cdb10,ceverino12,ceverino15_e,mandelker14}.
Like in other simulations, the treatments of star formation and feedback 
processes are rather simplified. The code may assume a realistic SFR efficiency 
per free fall time on the grid scale 
but it does not follow in detail the formation of 
molecules and the effect of metallicity on SFR. 
The feedback is treated in a crude way, where the resolution does not allow 
the capture of the Sedov-Taylor phase of supernova bubbles. 
The radiative stellar feedback assumed 
no infrared trapping, in the spirit of low trapping advocated by 
\citet{dk13} based on \citet{krum_thom13},
which makes the radiative feedback weaker than in other simulations
that assume more significant trapping \citep{murray10,hopkins12b}. 
Finally, AGN feedback, and feedback associated with cosmic rays and magnetic 
fields, are not yet implemented. Nevertheless, as shown in 
\citet{ceverino14}, the star formation rates, gas
fractions, and stellar-to-halo mass ratio are all in the ballpark of the 
estimates deduced from observations.

%-----------------
\subsection{The Galaxy Sample and Measurements}\label{subsec:sample}

\begin{table*}
\centering
\begin{tabular}{@{}lcccccccccccccc}
\multicolumn{15}{c}{{\bf Properties of the VELA galaxies}} \\
\hline
Galaxy & $\Mv$ & $\Ms$ & $\Mg$ & SFR & $\Re$ & $\Rd$ & $\Hd$ & & $n_{\rm thin}$
& SFR$_{\rm thin}$ & $f_{\rm hot,thin}$ & & 
$a_{\rm in,disc}$ & $a_{\rm fin,disc}$  \\
  & $10^{12}\msun$ & $10^{10}\msun$ & $10^{10}\msun$ & $\msun/\yr$ & kpc & kpc &
 kpc & & $ {\rm cm}^{-3}$ & $\msun /\yr$ & & & \\
\hline
\hline
V01 & 0.16 & 0.22 & 0.24 & 2.65 & 0.93 & 5.15 & 2.57 & & 0.115 & 1.39 & 0.69  
& & 0.38 & 0.38 \\
V02 & 0.13 & 0.19 & 0.31 & 1.83 & 1.81 & 6.37 & 3.57 & & 0.039 & 0.26 & 0.74  
& & - & - \\
V03 & 0.14 & 0.43 & 0.18 & 3.74 & 1.41 & 5.21 & 2.34 & & 0.212 & 2.92 & 0.58  
& & - & - \\
V04 & 0.12 & 0.10 & 0.17 & 0.48 & 1.73 & 5.71 & 2.79 & & 0.036 & 0.12 & 0.49  
& & 0.48 & 0.48 \\
V05 & 0.07 & 0.10 & 0.15 & 0.59 & 1.81 & 5.36 & 1.98 & & 0.014 & 0.11 & 0.60  
& & - & - \\
V06 & 0.55 & 2.22 & 0.52 & 20.63 & 1.05 & 2.53 & 0.42 & & 4.155 & 16.33 & 0.61 
& & 0 .17 & 0.33 \\
V07 & 0.90 & 6.37 & 1.98 & 26.66 & 2.85 & 12.59 & 2.06 & & 0.143 & 4.67 & 0.65
& & 0.20 & 0.54 \\
V08 & 0.28 & 0.36 & 0.32 & 5.76 & 0.74 & 4.03 & 1.53 & & 0.542 & 4.58 & 0.81  
& & 0.45 & 0.57 \\
V09 & 0.27 & 1.07 & 0.49 & 3.94 & 1.74 & 7.34 & 2.12 & & 0.121 & 0.91 & 0.57  
& & 0.29 & 0.40 \\
V10 & 0.13 & 0.64 & 0.21 & 3.27 & 0.46 & 4.51 & 1.19 & & 0.396 & 2.07 & 0.59  
& & 0.27 & 0.56 \\
V11 & 0.27 & 1.02 & 0.86 & 17.18 & 2.14 & 8.34 & 5.08 & & 0.115 & 1.85 & 0.79  
& & 0.29 & 0.46 \\
V12 & 0.27 & 2.06 & 0.30 & 2.90 & 1.13 & 6.53 & 1.72 & & 0.079 & 0.65 & 0.61  
& & 0.20 & 0.44 \\
V13 & 0.31 & 0.96 & 1.34 & 21.20 & 2.48 & 9.74 & 4.75 & & 0.088 & 1.47 & 0.66  
& & 0.36 & 0.39 \\
V14 & 0.36 & 1.40 & 0.79 & 27.50 & 0.32 &  -   &  -   & &   -   &   -  & -
& & 0.37 & 0.41 \\
V15 & 0.12 & 0.56 & 0.23 & 1.71 & 1.07 & 6.26 & 1.08 & & 0.096 & 0.82 & 0.57  
& & 0.30 & 0.51 \\
V16 & - & - & - & - & - & - & - & & - & - & - & & 0.14 & 0.24 \\
V17 & - & - & - & - & - & - & - & & - & - & - & & 0.15 & 0.31 \\
V18 & - & - & - & - & - & - & - & & - & - & - & & - & - \\
V19 & - & - & - & - & - & - & - & & - & - & - & & 0.14 & 0.29 \\
V20 & 0.53 & 3.92 & 0.73 & 7.26 & 1.72 & 9.57 & 2.75 & & 0.053 & 1.09 & 0.76  
& & 0.11 & 0.44 \\
V21 & 0.62 & 4.28 & 0.92 & 9.80 & 1.73 & 9.48 & 1.18 & & 0.191 & 2.78 & 0.56  
& & 0.20 & 0.49 \\
V22 & 0.49 & 4.57 & 0.26 & 12.08 & 1.31 & 4.70 & 0.40 & & 1.270 & 9.41 & 0.58  
& & 0.16 & 0.50 \\
V23 & 0.15 & 0.84 & 0.32 & 3.37 & 1.16 & 6.28 & 1.54 & & 0.167 & 2.03 & 0.60  
& & 0.33 & 0.46 \\
V24 & 0.28 & 0.95 & 0.49 & 4.39 & 1.68 & 7.29 & 1.95 & & 0.071 & 0.82 & 0.60  
& & 0.37 & 0.48 \\
V25 & 0.22 & 0.76 & 0.17 & 2.31 & 0.73 & 5.70 & 0.82 & & 0.173 & 2.09 & 0.62  
& & 0.32 & 0.50 \\
V26 & 0.36 & 1.63 & 0.44 & 9.66 & 0.74 & 5.42 & 1.30 & & 0.472 & 4.39 & 0.69  
& & 0.26 & 0.50 \\
V27 & 0.33 & 0.90 & 0.90 & 8.69 & 1.98 & 9.16 & 4.97 & & 0.085 & 3.57 & 0.74  
& & 0.37 & 0.50 \\
V28 & 0.20 & 0.27 & 0.38 & 5.72 & 2.32 & 5.66 & 2.97 & & 0.160 & 2.74 & 0.74  
& & 0.41 & 0.50 \\
V29 & 0.52 & 2.67 & 0.54 & 18.75 & 1.89 & 7.46 & 0.97 & & 0.270 & 6.37 & 0.78  
& & 0.26 & 0.50 \\
V30 & 0.31 & 1.71 & 0.66 & 3.85 & 1.43 & 9.32 & 1.67 & & 0.051 & 0.72 & 0.61  
& & 0.19 & 0.33 \\
V31 & - & - & - & - & - & - & - & & - & - & - & & 0.17 & 0.19 \\
V32 & 0.59 & 2.74 & 0.60 & 15.00 & 2.58 & 4.98 & 1.06 & & 0.822 & 4.27 & 0.60  
& & 0.16 & 0.33 \\
V33 & 0.83 & 5.17 & 0.59 & 32.74 & 1.23 & 4.59 & 0.88 & & 1.384 & 17.02 & 0.72
& & 0.19 & 0.39 \\
V34 & 0.52 & 1.73 & 0.67 & 14.69 & 1.84 & 5.29 & 1.87 & & 0.629 & 6.03 & 0.68  
& & - & - \\
V35 & - & - & - & - & - & - & - & & - & - & - & & - & - \\
\hline
\end{tabular}
 \caption{Global properties of the VELA 3 galaxies.
The quantities are quoted at $z=2$.
$\Mv$ is the total virial mass. 
The following four quantities are measured within $0.2\Rv$:
$\Ms$ is the stellar mass,  
$\Mg$ is the gas mass,
SFR is the star formation rate,
and $\Re$ is the half-stellar-mass radius. 
The disc outer volume, as defined in \citet{mandelker14}, is given by
$\Rd$ and $\Hd$, the disc radius and half height. 
The following three quantities refer to the thin disc as analyzed here,
a cylinder of radius $0.8\Rd$ and half-height $0.25\kpc$:
$n_{\rm thin}$ is the mean density of cold gas ($T<3\times 10^4$K), 
SFR$_{\rm thin}$ is the star formation rate,
and $f_{\rm hot,thin}$ is the volume filling factor of hot gas 
($T>3\times 10^4$K).
The simulations of most galaxies end near $z\sim 1$, 
but galaxies V16-19,31,and 35 do not reach $z=2$.
$a_{\rm in,disc}$ and $a_{\rm fin,disc}$ are the earliest and latest
cosmological expansion factors within which the galaxy has a cold disc with 
$\Rd/\Hd>4$. Galaxies V02,03,05,18,34, and 35 do not have a disc phase
within the simulated period.}
\label{tab:sample}
\end{table*}

The virial and stellar properties of the galaxies 
are listed in Table~\ref{tab:sample}. 
The virial mass $\Mv$ is the total mass within a sphere of radius 
$\Rv$ that encompasses an overdensity of $\Delta(z)=
[18\pi^2-82\oml(z)-39\oml(z)^2]/\omm(z)$, 
where $\oml(z)$ and $\omm(z)$ are the cosmological
parameters at $z$ \citep{bryan98,db06}. The stellar mass $\Ms$ 
is measured within a radius of $0.2\Rv$.

\smallskip
We start the analysis at the cosmological time corresponding to expansion factor
 $a=0.125$ (redshift $z=7$). 
As can be seen in Table~\ref{tab:sample}, most galaxies reach $a=0.50$ ($z=1$).
Each galaxy is analyzed at output times separated by a constant interval in 
$a$, $\Delta a=0.01$, corresponding at $z=2$ to $\sim100~\Myr$ 
(roughly half an orbital time at the disc edge). 
The sample consists of totally $\sim 1000$ snapshots in the redshift range 
$z=6-1$ from 35 galaxies that at $z = 2$ span the stellar mass range 
$(0.2-6.4)\times10^{11}\Msun$. The half-mass sizes 
$\Re$ are determined from the $\Ms$ that are measured within a 
radius of $0.2\Rv$ and they range $\Re\simeq0.4-3.2\kpc$ at $z=2$.

\smallskip
The SFR for a simulated galaxy is obtained by 
${\rm SFR}=\langle M_{\star}(t_ {\rm age}<t_{\rm max})/t_{\rm max} 
\rangle_{t_{\rm max}}$, where $\Ms(t_{\rm age}<t_{\rm max})$ is the mass 
at birth in stars younger than $t_{\rm max}$.  
The average $\langle\cdot\rangle_{t_{\rm max}}$ is obtained by averaging over 
all $t_{\rm max}$ in the interval $[40,80]\Myr$ in steps of $0.2\Myr$.
The $t_{\rm max}$ in this range are long enough to ensure good statistics. 
The SFR ranges from $\sim 1$ to $33 \Msun\yr^{-1}$
at $z\sim2$.

\smallskip
The instantaneous mass of each star particle is derived from its initial mass 
at birth and its age using a fitting formula for the mass loss from the stellar
population represented by the star particle,
according to which 10\%, 20\% and 30\% of the mass is lost after 
30 Myr, 260 Myr , and 2 Gyr from birth, respectively.
We consistently use here the instantaneous stellar mass, $\Ms$, and 
define the specific SFR by 
${\rm sSFR}={\rm SFR}/\Ms$.

\smallskip
The determination of the centre of the galaxy is outlined in detail in 
Appendix B of \citet{mandelker14}. 
Briefly, starting form the most bound star, the centre is refined iteratively
  by calculating the centre of mass of stellar particles in spheres of 
decreasing radii, updating the centre and decreasing the radius at each 
iteration.  We begin with an initial radius of 600 pc, and decrease the radius 
by a factor of 1.1 at each iteration. The iteration terminates when the radius 
reaches 130 pc or when the number of stellar particles in the sphere drops 
below 20. 

\smallskip
%Disk definition from Mandelker. Values in the table}
The disc plane and dimensions are determined iteratively, as detailed in 
\citet{mandelker14}. The disc axis is defined by the angular 
momentum of cold gas ($T < 1.5 \times 10^{4}$K), which on average accounts for 
$\sim 97\%$ of the total gas mass in the disc. 
The radius $\Rd$ is chosen to contain $85\%$ of the cold gas mass in the 
galactic mid-plane out to $0.15\Rv$, and the half-height $\Hd$ is defined to
encompass $85\%$ of the cold gas mass in a thick cylinder where both the
radius and half-height equal $\Rd$.  

\smallskip % table
Relevant global properties of the VELA 3 galaxies at $z=2$ are listed
in \tab{sample} and explained in the caption. It includes the global
masses and sizes of the different components, and the quantities within the
thin discs analyzed that are relevant for the KS relation and the hot filling
factor.

\label{lastpage}
\end{document}